\documentclass[10pt]{article} 
\usepackage{amssymb,amsmath,amsfonts,theorem,subfigure} 
\usepackage{graphicx}
\usepackage{epic,curves}
\usepackage{multirow}
\usepackage{pstricks}
\usepackage{array}
\usepackage[usenames]{pstcol}
\usepackage[normalem]{ulem}

\newcommand{\lw}{\psset{linewidth=0.5pt}}
\newcommand{\unlw}{\psset{linewidth=1pt}}
\newrgbcolor{myc}{1 0 0}
\newrgbcolor{myc2}{0 0 1}
\newrgbcolor{myc3}{1 0 1}
\newrgbcolor{myc4}{0.588632 0.133333 1.}
\newrgbcolor{myc5}{0.0791028 0.666667 0}

\theorembodyfont{\itshape} 
\theoremheaderfont{\scshape}
\theoremstyle{plain}  
\newtheorem{Lemme}{Lemma}[section]
\newtheorem{Theoreme}[Lemme]{Theorem}
\newtheorem{Proposition}[Lemme]{Proposition}
\newtheorem{Corollaire}[Lemme]{Corollary}
\newtheorem{Definition}{Definition}[section]

\newcommand{\wj}[1]{W\hskip -0.25em J_{#1}}

\definecolor{forestgreen}{rgb}{0.13,0.54,0.13}

\begin{document}

%%%%%%%%%%%%%%%%%%%%%
%
% Title page
%
%%%%%%%%%%%%%%%%%%%%%

\title{A homomorphism between link and XXZ modules \\ over the periodic Temperley-Lieb algebra}
\author{
{Alexi Morin-Duchesne}\footnote{\ttfamily a.morinduchesne{\char'100}uq.edu.au}, \\
\it School of Mathematics and Physics\\ 
\it University of Queensland, St Lucia, Brisbane\\
\it Queensland 4072, Australia\\[10pt]
{Yvan Saint-Aubin}\footnote{\ttfamily yvan.saint-aubin{\char'100}umontreal.ca}\\
\it D\'epartement de math\'ematiques et de statistique\\ 
\it Universit\'e de Montr\'eal, C.P.\ 6128, succ.\ centre-ville, Montr\'eal\\ 
\it Qu\'ebec, Canada, H3C 3J7\\[10pt]}

\maketitle 

%%%%%%%%%%%%%%%%%%%%%
%
% Abstract
%
%%%%%%%%%%%%%%%%%%%%%
 
\begin{abstract}

We study finite loop models on a lattice wrapped around a cylinder. A section of the cylinder has $N$ sites. We use a family of link modules over the periodic Temperley-Lieb algebra $\mathcal EPTL_N(\beta, \alpha)$ introduced by Martin and Saleur, and Graham and Lehrer. These are labeled by the numbers of sites $N$ and of defects $d$, and extend the standard modules of the original Temperley-Lieb algebra. Beside the defining parameters $\beta=u^2+u^{-2}$ with $u=e^{i\lambda/2}$ (weight of contractible loops) and $\alpha$ (weight of non-contractible loops), this family also depends on a {\em twist parameter} $v$ that keeps track of how the defects wind around the cylinder.  The transfer matrix $T_N(\lambda, \nu)$ depends on the anisotropy 
$\nu$ and the spectral parameter $\lambda$ that fixes the model. 
(The thermodynamic limit of $T_N$ is believed to describe a conformal field theory of central charge $c=1-6\lambda^2/(\pi(\lambda-\pi))$.)

The family of periodic XXZ Hamiltonians is extended to depend on this new parameter $v$ and the relationship between this family and the loop models is established. The Gram determinant for the natural bilinear form on these link modules is shown to factorize in terms of an intertwiner $\tilde i_N^d$ between these link representations and the eigenspaces of $S^z$ of the XXZ models. This map is shown to be an isomorphism for generic values of $u$ and $v$ and the critical curves in the plane of these parameters for which $\tilde i_N^d$ fails to be an isomorphism are given.
\\


\end{abstract}

%%%%%%%%%%%%%%%%%%%%%
%
% TOC
%
%%%%%%%%%%%%%%%%%%%%%

\tableofcontents

%%%%%%%%%%%%%%%%
%
\section{Introduction} \label{sec:intro}
%
%%%%%%%%%%%%%%%%

In part due to Onsager's remarkable solution of the two-dimensional Ising model \cite{Onsager}, the transfer matrix formalism has played a central role in the study of
many problems in statistical physics. The $Q$-state Potts model, ice-type models and the dimer model \cite{TL, Baxter, Lieb} are the first examples that spring to mind. Other models, like the XXZ spin chain,  
are described in terms of their 
Hamiltonian matrix, an evolution operator that shares many properties with transfer matrices. In these original formulations, the transfer matrix or Hamiltonian
is a linear operator that acts on states of $N$ spins in a chain, i.e.~on a tensor product of $N$ copies of $\mathbb C^2$.

For some of these models, the evolution operator turns out to be a matrix realization of a particular element of the Temperley-Lieb algebra $TL_N(\beta)$, a one-parameter family of associative
algebras. Recently, Pearce, Rasmussen and Zuber \cite{PRZ} have introduced a family of transfer matrices, the {\it double-row transfer matrices}, that live on the geometry of the strip with finitely many sites and are defined as elements of the abstract algebra $TL_N(\beta)$. There exists a natural action of the Temperley-Lieb algebra on link modules that defines physically relevant representations of this algebra. Partition functions for the $Q$-state Potts model are computable from the eigenvalues of the matrix realization of the transfer matrices at $\beta = \sqrt Q$ \cite{JacobsenSaleur}. For almost all values of $\beta$, finding the spectrum of the double-row transfer matrix is an open problem, but nevertheless, the double-row transfer matrix formulation has the advantage that the sizes of the representations are independent of $Q$. Whereas the representations in the spin models give rise to symmetric matrix realizations of the transfer matrices which are then diagonalizable, their representative in the link representations exhibit rank $2$ Jordan cells if certain conditions on $\beta$ are satisfied \cite{PRZ,AMDSA}. This is a remarkable feature, as Jordan cells in the transfer matrix of the finite model are a signature, in the scaling limit, of logarithmic conformal field theory. 

For the geometry of the cylinder, the relevant algebra is the enlarged periodic Temperley-Lieb algebra $\mathcal EPTL_N(\beta, \alpha)$, a two-parameter algebra whose representation theory was studied by Martin and Saleur \cite{MartinSaleur}, Graham and Lehrer \cite{Lehrer}, and Green and Erdmann \cite{Greenseul, ErdmannGreen}. For $\beta = 0$, Pearce, Rasmussen and Villani \cite{PRV} introduced the {\it single-row transfer matrix}, an element of this algebra, and extended the definition of link modules to this geometry. In these representations,
the single-row transfer matrix is non-symmetric, opening up the possibility of non-trivial Jordan structures. For critical dense polymers ($\beta = 0$), exact expressions for the eigenvalues were found, the existence of Jordan blocks was observed
for small system size
and a modular invariant was found that is believed to correspond to the partition function of critical dense polymers on the torus \cite{PRV}.

The partition function and Jordan structure of the transfer matrix are of course not the only relevant properties one can hope to extract from the transfer matrix. Periodic boundary conditions are of course relevant for cylindrical and toroidal geometries. Then winding properties of clusters as they evolve along the cylinder should be somehow computable from this transfer matrix. If the two ends of a finite cylinder are joined, the homotopy properties of the clusters on the torus may also be considered \cite{Pinson, Arguin}.  However the link modules used to describe loop models, 
also known as standard modules, do not keep track of the winding of loops. One way to circumvent this problem was proposed by Richard and Jacobsen \cite{RichardJacobsen}, but we prefer another one here. We choose to use modules \cite{MartinSaleur,Lehrer} that depend on one further parameter, the {\em twist parameter} $v$, which is related to the winding of loops. In this paper, we start putting together tools to study these modified link modules.

The spin and link modules over the Temperley-Lieb algebra, though different, are not completely unrelated. On the strip, a homomorphism $i_N^d$ that maps link modules into subspaces of the spin modules led to a proof of selection rules for critical dense polymers on the strip \cite{AMDtoutseul}. In this work, we will construct a homomorphism (or intertwiner), hereafter denoted $\tilde i_N^d$, from link modules into submodules 
of an extended XXZ spin module (that depends upon the parameter $v$). This will prove particularly useful because, for generic values of the parameters, this map is
an isomorphism of modules. This tool is intimately related to another one, the Gram matrix. There is a bilinear form on link modules, the Gram product, that is invariant under the action of the Temperley-Lieb algebra. The Gram matrix represents this bilinear form in the link state basis. The radical of this bilinear form, that is the set of vectors that have a zero pairing with any other, is easily shown to be a submodule. It is non-trivial if and only if the determinant of the Gram matrix vanishes \cite{GLcell, Westbury, DRYSA}.  
In fact, we will show that the Gram matrix can be written as a product of two copies of $I_N^d$, the matrix representing $\tilde i_N^d$ in the natural bases. 
A non-trivial Jordan structure is often related to the radical being non-zero and it is natural to ask whether a non-trivial kernel of $\tilde i_N^d$ may provide similar information. This question will be studied in a subsequent paper but, in the present one, the conditions on the parameters $\beta$ and $v$ for this kernel to be non-trivial are obtained. 

The paper is organized as follows. Section \ref{sec:TLP} defines the periodic Temperley-Lieb algebras, the transfer matrix for the loop models and the family of representations depending on the {\em twist parameter}. Section \ref{sec:XXZprime} extends the periodic XXZ Hamiltonians into a family that, like the representations, depends on one more parameter. The linear map $\tilde i_N^d$, between link modules and the vector spaces upon which the XXZ Hamiltonians act, is then introduced. The factorization of the Gram matrix in term of the map $\tilde i_N^d$ is proven there. Section \ref{sec:Gram} gives a new computation of the Gram determinant, one that leads to the identification of the critical curves where the map $\tilde i_N^d$ becomes singular. This is where the proof that $\tilde i_N^d$ is actually an homomorphism ends. We will soon come back to these tools to show how they allow to unravel, at least partially, the Jordan structure of the periodic loop transfer matrix.

%%%%%%%%%%%%%%%%
%
\section{Periodic Temperley-Lieb algebras and loop models} \label{sec:TLP}
%
%%%%%%%%%%%%%%%%

\subsection{Periodic Temperley-Lieb algebras}

On a vertical cylinder, we draw $N$ equidistant points (or entries) on each of two parallel sections and label them $1$ to $N$. A connectivity is a set of $N$ curves connecting points pairwise by non-intersecting curves. Two connectivities are equivalent (or equal) if the curves of the first can be continuously deformed into those of the second. Clearly a rotation of $2\pi/N$ of the cylinder maps a connectivity onto another; these two connectivities, the original and the rotated, are usually distinct as their patterns of connections are then different. 

Throughout this article we will depict connectivities by planar diagrams on a periodic strip, as in the following:
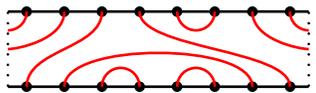
\begin{equation*}
\psset{unit=0.5}
\begin{pspicture}(-0.5,-0.2)(8.5,0.9)
\psline[linewidth=1pt]{-}(0.5,1)(8.5,1)
\psline[linewidth=1pt,linestyle=dotted]{-}(0.5,1)(0.5,-1)
\psline[linewidth=1pt,linestyle=dotted]{-}(8.5,1)(8.5,-1)
\psline[linewidth=1pt]{-}(8.5,-1)(0.5,-1)
\psset{linewidth=1pt}
\psdots(1,1)(2,1)(3,1)(4,1)(5,1)(6,1)(7,1)(8,1)(1,-1)(2,-1)(3,-1)(4,-1)(5,-1)(6,-1)(7,-1)(8,-1)
\psset{linecolor=myc}
\psarc(3.5,-1){0.5}{0}{180}
\psarc(5.5,-1){0.5}{0}{180}
\psbezier{-}(2,-1)(2,0.2)(7,0.2)(7,-1)
\psarc(0.5,1){0.5}{-90}{0}
\psarc(5.5,1){0.5}{180}{360}
\psarc(8.5,1){0.5}{180}{270}
\psbezier{-}(3,1)(3,0)(1,0)(1,-1)
\psbezier{-}(2,1)(2,0.25)(0.75,0)(0.5,0)
\psbezier{-}(7,1)(7,0.25)(8.25,0)(8.5,0)
\psbezier{-}(4,1)(4,0)(8,0)(8,-1)
\end{pspicture}\, \,. \vspace{0.3cm}
\end{equation*}
The leftmost point on the top and bottom slices bears the label $1$, the rightmost the label $N$. The cut along a line parallel to the cylinder axis that allows this planar representation is depicted using dotted lines at $x=1/2$ and $x=N+1/2$. These lines will be called {\em imaginary boundaries}.

A product between two connectivities $c_1$ and $c_2$ with $N$ entries is now defined.  The product $c_1c_2$ is obtained by drawing $c_2$ on top of $c_1$ and connecting the $N$ points on the bottom of $c_2$ with those at the top of $c_1$. The result is the connectivity obtained from this new diagram, with a multiplicative factor of $\beta^{n_\beta}\alpha^{n_\alpha}$ where $n_\beta$ and $n_\alpha$ are respectively the numbers of contractible and non-contractible loops closed in the process. For instance, \vspace{-0.3cm}
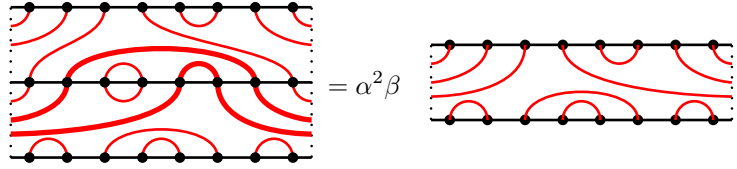
\begin{equation*}
\psset{unit=0.5}
\begin{pspicture}(-0.5,-1.3)(8.5,0.4)
\psline[linewidth=1pt]{-}(0.5,1)(8.5,1)
\psline[linewidth=1pt,linestyle=dotted]{-}(0.5,1)(0.5,-3)
\psline[linewidth=1pt,linestyle=dotted]{-}(8.5,1)(8.5,-3)
\psline[linewidth=1pt]{-}(8.5,-1)(0.5,-1)
\psline[linewidth=1pt]{-}(8.5,-3)(0.5,-3)
\psset{linewidth=1pt}
\psset{linecolor=myc}
\psarc(3.5,-1){0.5}{0}{180}
\psarc(0.5,1){0.5}{-90}{0}
\psarc(5.5,1){0.5}{180}{360}
\psarc(8.5,1){0.5}{180}{270}
\psbezier{-}(3,1)(3,0)(1,0)(1,-1)
\psbezier{-}(2,1)(2,0.25)(0.75,0)(0.5,0)
\psbezier{-}(4,1)(4,0)(8,0)(8,-1)
\psbezier{-}(7,1)(7,0.25)(8.25,0)(8.5,0)
\psarc(3.5,-1){-0.5}{0}{180}
\psarc(8.5,-1){0.5}{180}{270}
\psarc(0.5,-1){0.5}{-90}{0}
\psarc(1.5,-3){0.5}{0}{180}
\psarc(4.5,-3){0.5}{0}{180}
\psarc(7.5,-3){0.5}{0}{180}
\psbezier[linewidth=2pt]{-}(2,-1)(2,0.2)(7,0.2)(7,-1)
\psbezier[linewidth=2pt]{-}(2,-1)(2,-1.75)(0.75,-2)(0.5,-2)
\psbezier[linewidth=2pt]{-}(7,-1)(7,-1.75)(8.25,-2)(8.5,-2)
\psbezier{-}(3,-3)(3,-2)(6,-2)(6,-3)
\psarc[linewidth=2pt](5.5,-1){0.5}{0}{180}
\psbezier[linewidth=2pt]{-}(5,-1)(5,-2.25)(1,-2.375)(0.5,-2.375)
\psbezier[linewidth=2pt]{-}(6,-1)(6,-2.25)(8,-2.375)(8.5,-2.375)
\psset{linecolor=black}
\psdots(1,1)(2,1)(3,1)(4,1)(5,1)(6,1)(7,1)(8,1)(1,-1)(2,-1)(3,-1)(4,-1)(5,-1)(6,-1)(7,-1)(8,-1)(1,-3)(2,-3)(3,-3)(4,-3)(5,-3)(6,-3)(7,-3)(8,-3)
\end{pspicture}\, \, = \alpha^2 \beta 
\begin{pspicture}(-0.3,-2.3)(8.5,0.4)
\psline[linewidth=1pt,linestyle=dotted]{-}(0.5,-1)(0.5,-3)
\psline[linewidth=1pt,linestyle=dotted]{-}(8.5,-1)(8.5,-3)
\psline[linewidth=1pt]{-}(8.5,-1)(0.5,-1)
\psline[linewidth=1pt]{-}(8.5,-3)(0.5,-3)
\psset{linewidth=1pt}
\psdots(1,-1)(2,-1)(3,-1)(4,-1)(5,-1)(6,-1)(7,-1)(8,-1)(1,-3)(2,-3)(3,-3)(4,-3)(5,-3)(6,-3)(7,-3)(8,-3)
\psset{linecolor=myc}
\psarc(5.5,-1){-0.5}{0}{180}
\psarc(8.5,-1){0.5}{180}{270}
\psarc(0.5,-1){0.5}{-90}{0}
\psarc(1.5,-3){0.5}{0}{180}
\psarc(4.5,-3){0.5}{0}{180}
\psarc(7.5,-3){0.5}{0}{180}
\psbezier{-}(2,-1)(2,-1.75)(0.75,-2)(0.5,-2)
\psbezier{-}(7,-1)(7,-1.75)(8.25,-2)(8.5,-2)
\psbezier{-}(3,-3)(3,-2)(6,-2)(6,-3)
\psbezier{-}(3,-1)(3,-2.25)(1,-2.375)(0.5,-2.375)
\psbezier{-}(4,-1)(4,-2.25)(8,-2.375)(8.5,-2.375)
\end{pspicture}\, \, . 
\end{equation*} \vspace{0.2cm}

\noindent The two non-contractible loops, responsible for the factor $\alpha^2$, are drawn thicker here. Because curves can wind around the cylinder indefinitely, the number of connectivities is infinite.

\begin{Definition}
The algebra $\mathcal{PTL}_N(\beta,\alpha)$ is the vector space generated by connectivities and endowed with the product just defined and extended linearly to linear combinations of connectivities. The unit in $\mathcal{PTL}_N(\beta,\alpha)$ is the connectivity that connects the point $i$ on the bottom to the point $i$ on the top, for all $i$, with no winding.
\end{Definition}

We now define an abstract algebra, the periodic Temperley-Lieb algebra. Note that we use capital letters to denote it, instead of calligraphic ones for the algebra of connectivities.

\begin{Definition} The periodic Temperley-Lieb algebra $PTL_N(\beta)$ is the algebra generated by a unit $id$ and the generators $e_i$, $i = 1, ..., N$, constrained by the following relations
\begin{alignat}{3}
e_i^2&= \beta e_i,&\qquad & \nonumber\\
e_ie_j&=e_je_i, &&\text{\rm for }|i-j|>1,\label{eq:TLPN}\\
e_ie_{i\pm 1}e_i&=e_i.&& \nonumber \end{alignat}
The indices are understood to be taken, modulo $N$, in the range $1, ..., N$, and therefore $e_0 \equiv e_N$ and $e_{N+1}\equiv e_{1}$.
\end{Definition}

\begin{Definition}
The enlargement of $PTL_N(\beta)$, denoted $\mathcal E PTL_N(\beta)$, is generated by $id$, the $e_i$s and two more generators, $\Omega$ and $\Omega^{-1}$, satisfying \eqref{eq:TLPN} together with
\begin{align}
\Omega e_i \Omega^{-1} &= e_{i-1}, \nonumber\\
\Omega \Omega^{-1} &= \Omega^{-1} \Omega = id, \label{eq:ETLPN}\\
(\Omega^{\pm 1} e_N)^{N-1} &= \Omega^{\pm N} (\Omega ^{\pm 1} e_N) \nonumber
\end{align}
with, again, the indices $i$ of the $e_i$s taken modulo $N$. The last relation can also be written as $e_{N-1} ... e_{1} = \Omega^{2} e_{1}$ and $e_1 ... e_{N-1} = \Omega^{-2} e_{N-1}$. 
\end{Definition}

The generator $\Omega$ will be referred to as the {\it translation operator}. The relations (\ref{eq:TLPN}) can be translated in terms of the generators $e_N, \Omega$ and $\Omega^{-1}$ of $\mathcal E PTL_N(\beta)$:
\begin{alignat}{3}
e_N^2&= \beta e_N,&\qquad & \nonumber\\
e_N \Omega^{j} e_N \Omega^{-j} &=  \Omega^{j} e_N \Omega^{-j} e_N, &&\text{\rm for }2\le j\le N-2\label{eq:ETLPN0},\\
e_N \Omega^{\mp 1}e_N \Omega^{\pm 1}e_N&=e_N.&& \nonumber 
\end{alignat} 
The identification of $e_N$ and $e_0$ is simply $\Omega^N e_N \Omega^{-N} = e_N$. The algebra $\mathcal E PTL_N(\beta)$ is then just $\langle e_N, \Omega, \Omega^{-1}\rangle$ constrained by these relations. Finally a last algebra will be used. 

\begin{Definition} For $N$ even, we define $\mathcal E PTL_N(\beta,\alpha)$ to be the quotient of $\mathcal E PTL_N(\beta)$ by the relation 
\begin{equation}
E \Omega^{\pm 1} E = \alpha E, \qquad \textrm{where\ }E = e_2e_4...e_{N-2}e_N.
\label{eq:elementE}\end{equation}
\label{sec:quot}
\end{Definition}
From \eqref{eq:elementE}, it also follows that $F \Omega^{\pm 1} F = \alpha F$, where $F = e_1e_3e_5...e_{N-1}$.
Despite this quotient, the algebra $\mathcal E PTL_N(\beta,\alpha)$ is still infinite as it contains the infinite subalgebra $\langle \Omega\rangle$. For $N$ odd, we take no further quotient but, for simplicity, we will still write $\mathcal EPTL_N(\beta,\alpha)$ for the enlargement of the $PTL_N(\beta)$ algebra.

To each $g \in \mathcal E PTL_N(\beta,\alpha)$, we associate a connectivity $c = \phi(g)\in \mathcal{PTL}(\beta,\alpha)$ as follows: To each generator we associate
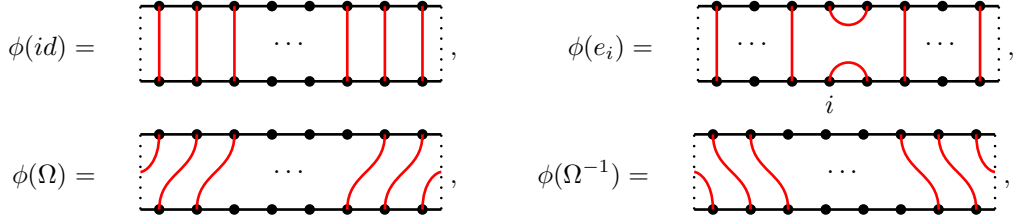
\begin{equation*} \phi(id) =
\psset{unit=0.5}
\begin{pspicture}(-0.5,-0.3)(8.5,0.8)
\psline[linewidth=1pt]{-}(0.5,1)(8.5,1)
\psline[linewidth=1pt,linestyle=dotted]{-}(0.5,1)(0.5,-1)
\psline[linewidth=1pt,linestyle=dotted]{-}(8.5,1)(8.5,-1)
\psline[linewidth=1pt]{-}(8.5,-1)(0.5,-1)
\psset{linewidth=1pt}
\psdots(1,1)(2,1)(3,1)(4,1)(5,1)(6,1)(7,1)(8,1)(1,-1)(2,-1)(3,-1)(4,-1)(5,-1)(6,-1)(7,-1)(8,-1)
\psset{linecolor=myc}
%\psarc(3.5,-1){0.5}{0}{180}
\psline(1,1)(1,-1)
\psline(2,1)(2,-1)
\rput(4.5,0){\dots}
\psline(3,1)(3,-1)
%\psline(4,1)(4,-1)
%\psline(5,1)(5,-1)
\psline(6,1)(6,-1)
\psline(7,1)(7,-1)
\psline(8,1)(8,-1)
\end{pspicture}\, \,, \qquad \qquad 
\phi(e_i) =
\begin{pspicture}(-0.5,-0.3)(8.5,0.8)
\psline[linewidth=1pt]{-}(0.5,1)(8.5,1)
\psline[linewidth=1pt,linestyle=dotted]{-}(0.5,1)(0.5,-1)
\psline[linewidth=1pt,linestyle=dotted]{-}(8.5,1)(8.5,-1)
\psline[linewidth=1pt]{-}(8.5,-1)(0.5,-1)
\psset{linewidth=1pt}
\psdots(1,1)(2,1)(3,1)(4,1)(5,1)(6,1)(7,1)(8,1)(1,-1)(2,-1)(3,-1)(4,-1)(5,-1)(6,-1)(7,-1)(8,-1)
\psset{linecolor=myc}
\psarc(4.5,-1){0.5}{0}{180}\psarc(4.5,1){0.5}{180}{360}
\rput(2,0){\dots}\rput(7,0){\dots}
\rput(4,-1.6){$i$}
\psline(1,1)(1,-1)
\psline(3,1)(3,-1)
\psline(6,1)(6,-1)
\psline(8,1)(8,-1)
\end{pspicture}
\, \,, \vspace{0.8cm}
\end{equation*}
\begin{equation*} \phi(\Omega) =
\psset{unit=0.5}
\begin{pspicture}(-0.5,-0.3)(8.5,0.8)
\psline[linewidth=1pt]{-}(0.5,1)(8.5,1)
\psline[linewidth=1pt,linestyle=dotted]{-}(0.5,1)(0.5,-1)
\psline[linewidth=1pt,linestyle=dotted]{-}(8.5,1)(8.5,-1)
\psline[linewidth=1pt]{-}(8.5,-1)(0.5,-1)
\psset{linewidth=1pt}
\psdots(1,1)(2,1)(3,1)(4,1)(5,1)(6,1)(7,1)(8,1)(1,-1)(2,-1)(3,-1)(4,-1)(5,-1)(6,-1)(7,-1)(8,-1)
\psset{linecolor=myc}
\psbezier{-}(1,1)(1,0.25)(0.625,0)(0.5,0)
\psbezier{-}(2,1)(2,0)(1,0)(1,-1)
\psbezier{-}(3,1)(3,0)(2,0)(2,-1)
\psbezier{-}(8,1)(8,0)(7,0)(7,-1)
\psbezier{-}(7,1)(7,0)(6,0)(6,-1)
\psbezier{-}(8,-1)(8,-0.25)(8.375,0)(8.5,0)
\rput(4.5,0){\dots}
\end{pspicture}\, \,, \qquad \hspace{0.3cm}%\qquad 
\phi(\Omega^{-1}) =
\begin{pspicture}(-0.5,-0.3)(8.5,0.8)
\psline[linewidth=1pt]{-}(0.5,1)(8.5,1)
\psline[linewidth=1pt,linestyle=dotted]{-}(0.5,1)(0.5,-1)
\psline[linewidth=1pt,linestyle=dotted]{-}(8.5,1)(8.5,-1)
\psline[linewidth=1pt]{-}(8.5,-1)(0.5,-1)
\psset{linewidth=1pt}
\psdots(1,1)(2,1)(3,1)(4,1)(5,1)(6,1)(7,1)(8,1)(1,-1)(2,-1)(3,-1)(4,-1)(5,-1)(6,-1)(7,-1)(8,-1)
\psset{linecolor=myc}
\rput(4.5,0){\dots}
\psbezier{-}(1,-1)(1,-0.25)(0.625,0)(0.5,0)
\psbezier{-}(2,-1)(2,-0)(1,0)(1,1)
\psbezier{-}(3,-1)(3,-0)(2,0)(2,1)
\psbezier{-}(8,-1)(8,-0)(7,0)(7,1)
\psbezier{-}(7,-1)(7,-0)(6,0)(6,1)
\psbezier{-}(8,1)(8,0.25)(8.375,0)(8.5,0)
\end{pspicture}
\, \,, \vspace{0.5cm}
\end{equation*}
and for any $g = \prod_i f_i$ a product of the generators of $\mathcal E PTL_N(\beta, \alpha)$ (with $f_i \in \{ e_1, ..., e_{N}, \Omega, \Omega^{-1} \}$), we set $\phi(g) = \prod_i \phi(f_i)$, with the same product of diagrams defined for $\mathcal{PTL}_N(\beta,\alpha)$. One can verify that equations (\ref{eq:TLPN}), (\ref{eq:ETLPN}) and \eqref{eq:elementE} are satisfied if we replace $e_i$, $\Omega$ and $\Omega^{-1}$ by $\phi(e_i), \phi(\Omega)$ and $\phi(\Omega^{-1})$, so 
$\phi:\mathcal EPTL_N(\beta,\alpha)\rightarrow \mathcal{PTL}_N(\beta,\alpha)$
is a homomorphism of algebras.  It is surjective as all generators of $\mathcal{PTL}_N(\beta,\alpha)$ have preimages. Moreover, Green and Fan \cite{GreenFan} have shown that $\phi$ is injective. In this sense, $\mathcal E PTL_N(\beta,\alpha)$ and $\mathcal{PTL}_N(\beta,\alpha)$ are isomorphic, and throughout the rest of this paper we will use $\mathcal E PTL_N(\beta,\alpha)$ to denote both the algebra of connectivities and $\langle e_N, \Omega, \Omega^{-1} \rangle / (\textrm{relations\ }\eqref{eq:ETLPN}, \eqref{eq:ETLPN0}, \eqref{eq:elementE})$.

\subsection{The loop transfer matrix $T_N(\lambda,\nu)$}\label{sec:transferMatrix}
\begin{Definition}\label{def:laTN} The loop transfer matrix $T_N(\lambda, \nu)$ is an element of $\mathcal E PTL_N(\beta,\alpha)$ defined by
\begin{equation*} T_N(\lambda,\nu) = \quad
\psset{unit=1}
\psset{linewidth=1pt}
\overbrace{
\begin{pspicture}(-0,0.375)(5,1.2)
\psdots(0.5,0)(1.5,0)(4.5,0)
\psdots(0.5,1)(1.5,1)(4.5,1)
\lw
\psline{-}(-0.15,0.5)(0.0,0.5)
\psline{-}(5.15,0.5)(5.0,0.5)
\unlw
\psline{-}(0,0)(1,0)(1,1)(0,1)(0,0)\psarc[linewidth=0.5pt]{-}(0,0){0.25}{0}{90}\rput(0.5,0.5){$\nu$}
\psline{-}(1,0)(2,0)(2,1)(1,1)(1,0)\psarc[linewidth=0.5pt]{-}(1,0){0.25}{0}{90}\rput(1.5,0.5){$\nu$}
\psline{-}(4,0)(5,0)(5,1)(4,1)(4,0)\psarc[linewidth=0.5pt]{-}(4,0){0.25}{0}{90}\rput(4.5,0.5){$\nu$}
\psline{-}(2,0)(2.5,0)\psline[linestyle=dashed,dash=2pt 2pt]{-}(2.5,0)(3.5,0)\psline{-}(3.5,0)(4,0)
\psline{-}(2,1)(2.5,1)\psline[linestyle=dashed,dash=2pt 2pt]{-}(2.5,1)(3.5,1)\psline{-}(3.5,1)(4,1)
\psset{linecolor=myc}\unlw
\end{pspicture}}^N
\end{equation*}
where the boxes are given by
\begin{equation*}
\psset{unit=1}
\psset{linewidth=1pt}
\begin{pspicture}(-0.5,-0.1)(0.5,0.5)
\psline{-}(-0.5,-0.5)(0.5,-0.5)(0.5,0.5)(-0.5,0.5)(-0.5,-0.5)
\psarc[linewidth=0.5pt]{-}(-0.5,-0.5){0.25}{0}{90}
\rput(0,0){$\nu$}
\end{pspicture}\ =\ \sin(\lambda-\nu)\ \ 
\begin{pspicture}(-0.5,-0.1)(0.5,0.5)
\psline{-}(-0.5,-0.5)(0.5,-0.5)(0.5,0.5)(-0.5,0.5)(-0.5,-0.5)
\psset{linecolor=myc}
\psarc{-}(0.5,-0.5){0.5}{90}{180}
\psarc{-}(-0.5,0.5){0.5}{270}{360}
\end{pspicture}\ +\ \sin \nu\ \ 
\begin{pspicture}(-0.5,-0.1)(0.5,0.5)
\psline{-}(-0.5,-0.5)(0.5,-0.5)(0.5,0.5)(-0.5,0.5)(-0.5,-0.5)
\psset{linecolor=myc}
\psarc{-}(-0.5,-0.5){0.5}{0}{90}
\psarc{-}(0.5,0.5){0.5}{180}{270}
\end{pspicture}
\ \ =\ \ 
\begin{pspicture}(-0.5,-0.1)(0.5,0.5)
\psline{-}(-0.5,-0.5)(0.5,-0.5)(0.5,0.5)(-0.5,0.5)(-0.5,-0.5)
\psarc[linewidth=0.5pt]{-}(0.5,-0.5){0.25}{90}{180}
\rput(0,0){$\lambda\!-\!\nu$}
\end{pspicture}\, \, ,
\end{equation*}
\\
$\beta = 2 \cos \lambda$, $\nu$ is the anisotropy and the leftmost and rightmost boxes are connected.
\end{Definition}
The loop transfer matrix, or simply {\em transfer matrix}, is related to the Fortuin-Kasteleyn description of two-dimensional lattice models and has many crucial mathematical properties. (Several of the following properties were proved in a general context in \cite{BPOB}. Proofs and discussion of these properties in a context similar to the present one can be found in \cite{PRZ, PRV}. The tie with lattice models is found in \cite{BKW} or, for a presentation similar to the one here, in \cite{AMDSA} for example.)
\begin{enumerate}
\item[{(i)}] 
It forms a communing family: $[T_N(\lambda,\nu_1),T_N(\lambda,\nu_2)]=0$ for all $\nu_1$ and $\nu_2$.
\item[{(ii)}] It satisfies a crossing-reflection symmetry: $T_N(\lambda, \lambda-\nu) = R^{-1} T_N(\lambda, \nu) R$ where $R$ is the left-right reflection: $e_i = R^{-1} e_{N-i} R$.
\item[{(iii)}] It is invariant under translation: $[T_N(\lambda, \nu),\Omega] =0$.
\item[{(iv)}] Its expansion around $\nu=0$ is 
\begin{equation}T_N(\lambda,\nu) \simeq \Omega \sin^N\lambda \, [(1-\nu N \cot \lambda)id + \nu \mathcal H/\sin\lambda] + \mathcal O(\nu^2)
\end{equation}
where 
\begin{equation}\mathcal H = \sum_{i=1}^{N}e_i.\label{eq:hamiltonienBoucle}
\end{equation} 
\end{enumerate}

%It is this transfer matrix whose properties we seek to elucidate. In many situations, we shall have to settle for the simpler task of studying the linear term $\mathcal H$ which will be called the Hamiltonian for loop models.

\subsection{Link states and representations of $\mathcal E PTL_N(\beta,\alpha)$}\label{sec:jemanquedidees}

In the following, we will work with link representations of $\mathcal E PTL_N(\beta, \alpha)$ similar to those introduced in \cite{PRV}. They extend to the periodic case the link representations of the (original) Temperley-Lieb algebra introduced for models on a strip \cite{Martin, Westbury}. (See also \cite{PRZ, AMDSA}.)

Again $N$ points (sometimes called {\it entries}) are aligned equidistant on a section of a vertical cylinder. A {\it link state} $w$ (or {\it link pattern}) is a graph where the $N$ points are either connected pairwise by non-intersecting curves or connected by a straight line to $+\infty$. The non-intersecting curves and half-lines are drawn above the section, that is, on one side of it. A point connected to infinity will be called a {\it defect} and the number of defects of a link state will usually be denoted by $d$. The set  $\tilde B_N$ of link states with $N$ entries is naturally partitioned in the subsets $\tilde B_N^d$s of link states with $N$ entries and $d$ defects (with $N \equiv d\,\, \textrm{mod} \,\, 2$), each containing $\left(\begin{smallmatrix}N\\  (N-d)/2  \end{smallmatrix}\right)$ elements. The vector spaces generated by $\tilde B_N$ and $\tilde B_N^d$ are denoted $\tilde V_N$ and $\tilde V_N^d$ respectively.

By convention, the elements of $\tilde B_N$ are ordered in ascending number of defects. When drawn in the plane, entries are placed on the horizontal axis at points of coordinates $(1,0)$, \,$(2,0)$, ..., $(N,0)$, and the diagram for link states is taken to be periodic in the $x$ direction with $x+N \equiv x$. The curves connecting the entries can connect through the imaginary boundary line at $x=1/2$ and $x = N+1/2$. We will call these curves {\it boundary curves} or {\it boundary bubbles}. Here are the three subsets $\tilde B_N^d$ for $N=4$:
\begin{equation}
\psset{unit=0.6}
 \qquad  \tilde B_4^0 = \big\{
 \begin{pspicture}(-0.3,0)(1.8,0.4)
\psdots(0,0)(0.5,0)(1,0)(1.5,0)
\psset{linewidth=1pt}
\psset{linecolor=myc2}
\psarc{-}(0.25,0){0.25}{0}{180}
\psarc{-}(1.25,0){0.25}{0}{180}
\end{pspicture},
 \begin{pspicture}(-0.3,0)(1.8,0.64)
\psdots(0,0)(0.5,0)(1,0)(1.5,0)
\psset{linewidth=1pt}
\psset{linecolor=myc2}
\psarc{-}(0.75,0){0.25}{0}{180}
\psbezier{-}(0,0)(0,0.75)(1.5,0.75)(1.5,0)
\end{pspicture},
 \begin{pspicture}(-0.3,0)(1.8,0.4)
\psdots(0,0)(0.5,0)(1,0)(1.5,0)
\psset{linewidth=1pt}
\psset{linecolor=myc2}
\psbezier{-}(0.5,0)(0.5,0.65)(1.5,0.65)(1.75,0.4)
\psbezier{-}(0,0)(0,0.275)(-0.125,0.4)(-0.25,0.4)
\psarc{-}(1.25,0){0.25}{0}{180}
\end{pspicture},
  \begin{pspicture}(-0.3,0)(1.8,0.4)
\psdots(0,0)(0.5,0)(1,0)(1.5,0)
\psset{linewidth=1pt}
\psset{linecolor=myc2}
\psarc{-}(-0.25,0){0.25}{0}{90}
\psarc{-}(1.75,0){0.25}{90}{180}
\psarc{-}(0.75,0){0.25}{0}{180}
\end{pspicture},
\begin{pspicture}(-0.3,0)(1.8,0.4)
\psdots(0,0)(0.5,0)(1,0)(1.5,0)
\psset{linewidth=1pt}
\psset{linecolor=myc2}
\psbezier{-}(1,0)(1,0.65)(0,0.65)(-0.25,0.4)
\psbezier{-}(1.5,0)(1.5,0.275)(1.75,0.4)(1.75,0.4)
\psarc{-}(0.25,0){0.25}{0}{180}
\end{pspicture},
   \begin{pspicture}(-0.3,0)(1.8,0.4)
\psdots(0,0)(0.5,0)(1,0)(1.5,0)
\psset{linewidth=1pt}
\psset{linecolor=myc2}
\psarc{-}(-0.25,0){0.25}{0}{90}
\psarc{-}(1.75,0){0.25}{90}{180}
%\psarc{-}(0.75,0){0.25}{0}{180}
\psbezier{-}(0.5,0)(0.5,0.5)(0,0.6)(-0.25,0.6)
\psbezier{-}(1,0)(1,0.5)(1.5,0.6)(1.75,0.6)
\end{pspicture}
 \big\} ,
\label{eq:linkbasis40}
\end{equation}
\begin{equation*}
\psset{unit=0.6}
 \qquad  \tilde B_4^2 = \big\{
 \begin{pspicture}(-0.3,0)(1.8,0.4)
\psdots(0,0)(0.5,0)(1,0)(1.5,0)
\psset{linewidth=1pt}
\psset{linecolor=myc2}
\psarc{-}(0.25,0){0.25}{0}{180}
\psline{-}(1,0)(1,0.75)
\psline{-}(1.5,0)(1.5,0.75)
\end{pspicture},
 \begin{pspicture}(-0.3,0)(1.8,0.4)
\psdots(0,0)(0.5,0)(1,0)(1.5,0)
\psset{linewidth=1pt}
\psset{linecolor=myc2}
\psarc{-}(0.75,0){0.25}{0}{180}
\psline{-}(0,0)(0,0.75)
\psline{-}(1.5,0)(1.5,0.75)
\end{pspicture},
 \begin{pspicture}(-0.3,0)(1.8,0.4)
\psdots(0,0)(0.5,0)(1,0)(1.5,0)
\psset{linewidth=1pt}
\psset{linecolor=myc2}
\psarc{-}(1.25,0){0.25}{0}{180}
\psline{-}(0,0)(0,0.75)
\psline{-}(0.5,0)(0.5,0.75)
\end{pspicture},
\begin{pspicture}(-0.3,0)(1.8,0.4)
\psdots(0,0)(0.5,0)(1,0)(1.5,0)
\psset{linewidth=1pt}
\psset{linecolor=myc2}
\psarc{-}(-0.25,0){0.25}{0}{90}
\psarc{-}(1.75,0){0.25}{90}{180}
\psline{-}(1,0)(1,0.75)
\psline{-}(0.5,0)(0.5,0.75)
\end{pspicture}
\big\} , \qquad  \tilde B_4^4 = \big\{
 \begin{pspicture}(-0.3,0)(1.8,0.4)
\psdots(0,0)(0.5,0)(1,0)(1.5,0)
\psset{linewidth=1pt}
\psset{linecolor=myc2}
\psline{-}(0,0)(0,0.75)
\psline{-}(0.5,0)(0.5,0.75)
\psline{-}(1,0)(1,0.75)
\psline{-}(1.5,0)(1.5,0.75)
\end{pspicture}\big\} .
\end{equation*}

The sets $B_N^d$ and $B_N$, without ``\~ \,'', will refer to the subsets of $\tilde B_N^d$ and $\tilde B_N$ containing link patterns with no boundary curves (and similarly for $V_N$ and $V_N^d$, the vector spaces they span).  These form the sets of link states used for representations of the (original) Temperley-Lieb algebra. We note that the set $\tilde B_N^d$ corresponds to the set of {\em distinct link states} used by Pearce, Rasmussen and Villani \cite{PRV}. (We shall not use their set of {\em identified link states}.)

\begin{Definition} [The map $\omega_d: \mathcal EPTL_N(\beta,\alpha) \rightarrow \textrm{End}(\tilde V_N^d)$] Let $c$ be a connectivity in $\mathcal E PTL_N(\beta,\alpha)$ and $w \in \tilde B_N^d$. An action of the diagram $c$ on $w$ is defined by joining the $N$ entries of the link state $w$ to the $N$ top entries of $c$. The resulting link pattern is found by reading the new connections at the bottom $N$ points of $c$. The result is then multiplied by the following factors: weights related to closed loops (a factor of $\beta$ for each contractible loop and one of $\alpha$ for each non-contractible one) and weights due to the lateral twist of defects. These are computed as follows. First, if two defects are connected in the diagram $cw$, the result is set to $0$. Second, for each defect in $w$, a multiplicative factor of $v^{\Delta}$ is added, where $\Delta$ is the distance the defect has traveled toward the left, that is, its position in the original state $w$ minus its new one in the resulting $cw$. (Again, consecutive positions in $w$ are at distance of $1$.) The constant v is the {\it twist parameter}. The map $\omega_d$ is obtained by extending this action linearly to all elements in $\mathcal EPTL_N(\beta, \alpha)$ and depends on $v$: $\omega_d=\omega_d(v)$.
\label{sec:omegarep}
\end{Definition}

\noindent Examples are useful in understanding the product just defined. Here are computations for $\omega_d$.\vspace{0.3cm}

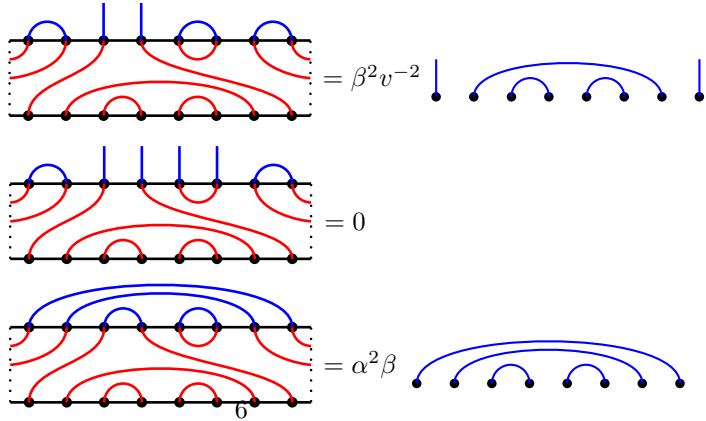
\begin{equation*}
\psset{unit=0.5}
\begin{pspicture}(-0.5,-0.3)(8.5,0.8)
\psline[linewidth=1pt]{-}(0.5,1)(8.5,1)
\psline[linewidth=1pt,linestyle=dotted]{-}(0.5,1)(0.5,-1)
\psline[linewidth=1pt,linestyle=dotted]{-}(8.5,1)(8.5,-1)
\psline[linewidth=1pt]{-}(8.5,-1)(0.5,-1)
\psset{linewidth=1pt}
\psdots(1,1)(2,1)(3,1)(4,1)(5,1)(6,1)(7,1)(8,1)(1,-1)(2,-1)(3,-1)(4,-1)(5,-1)(6,-1)(7,-1)(8,-1)
\psset{linecolor=myc}
\psarc(3.5,-1){0.5}{0}{180}
\psarc(5.5,-1){0.5}{0}{180}
\psbezier{-}(2,-1)(2,0.2)(7,0.2)(7,-1)
\psarc(0.5,1){0.5}{-90}{0}
\psarc(5.5,1){0.5}{180}{360}
\psarc(8.5,1){0.5}{180}{270}
\psbezier{-}(3,1)(3,0)(1,0)(1,-1)
\psbezier{-}(2,1)(2,0.25)(0.75,0)(0.5,0)
\psbezier{-}(7,1)(7,0.25)(8.25,0)(8.5,0)
\psbezier{-}(4,1)(4,0)(8,0)(8,-1)
\psset{linecolor=myc2}
\psline{-}(3,1)(3,2)
\psline{-}(4,1)(4,2)
%\psline{-}(5,1)(5,2)
%\psline{-}(6,1)(6,2)
\psarc(1.5,1){0.5}{0}{180}
\psarc(5.5,1){0.5}{0}{180}
\psarc(7.5,1){0.5}{0}{180}
\psset{linecolor=black}
%\psarc(6.5,1){1.5}{0}{180}
\end{pspicture} \qquad \, \,
\begin{pspicture}(9.0,-0.3)(12.7,0.8)
\rput(9.0,0){$= \beta^2 v^{-2}$}
\psdots(17.7,-0.5)(16.7,-0.5)(15.7,-0.5)(14.7,-0.5)(13.7,-0.5)(12.7,-0.5)(11.7,-0.5)(10.7,-0.5)
\psset{linecolor=myc2}
\psarc(13.2,-0.5){0.5}{0}{180}
\psarc(15.2,-0.5){0.5}{0}{180}
\psbezier{-}(11.7,-0.5)(11.7,0.7)(16.7,0.7)(16.7,-0.5)
%
%\psarc(12.2,-0.5){0.5}{0}{180}
%\psbezier{-}(10.7,-0.5)(10.7,0.5)(13.7,0.5)(13.7,-0.5)
%%\psarc(12.0,-1){1.5}{0}{180}
\psline{-}(10.7,-0.5)(10.7,0.5)
\psline{-}(17.7,-0.5)(17.7,0.5)
\end{pspicture}
\end{equation*} \vspace{0.2cm}

\begin{equation*}
\psset{unit=0.5}
\begin{pspicture}(-0.5,-0.3)(8.5,0.8)
\psline[linewidth=1pt]{-}(0.5,1)(8.5,1)
\psline[linewidth=1pt,linestyle=dotted]{-}(0.5,1)(0.5,-1)
\psline[linewidth=1pt,linestyle=dotted]{-}(8.5,1)(8.5,-1)
\psline[linewidth=1pt]{-}(8.5,-1)(0.5,-1)
\psset{linewidth=1pt}
\psdots(1,1)(2,1)(3,1)(4,1)(5,1)(6,1)(7,1)(8,1)(1,-1)(2,-1)(3,-1)(4,-1)(5,-1)(6,-1)(7,-1)(8,-1)
\psset{linecolor=myc}
\psarc(3.5,-1){0.5}{0}{180}
\psarc(5.5,-1){0.5}{0}{180}
\psbezier{-}(2,-1)(2,0.2)(7,0.2)(7,-1)
\psarc(0.5,1){0.5}{-90}{0}
\psarc(5.5,1){0.5}{180}{360}
\psarc(8.5,1){0.5}{180}{270}
\psbezier{-}(3,1)(3,0)(1,0)(1,-1)
\psbezier{-}(2,1)(2,0.25)(0.75,0)(0.5,0)
\psbezier{-}(7,1)(7,0.25)(8.25,0)(8.5,0)
\psbezier{-}(4,1)(4,0)(8,0)(8,-1)
\psset{linecolor=myc2}
\psline{-}(3,1)(3,2)
\psline{-}(4,1)(4,2)
\psline{-}(5,1)(5,2)
\psline{-}(6,1)(6,2)
\psarc(1.5,1){0.5}{0}{180}
\psarc(7.5,1){0.5}{0}{180}
\psset{linecolor=black}
%\psarc(6.5,1){1.5}{0}{180}
\end{pspicture} \qquad
\begin{pspicture}(9.3,-0.3)(13.2,0.8)
\rput(8.8,0){$= 0$}
%\psdots(17.7,-0.5)(16.7,-0.5)(15.7,-0.5)(14.7,-0.5)(13.7,-0.5)(12.7,-0.5)(11.7,-0.5)(10.7,-0.5)
%\psset{linecolor=myc2}
%\psarc(13.2,-0.5){0.5}{0}{180}
%\psarc(15.2,-0.5){0.5}{0}{180}
%\psbezier{-}(11.7,-0.5)(11.7,0.7)(16.7,0.7)(16.7,-0.5)
%%
%%\psarc(12.2,-0.5){0.5}{0}{180}
%%\psbezier{-}(10.7,-0.5)(10.7,0.5)(13.7,0.5)(13.7,-0.5)
%%%\psarc(12.0,-1){1.5}{0}{180}
%\psline{-}(10.7,-0.5)(10.7,0.5)
%\psline{-}(17.7,-0.5)(17.7,0.5)
\end{pspicture}
\end{equation*} \vspace{0.2cm}

\begin{equation*}
\psset{unit=0.5}
\begin{pspicture}(-0.5,-0.3)(8.5,0.8)
\psline[linewidth=1pt]{-}(0.5,1)(8.5,1)
\psline[linewidth=1pt,linestyle=dotted]{-}(0.5,1)(0.5,-1)
\psline[linewidth=1pt,linestyle=dotted]{-}(8.5,1)(8.5,-1)
\psline[linewidth=1pt]{-}(8.5,-1)(0.5,-1)
\psset{linewidth=1pt}
\psdots(1,1)(2,1)(3,1)(4,1)(5,1)(6,1)(7,1)(8,1)(1,-1)(2,-1)(3,-1)(4,-1)(5,-1)(6,-1)(7,-1)(8,-1)
\psset{linecolor=myc}
\psarc(3.5,-1){0.5}{0}{180}
\psarc(5.5,-1){0.5}{0}{180}
\psbezier{-}(2,-1)(2,0.2)(7,0.2)(7,-1)
\psarc(0.5,1){0.5}{-90}{0}
\psarc(5.5,1){0.5}{180}{360}
\psarc(8.5,1){0.5}{180}{270}
\psbezier{-}(3,1)(3,0)(1,0)(1,-1)
\psbezier{-}(2,1)(2,0.25)(0.75,0)(0.5,0)
\psbezier{-}(7,1)(7,0.25)(8.25,0)(8.5,0)
\psbezier{-}(4,1)(4,0)(8,0)(8,-1)
\psset{linecolor=myc2}
\psbezier{-}(1,1)(1,2.5)(8,2.5)(8,1)
\psbezier{-}(2,1)(2,2.2)(7,2.2)(7,1)
\psarc(3.5,1){0.5}{0}{180}
\psarc(5.5,1){0.5}{0}{180}
%\psarc(7.5,1){0.5}{0}{180}
\psset{linecolor=black}
%\psarc(6.5,1){1.5}{0}{180}
\end{pspicture} \qquad
\begin{pspicture}(9.3,-0.3)(13.2,0.8)
\rput(9.2,0){$= \alpha^2 \beta $}
\psdots(17.7,-0.5)(16.7,-0.5)(15.7,-0.5)(14.7,-0.5)(13.7,-0.5)(12.7,-0.5)(11.7,-0.5)(10.7,-0.5)
\psset{linecolor=myc2}
\psarc(13.2,-0.5){0.5}{0}{180}
\psarc(15.2,-0.5){0.5}{0}{180}
\psbezier{-}(11.7,-0.5)(11.7,0.7)(16.7,0.7)(16.7,-0.5)
\psbezier{-}(10.7,-0.5)(10.7,1)(17.7,1)(17.7,-0.5)
\end{pspicture}
\end{equation*}\vspace{0.2cm}\\
\noindent In the first example, the first defect has $\Delta_1  = 2$ and the second $\Delta_2 = -4$, resulting in an overall factor of $v^{\Delta_1+\Delta_2}= v^{-2}$. As a last example, note that $\Omega$ shifts the link pattern $w$ one position to the left and therefore multiplies the resulting state by a factor $v^d$. 

\begin{Proposition}\label{prop:omegaRep} The map $\omega_d$ is a 
representation of $\mathcal E PTL_N(\beta,\alpha)$.
\end{Proposition}

We do not give the details of this proof as the verification is standard, though tedious. It suffices to show that the defining relations (\ref{eq:TLPN}), (\ref{eq:ETLPN}) and \eqref{eq:elementE} hold for the action on any link state $w\in\tilde B_N^d$. The relations (\ref{eq:TLPN}) and (\ref{eq:ETLPN}) involve at most four entries of the link state (except for the last one of (\ref{eq:ETLPN})) so that, for these, one may concentrate on the two, three or four connections changed. The other two equations can also be seen to hold, and we leave the verifications to the reader. The result is nevertheless non-trivial. Indeed, imposing that connecting defects give zero is essential, as the map $\omega_d$ would not be a representation of $\mathcal E PTL_N(\beta,\alpha)$ otherwise. For example, \vspace{-0.2cm}

\begin{equation*} e_3 e_1 \Big(
\psset{unit=0.5}
\begin{pspicture}(0.5,-1)(4.5,2)
\psdots(1,-1)(2,-1)(3,-1)(4,-1)
\psset{linecolor=myc2}
\psarc(2.5,-1){0.5}{0}{180}
\psline{-}(1,-1)(1,0)
\psline{-}(4,-1)(4,0)
\end{pspicture} \Big)
 = 
\begin{pspicture}(0.25,-1)(4.5,2)
\psline[linewidth=1pt]{-}(0.5,1)(4.5,1)
\psline[linewidth=1pt]{-}(0.5,-3)(4.5,-3)
\psline[linewidth=1pt,linestyle=dotted]{-}(0.5,1)(0.5,-3)
\psline[linewidth=1pt,linestyle=dotted]{-}(4.5,1)(4.5,-3)
\psline[linewidth=1pt]{-}(4.5,-1)(0.5,-1)
\psset{linewidth=1pt}
\psdots(1,1)(2,1)(3,1)(4,1)(1,-1)(2,-1)(3,-1)(4,-1)(1,-3)(2,-3)(3,-3)(4,-3)
\psset{linecolor=myc}
\psline{-}(3,1)(3,-1)
\psline{-}(4,1)(4,-1)
\psline{-}(1,-3)(1,-1)
\psline{-}(2,-3)(2,-1)
\psarc(1.5,1){0.5}{180}{360}
\psarc(1.5,-1){0.5}{0}{180}
\psarc(3.5,-1){0.5}{180}{360}
\psarc(3.5,-3){0.5}{0}{180}
\psset{linecolor=myc2}
\psarc(2.5,1){0.5}{0}{180}
\psline{-}(1,1)(1,2)
\psline{-}(4,1)(4,2)
\end{pspicture} \, = v^{-2}
\begin{pspicture}(-0.0,0)(4.5,2)
\psline[linewidth=1pt]{-}(0.5,1)(4.5,1)
\psline[linewidth=1pt,linestyle=dotted]{-}(0.5,1)(0.5,-1)
\psline[linewidth=1pt,linestyle=dotted]{-}(4.5,1)(4.5,-1)
\psline[linewidth=1pt]{-}(4.5,-1)(0.5,-1)
\psset{linewidth=1pt}
\psdots(1,1)(2,1)(3,1)(4,1)(1,-1)(2,-1)(3,-1)(4,-1)
\psset{linecolor=myc}
%\psline{-}(3,1)(3,-1)
%\psline{-}(4,1)(4,-1)
\psline{-}(1,1)(1,-1)
\psline{-}(2,1)(2,-1)
%\psarc(1.5,1){0.5}{180}{360}
%\psarc(1.5,-1){0.5}{0}{180}
\psarc(3.5,1){0.5}{180}{360}
\psarc(3.5,-1){0.5}{0}{180}
\psset{linecolor=myc2}
\psarc(1.5,1){0.5}{0}{180}
\psline{-}(3,1)(3,2)
\psline{-}(4,1)(4,2)
\end{pspicture} \, = v^{-2} 
\begin{pspicture}(0.5,-1)(4.5,2)
\psdots(1,-1)(2,-1)(3,-1)(4,-1)
\psset{linecolor=myc2}
\psarc(1.5,-1){0.5}{0}{180}
\psarc(3.5,-1){0.5}{0}{180}
\end{pspicture}
\end{equation*}\\

\begin{equation*} e_1 e_3 \Big(
\psset{unit=0.5}
\begin{pspicture}(0.5,-1)(4.5,2)
\psdots(1,-1)(2,-1)(3,-1)(4,-1)
\psset{linecolor=myc2}
\psarc(2.5,-1){0.5}{0}{180}
\psline{-}(1,-1)(1,0)
\psline{-}(4,-1)(4,0)
\end{pspicture} \Big)
 = 
\begin{pspicture}(0.25,-1)(4.5,2)
\psline[linewidth=1pt]{-}(0.5,1)(4.5,1)
\psline[linewidth=1pt]{-}(0.5,-3)(4.5,-3)
\psline[linewidth=1pt,linestyle=dotted]{-}(0.5,1)(0.5,-3)
\psline[linewidth=1pt,linestyle=dotted]{-}(4.5,1)(4.5,-3)
\psline[linewidth=1pt]{-}(4.5,-1)(0.5,-1)
\psset{linewidth=1pt}
\psdots(1,1)(2,1)(3,1)(4,1)(1,-1)(2,-1)(3,-1)(4,-1)(1,-3)(2,-3)(3,-3)(4,-3)
\psset{linecolor=myc}
\psline{-}(3,-3)(3,-1)
\psline{-}(4,-3)(4,-1)
\psline{-}(1,1)(1,-1)
\psline{-}(2,1)(2,-1)
\psarc(1.5,-1){0.5}{180}{360}
\psarc(1.5,-3){0.5}{0}{180}
\psarc(3.5,1){0.5}{180}{360}
\psarc(3.5,-1){0.5}{0}{180}
\psset{linecolor=myc2}
\psarc(2.5,1){0.5}{0}{180}
\psline{-}(1,1)(1,2)
\psline{-}(4,1)(4,2)
\end{pspicture} \, = v^{2}
\begin{pspicture}(-0.0,0)(4.5,2)
\psline[linewidth=1pt]{-}(0.5,1)(4.5,1)
\psline[linewidth=1pt,linestyle=dotted]{-}(0.5,1)(0.5,-1)
\psline[linewidth=1pt,linestyle=dotted]{-}(4.5,1)(4.5,-1)
\psline[linewidth=1pt]{-}(4.5,-1)(0.5,-1)
\psset{linewidth=1pt}
\psdots(1,1)(2,1)(3,1)(4,1)(1,-1)(2,-1)(3,-1)(4,-1)
\psset{linecolor=myc}
%\psline{-}(3,1)(3,-1)
%\psline{-}(4,1)(4,-1)
\psline{-}(3,1)(3,-1)
\psline{-}(4,1)(4,-1)
%\psarc(1.5,1){0.5}{180}{360}
%\psarc(1.5,-1){0.5}{0}{180}
\psarc(1.5,1){0.5}{180}{360}
\psarc(1.5,-1){0.5}{0}{180}
\psset{linecolor=myc2}
\psarc(3.5,1){0.5}{0}{180}
\psline{-}(1,1)(1,2)
\psline{-}(2,1)(2,2)
\end{pspicture} \, = v^{2} 
\begin{pspicture}(0.5,-1)(4.5,2)
\psdots(1,-1)(2,-1)(3,-1)(4,-1)
\psset{linecolor=myc2}
\psarc(1.5,-1){0.5}{0}{180}
\psarc(3.5,-1){0.5}{0}{180}
\end{pspicture}
\end{equation*}\vspace{0.5cm}\\
Because $e_1e_3=e_3e_1$ in $\mathcal EPTL_N, N\ge 4$, this would not have been a representation. This also means that, to define an action of the whole link space $\tilde V_N$ where connecting defects are not given weight $0$, the twist $v$ should solve some algebraic constraints, like the above $v^4=1$. 

The usefulness of the representation $\omega_d$ with its twist parameter $v$ stems form the fact that $\omega_d(\Omega^N) = v^{Nd}\, \omega_d(id)$ and not simply $\omega_d(id)$. It therefore allows one to keep track of the winding number of Fortuin-Kasteleyn clusters along the cylinder, a physical property that plays an important role in the mathematical description of these models \cite{Pinson, Arguin, AMDYSA0}. Another appropriate name for the parameter $v$ would have been the {\em momentum parameter}. Indeed, $\Omega$ acts as a translation operator around the cylinder, or a rotation operator. Clearly its eigenvalues are expressed in terms of $v$ and should be interpreted as the possible values of the momentum.

%%%%%%%%%%%%%%%%
%
\section{The XXZ model and the intertwiner $\tilde i_N^d$} \label{sec:XXZprime}
%
%%%%%%%%%%%%%%%%

On the strip, the loop models are intimately related to another family of physically relevant ones. Both models are defined by an {\it evolution operator}: the transfer matrix for the loop models and a Hamiltonian for the XXZ models. Even though the vector spaces upon which the evolution operators act are different, they both carry a representation of the Temperley-Lieb algebra. Often properties of one of the models can teach us something about the properties of the other and, with this objective in mind, we give in this section the precise relation between loop models and the XXZ Hamiltonians on the cylinder.
 
\subsection{An extended family of XXZ models}\label{sec:extendedXXZ}

Instead of working with the full transfer matrix, we concentrate on the first non-trivial term in its expansion around $u=0$, given in \eqref{eq:hamiltonienBoucle}: $\mathcal H=\sum_{1\le i\le N}e_i$. Since the XXZ Hamiltonians are maps of $(\mathbb C^2)^{\otimes N}$ onto itself, we need to define a representation of $\mathcal EPTL_N(\beta,\alpha)$ on this vector space, or at least of $PTL_N(\beta)$. Then the XXZ Hamiltonians will simply be $H=\sum_{1\le i\le N}\bar e_i$ where $\bar e_i$ are the matrices representing the generators $e_i$ of $\mathcal EPTL_N(\beta,\alpha)$.

We use the usual notation $$\sigma_j^a=\underbrace{id_2\otimes \dots\otimes id_2}_{j-1} \otimes \, \, \sigma^a\otimes \underbrace{id_2\otimes\dots\otimes id_2}_{N-j}$$ for $1\le j\le N$ and $a\in\{x,y,z,+,-\}$, and we set $\sigma^a_{N+1} \equiv \sigma^a_{1}$. The tensor product contains $N$ two-by-two matrices and $\sigma^a$ is the $j$-th factor in this product. The matrices $\bar e_j\in \textrm{End}\big((\mathbb C^2)^{\otimes N}\big)$ are 
\begin{align} \bar e_j &= \frac12 \Big( \frac{v^2+v^{-2}}2 (\sigma_j^x \sigma_{j+1}^x + \sigma_j^y \sigma_{j+1}^y) +  \frac{v^2-v^{-2}}{2i} (\sigma_j^x \sigma_{j+1}^y - \sigma_j^y \sigma_{j+1}^x) \nonumber 
\\ & \hspace{5cm} - \frac{u^2+u^{-2}}2 (\sigma_j^z \sigma_{j+1}^z - id) + \frac{u^2-u^{-2}}2 (\sigma_j^z - \sigma_{j+1}^z) \Big) \nonumber \\
&= v^{-2}\sigma^-_j\sigma^+_{j+1} +v^2 \sigma^+_j\sigma^-_{j+1} - (u^2+u^{-2}) \sigma^+_j \sigma^-_j\sigma^+_{j+1}\sigma^-_{j+1} + u^2 \sigma^+_j \sigma^-_j + u^{-2}\sigma^+_{j+1}\sigma^-_{j+1} \label{eq:ebar} \\
  &=  \underbrace{id_2 \otimes id_2 \otimes \dots \otimes id_2}_{j-1} \otimes \, \bar{e} \otimes \underbrace{id_2 \otimes id_2 \otimes \dots \otimes id_2}_{N-j-1} \nonumber
\end{align}
with
\begin{equation}\bar{e} = \begin{pmatrix} 
0 & 0 & 0 & 0 \\
0 & u^2 & v^2 & 0 \\
0 & v^{-2}  & u^{-2} & 0 \\
0 & 0 & 0 & 0
\end{pmatrix},\end{equation}
where the allowed values for $j$ are from $1$ to $N$ for the first two forms and from $1$ to $N-1$ for the last. The periodic XXZ Hamiltonian found in \cite{PasquierSaleur} also depends upon a twist parameter (named $e^{i \varphi}$ therein) which only enters in the definition 
of the first generator $\bar e_N$. We note however that a similarity transformation $\mathcal O \bar e_i \mathcal O^{-1}$, with $\mathcal O = v^{\sum_{j=1}^N j \sigma^z_j}$, maps our generators to theirs if $e^{i\varphi} = v^{2N}$.

It is clear from the second form that each $\bar e_j$ commutes with the total spin $S^z=\frac12\sum_{1\le i\le N}\sigma_i^z$. The matrices $\bar e_j$ are not hermitian. But, if $u$ and $v$ are on the unit circle, the first three terms of the first form in \eqref{eq:ebar} are clearly hermitian. Only the term $\frac12(u^2-u^{-2})(\sigma^z_j-\sigma^z_{j+1})$ is not. Finally one can verify that these matrices satisfy the relations \eqref{eq:TLPN}, with $\bar{e}_{N+1} \equiv \bar{e}_1$ and $\beta = u^2 + u^{-2}$. Therefore
\begin{Proposition} The matrices $\bar e_i, 1\le i\le N$, form a representation of $PTL_N(\beta=u^2+u^{-2})$ for all $v\in\mathbb C^\times$.
\end{Proposition}
We shall often use the following parametrization for $u$ and $v$: $v = e^{i \mu}, u = e^{i \lambda/2}$. If $\mu$ and $\lambda$ are real, the Hamiltonian $H =H(u,v)= \sum_{i=1}^{N} \bar e_i$ is hermitian, since the sum $\sum_j (\sigma^z_j-\sigma^z_{j+1})$ vanishes. The usual XXZ model corresponds to the case $v^2=1$ (for the case with boundary see for example \cite{PasquierSaleur} and also \cite{AMDtoutseul} where the interplay between loop models and XXZ Hamiltonian has been exploited).

We finally introduce the matrices $t^{\pm 1}$ and $\bar \Omega^{\pm1}$. The operators $t$ and $t^{-1}$ are left and right translations around the cylinder. In the basis $|x_1 x_2 \dots x_{N} \rangle$ where every $x_i \in \{+1, -1 \}$,  they act as
\begin{align*}t \, |x_1 x_2 \dots x_{N} \rangle &=  |x_2 x_3\dots x_{N} x_1 \rangle \\
t^{-1} |x_1 x_2 \dots x_{N} \rangle &=  |x_{N} x_1 x_2 \dots x_{N-1} \rangle
\end{align*}
and satisfy $t^{\pm 1} \sigma_j^a = \sigma^a_{j\mp 1} t^{\pm 1}$. Then we define $\bar \Omega^{\pm 1} = v^{\pm 2S^z} t^{\pm 1}$. Because $t \bar e_j = \bar e_{j-1} t$ and $[v^{2S^z}, \bar e_j] = 0$, the first and second equations of (\ref{eq:ETLPN}), $\bar \Omega \bar e_j \bar \Omega^{-1}= \bar e_{j-1}$ and $\bar \Omega \bar \Omega^{-1} = \bar \Omega^{-1} \bar \Omega = id$, are both satisfied. That 
\begin{equation} (\bar \Omega^{\pm 1} \bar e_N)^{N-1} = \bar \Omega^{\pm N} (\bar \Omega ^{\pm 1} \bar  e_N) 
\label{eq:nontrivial}\end{equation}
holds is far less trivial. Moreover, to check that the matrices $\bar e_i$ and $\bar\Omega^{\pm 1}$ generate a representation of $\mathcal EPTL_N(\beta,\alpha)$, we would have also to show that, for $N$ even, 
\begin{equation}\label{eq:conjuOmega}
\bar E \bar \Omega^{\pm 1} \bar E = \alpha \bar E
\end{equation}
for $\bar E = \bar{e}_2 \bar{e}_4 \bar{e}_6 ... \bar{e}_{N}$ and some $\alpha = \alpha(u,v)$. There might be a way to prove \eqref{eq:nontrivial} and \eqref{eq:conjuOmega} by direct computation. We prefer a more roundabout way. Let us denote by $\tau$ the map defined on the generators by $e_i\mapsto \bar e_i$ and $\Omega^\pm\mapsto \bar\Omega^\pm$ and, in general, by $\prod_i f_i \mapsto \prod_i \bar f_i$ for $f_i \in \{e_1, ..., e_N, \Omega, \Omega^{-1}\}$. This map is a representation of $PTL_N(\beta)$. Moreover the two first relations in \eqref{eq:ETLPN} are satisfied if the generators are replaced by their barred representatives. In section \ref{sub:ind} we shall show that for generic values of $u$ and $v$, there exists an isomorphism $\tilde i_N^d$ of vector spaces between the XXZ and the link state representations $\omega_d$
that intertwines the $\Omega^\pm$ and $e_i$s, and their barred representatives: $\tilde i_N^d \circ \Omega^{\pm 1} = \bar \Omega^{\pm 1} \circ \tilde i_N^d $ and $\tilde i_N^d \circ e_i = \bar e_i \circ \tilde i_N^d $. It will then follow that equations \eqref{eq:nontrivial} and \eqref{eq:conjuOmega} are satisfied for {\em  all} values of $u$ and $v$ and that $\tau$ is a representation of $\mathcal EPTL_N(\beta,\alpha)$, with the parameter $\alpha$ equal to $v^N+v^{-N}$. Thus $v$ comes into play both as a twist parameter and in the weight of non-contractible loops wrapping around the cylinder.

\subsection{The map $\tilde i_N^d$ between link and spin states}\label{sec:linkAndSpin}

Let  $w \in \tilde B_N^d$ be a link state containing $n = (N-d)/2$ bubbles and let $\psi(w) = \{ (i_1, j_1), (i_2, j_2), ..., (i_n, j_n)\}$, where the $i_m$s are positions where bubbles start and the $j_m$s the positions where they end. The $i_m$s are chosen in the interval $1, ..., N$ while the $j_m$s satisfy $i_m+1 \le j_m \le N+i_m-1$. In $\psi(w)$, the pairs $(i_m, j_m)$ are ordered such that the $i_m$s increase, even though this will play no role.

\begin{Definition}
The linear transformation $\tilde{i}^d_N$: $\tilde{V}_N^d \rightarrow  \left.(\mathbb{C}^2)^{\otimes N}\right|_{S^z = d/2} $ is defined by its action on elements $w$ of the basis $\tilde B_N^d$:
\begin{equation}
\tilde{i}_N^d(w) = \prod_{(i,j) \in \psi(w)} \hspace{-0.4cm}\tilde{T}_{i,j} \hspace{0.1cm}| 0 \rangle \qquad \textrm{where} \qquad \tilde{T}_{i,j} = v^{j-i} u \sigma^-_j + v^{-(j-i)}u^{-1} \sigma^-_i, 
\label{eq:homomorphisme}\end{equation}
and $ | 0 \rangle =  | ++\dots + \, \rangle$ is the unique state with all spins up. \end{Definition}

%Some states have boundary bubbles, that is bubbles that cross the imaginary boundaries at $x=1/2$ and $x=N+1/2$. This happens if a point $i$ is connected to a point $j\ge N+1$. 
If a boundary bubble is present, its starting point $i$ is connected to a point $j\ge N+1$ and we then use the convention that  $\sigma^\pm_j \equiv \sigma^\pm_{j\textrm{mod} \vspace{-0.1cm} N}$. 

\begin{Proposition}
For all $c \in \mathcal E PTL_N(\beta,\alpha)$ %$u^2 + u^{-2}, v^N + v^{-N})$ 
and all $w \in \tilde{V}_N^d$, $\tilde{i}_N^d(c w) = \tau(c) \tilde i_N^d(w)$, where  $\beta = u^2 + u^{-2}$, $\alpha = v^N + v^{-N}$ and the action of $c$ on $w$ is the one defined for the 
representation $\omega_d$, with twist parameter $v$.
\label{sec:fermeture}
\end{Proposition}

Once equations \eqref{eq:nontrivial} and \eqref{eq:conjuOmega} have been established, this proposition will simply state that $\tilde i_N^d$ is an intertwiner between the representations $\omega_d$ and $\tau$ if the parameters of $\mathcal E PTL_N(\beta,\alpha)$ are chosen to be $\beta=u^2+u^{-2}$ and $\alpha=v^N+v^{-N}$.

\medskip

\noindent{\scshape Proof\ \ } It actually suffices to check that $\tilde{i}_N^d(c w) = \tau(c) \tilde i_N^d(w)$ holds if $c$ is one of the generators $e_i$, $\Omega$ or $\Omega^{-1}$, and $w \in \tilde B_N^d$. Let $Y(w) =  \prod_{(m,n)\in \psi'(w)} \tilde T_{m,n}$ and $\psi'(w)$ be the part of $\psi(w)$ that does not touch points $i$, $i+1$, $j$ and $k$ (see the diagrams below for the meaning of these indices). We give below a list of relations that are sufficient to establish the result. For each element of this list, we give the diagrammatic relation and its algebraic counterpart to be checked explicitly. For the diagrammatic relations, we draw in $w$ only the positions that play a role in the verification. For example, in the first, the check is for all vectors $w$ whose positions $i$ and $i+1$ are defects and, because $e_i$ acts only on these positions, they are the only ones drawn.
\begin{itemize}
\item[1)]
$ \tau\hspace{-0.1cm}\left(
\psset{unit=0.4}
\begin{pspicture}(-0.2,-1)(1.2,0.4)
\psset{linewidth=1pt}
\rput(0,-2.1){$i$}
\psdots[{linewidth=0.7pt}](0,0)(1,0)
\psdots[{linewidth=0.7pt}](0,-1.5)(1,-1.5)
\psset{linecolor=myc2}
\psarc{-}(0.5,0){0.5}{180}{360}
\psarc{-}(0.5,-1.5){0.5}{0}{180}
\end{pspicture} \right)
 \tilde i_N^d \hspace{-0.1cm}\left(
 \psset{unit=0.4}
 \begin{pspicture}(-0.2,0.0)(1.2,0.8)
\psset{linewidth=1pt}
\rput(0,-0.6){$i$}
\psdots[{linewidth=0.7pt}](0,0)(1,0)
\psset{linecolor=myc2}
\psline{-}(0,1)(0,0)
\psline{-}(1,1)(1,0)
\end{pspicture}
\right)
 = \tilde i_N^d\hspace{-0.1cm} \left(
\psset{unit=0.4}
\begin{pspicture}(-0.2,-1)(1.2,0.4)
\psset{linewidth=1pt}
\rput(0,-2.1){$i$}
\psdots[{linewidth=0.7pt}](0,0)(1,0)
\psdots[{linewidth=0.7pt}](0,-1.5)(1,-1.5)
\psset{linecolor=myc2}
\psline{-}(0,1)(0,0)
\psline{-}(1,1)(1,0)
\psarc{-}(0.5,0){0.5}{180}{360}
\psarc{-}(0.5,-1.5){0.5}{0}{180}
\end{pspicture}\right) =0 $ \vspace{0.3cm}\\$\,\rightarrow \,  \bar e_i  Y(w) |0\rangle = 0, $
\item[2)]
$ \tau\hspace{-0.1cm}\left(
\psset{unit=0.4}
\begin{pspicture}(-0.2,-1)(1.2,0.4)
\psset{linewidth=1pt}
\rput(0,-2.1){$i$}
\psdots[{linewidth=0.7pt}](0,0)(1,0)
\psdots[{linewidth=0.7pt}](0,-1.5)(1,-1.5)
\psset{linecolor=myc2}
\psarc{-}(0.5,0){0.5}{180}{360}
\psarc{-}(0.5,-1.5){0.5}{0}{180}
\end{pspicture} \right)
 \tilde i_N^d \hspace{-0.1cm}\left(
 \psset{unit=0.4}
 \begin{pspicture}(-0.2,0.0)(1.2,0.8)
\psset{linewidth=1pt}
\rput(0,-0.6){$i$}
\psdots[{linewidth=0.7pt}](0,0)(1,0)
\psset{linecolor=myc2}
\psarc{-}(0.5,0){0.5}{0}{180}
\end{pspicture}
\right)
  = (u^{2}+u^{-2}) \,
  \tilde i_N^d \hspace{-0.1cm}\left(
 \psset{unit=0.4}
 \begin{pspicture}(-0.2,0.0)(1.2,0.8)
\psset{linewidth=1pt}
\rput(0,-0.6){$i$}
\psdots[{linewidth=0.7pt}](0,0)(1,0)
\psset{linecolor=myc2}
\psarc{-}(0.5,0){0.5}{0}{180}
\end{pspicture}
\right) $ \vspace{0.3cm}\\$
   \rightarrow  \bar e_i  \tilde T_{i,i+1} Y(w) |0\rangle = (u^{2}+u^{-2}) \, \tilde T_{i,i+1} Y(w) |0\rangle, $
% \\
\item[3)]
$
\tau\hspace{-0.1cm}\left(
\psset{unit=0.4}
\begin{pspicture}(-0.2,-1.0)(3.4,0.4)
\psset{linewidth=1pt}
\rput(0,-2.1){$i$}
\rput(3,-2.1){$j$}
\rput(2.1,0){$\dots$}
\rput(2.1,-1.5){$\dots$}
\psdots[{linewidth=0.7pt}](0,0)(1,0)(3,0)
\psdots[{linewidth=0.7pt}](0,-1.5)(1,-1.5)(3,-1.5)
\psset{linecolor=myc2}
\psline{-}(3,0)(3,-1.5)
\psarc{-}(0.5,0){0.5}{180}{360}
\psarc{-}(0.5,-1.5){0.5}{0}{180}
\end{pspicture} 
\right)
 \tilde i_N^d \hspace{-0.1cm}\left(
 \psset{unit=0.4}  
 \begin{pspicture}(-0.4,0)(3.2,0.8)
 \rput(2.1,0){$\dots$}
 \rput(0,-0.6){$i$}
\rput(3,-0.6){$j$}
 \psdots[{linewidth=0.7pt}](0,0)(1,0)(3,0)
 \psset{linecolor=myc2}
 \psbezier{-}(1,0)(1,1.2)(3,1.2)(3,0)
 \psline{-}(0,0)(0,1)
  \end{pspicture}\right)
  = v^{i-j} \tilde i_N^d \hspace{-0.1cm}\left(
\psset{unit=0.4}  
\begin{pspicture}(-0.4,0)(3.2,0.8)
\psset{linewidth=1pt}
\rput(0,-0.6){$i$}
\rput(3,-0.6){$j$}
\psdots[{linewidth=0.7pt}](0,0)(1,0)(3,0)
\rput(2,0){$\dots$}
\psset{linecolor=myc2}
\psline{-}(3,0)(3,1)
\psarc{-}(0.5,0){0.5}{0}{180}
\end{pspicture} \right) $ \vspace{0.3cm}\\$
  \rightarrow \bar e_i \tilde T_{i+1,j} Y(w) |0\rangle =  v^{i-j} \tilde T_{i,i+1} Y(w) |0\rangle, 
$
%\\
\item[4)]
$
\tau\hspace{-0.1cm}\left(
\psset{unit=0.4}
\begin{pspicture}(-0.2,-1.)(3.4,0.4)
\psset{linewidth=1pt}
\psset{linewidth=1pt}
\rput(2,-2.1){$i$}
\rput(0,-2.1){$j$}
\rput(1.1,0){$\dots$}
\rput(1.1,-1.5){$\dots$}
\psdots[{linewidth=0.7pt}](0,0)(2,0)(3,0)
\psdots[{linewidth=0.7pt}](0,-1.5)(2,-1.5)(3,-1.5)
\psset{linecolor=myc2}
\psline{-}(0,0)(0,-1.5)
\psarc{-}(2.5,0){0.5}{180}{360}
\psarc{-}(2.5,-1.5){0.5}{0}{180}
\end{pspicture} 
\right)
\tilde  i_N^d \hspace{-0.1cm}\left(
 \psset{unit=0.4}  
 \begin{pspicture}(-0.4,0)(3.2,0.8)
 \rput(2,-0.6){$i$}
 \rput(0,-0.6){$j$}
 \rput(1.1,0){$\dots$}
 \psdots[{linewidth=0.7pt}](0,0)(2,0)(3,0)
 \psset{linecolor=myc2}
\psline{-}(3,0)(3,1)
\psbezier{-}(0,0)(0,1.2)(2,1.2)(2,0)
  \end{pspicture}\right)
  =  v^{i+1-j}\tilde i_N^d \hspace{-0.1cm}\left(
 \psset{unit=0.4}  
 \begin{pspicture}(-0.4,0)(3.2,0.8)
 \rput(2,-0.6){$i$}
 \rput(0,-0.6){$j$}
 \rput(1.1,0){$\dots$}
 \psdots[{linewidth=0.7pt}](0,0)(2,0)(3,0)
 \psset{linecolor=myc2}
\psline{-}(0,0)(0,1)
\psarc{-}(2.5,0){0.5}{0}{180}
  \end{pspicture}\right) $ \vspace{0.3cm}\\$
   \rightarrow \bar e_i  \tilde T_{j,i} Y(w) |0\rangle    = v^{i+1-j}  \tilde T_{i,i+1} Y(w) |0\rangle, 
$
%\\
\item[5)]
$
\tau\hspace{-0.1cm}\left(
\psset{unit=0.4}
\begin{pspicture}(0.8,-1.0)(6.4,0.4)
\psset{linewidth=1pt}
\psset{linewidth=1pt}
\rput(3.1,0){$\dots$}
\rput(3.1,-1.5){$\dots$}
\rput(5.1,0){$\dots$}
\rput(5.1,-1.5){$\dots$}
\rput(1,-2.1){$i$}
\rput(4,-2.1){$j$}
\rput(6,-2.1){$k$}
\psdots[{linewidth=0.7pt}](1,0)(2,0)(4,0)(6,0)
\psdots[{linewidth=0.7pt}](1,-1.5)(2,-1.5)(4,-1.5)(6,-1.5)
\psset{linecolor=myc2}
\psline{-}(4,0)(4,-1.5)
\psline{-}(6,0)(6,-1.5)
\psarc{-}(1.5,0){0.5}{180}{360}
\psarc{-}(1.5,-1.5){0.5}{0}{180}
\end{pspicture} 
\right)
\tilde  i_N^d \hspace{-0.1cm}\left(
 \psset{unit=0.4}  
 \begin{pspicture}(0.6,0)(6.4,0.8)
\rput(1,-0.6){$i$}
\rput(4,-0.6){$j$}
\rput(6,-0.6){$k$}
\rput(3.1,0){$\dots$}
\rput(5.1,0){$\dots$}
 \psdots[{linewidth=0.7pt}](1,0)(2,0)(4,0)(6,0)
 \psset{linecolor=myc2}
\psbezier{-}(2,0)(2,1.2)(4,1.2)(4,0)
\psbezier{-}(1,0)(1,2)(6,2)(6,0)
  \end{pspicture}\right)
  = \tilde  i_N^d \hspace{-0.1cm}\left(
 \psset{unit=0.4}  
 \begin{pspicture}(0.6,0)(6.4,0.8)
\rput(1,-0.6){$i$}
\rput(4,-0.6){$j$}
\rput(6,-0.6){$k$}
\rput(3.1,0){$\dots$}
\rput(5.1,0){$\dots$}
 \psdots[{linewidth=0.7pt}](1,0)(2,0)(4,0)(6,0)
 \psset{linecolor=myc2}
\psarc{-}(1.5,0){0.5}{0}{180}
\psbezier{-}(4,0)(4,1.2)(6,1.2)(6,0)
  \end{pspicture}\right)$ \vspace{0.3cm}\\$
     \rightarrow \bar e_i \tilde T_{i,k} \tilde T_{i+1,j} Y(w) |0\rangle =\tilde T_{i,i+1} \tilde T_{j,k} Y(w) |0\rangle, 
$ 
%\\
\item[6)]
$
\tau\hspace{-0.1cm}\left(
\psset{unit=0.4}
\begin{pspicture}(-0.2,-1.0)(5.4,0.4)
\psset{linewidth=1pt}
\psset{linewidth=1pt}
\rput(1.1,0){$\dots$}
\rput(1.1,-1.5){$\dots$}
\rput(4.1,0){$\dots$}
\rput(4.1,-1.5){$\dots$}
\rput(2,-2.1){$i$}
\rput(0,-2.1){$j$}
\rput(5,-2.1){$k$}
\psdots[{linewidth=0.7pt}](0,0)(2,0)(3,0)(5,0)
\psdots[{linewidth=0.7pt}](0,-1.5)(2,-1.5)(3,-1.5)(5,-1.5)
\psset{linecolor=myc2}
\psline{-}(0,0)(0,-1.5)
\psline{-}(5,0)(5,-1.5)
\psarc{-}(2.5,0){0.5}{180}{360}
\psarc{-}(2.5,-1.5){0.5}{0}{180}
\end{pspicture} 
\right)
\tilde  i_N^d \hspace{-0.1cm}\left(
 \psset{unit=0.4}  
 \begin{pspicture}(-0.4,0)(5.4,0.8)
\rput(1.1,0){$\dots$}
\rput(4.1,0){$\dots$}
\rput(2,-0.6){$i$}
\rput(0,-0.6){$j$}
\rput(5,-0.6){$k$}
\psdots[{linewidth=0.7pt}](0,0)(2,0)(3,0)(5,0)
 \psset{linecolor=myc2}
\psbezier{-}(0,0)(0,1.2)(2,1.2)(2,0)
\psbezier{-}(3,0)(3,1.2)(5,1.2)(5,0)  \end{pspicture}\right)
  = \tilde i_N^d \hspace{-0.1cm}\left(
 \psset{unit=0.4}  
 \begin{pspicture}(-0.4,0)(5.4,0.8)
\rput(1.1,0){$\dots$}
\rput(4.1,0){$\dots$}
\rput(2,-0.6){$i$}
\rput(0,-0.6){$j$}
\rput(5,-0.6){$k$}
\psdots[{linewidth=0.7pt}](0,0)(2,0)(3,0)(5,0)
 \psset{linecolor=myc2}
\psarc{-}(2.5,0){0.5}{0}{180}
\psbezier{-}(0,0)(0,1.6)(5,1.6)(5,0)  \end{pspicture}\right)$ \vspace{0.3cm}\\$
    \rightarrow  \bar e_i \tilde T_{j,i}  \tilde T_{i+1,k} Y(w) |0\rangle = \tilde T_{i,i+1} \tilde T_{j,k} Y(w) |0\rangle, 
$ 
%\\
\item[7)]
$
\tau\hspace{-0.1cm}\left(
\psset{unit=0.4}
\begin{pspicture}(-0.2,-1.0)(5.4,0.4)
\psset{linewidth=1pt}
\psset{linewidth=1pt}
\rput(1.1,0){$\dots$}
\rput(1.1,-1.5){$\dots$}
\rput(3.1,0){$\dots$}
\rput(3.1,-1.5){$\dots$}
\rput(4,-2.1){$i$}
\rput(0,-2.1){$j$}
\rput(2,-2.1){$k$}
\psdots[{linewidth=0.7pt}](0,0)(2,0)(4,0)(5,0)
\psdots[{linewidth=0.7pt}](0,-1.5)(2,-1.5)(4,-1.5)(5,-1.5)
\psset{linecolor=myc2}
\psline{-}(0,0)(0,-1.5)
\psline{-}(2,0)(2,-1.5)
\psarc{-}(4.5,0){0.5}{180}{360}
\psarc{-}(4.5,-1.5){0.5}{0}{180}
\end{pspicture} 
\right)
\tilde  i_N^d \hspace{-0.1cm}\left(
 \psset{unit=0.4}  
 \begin{pspicture}(-0.4,0)(5.4,0.8)
\rput(1.1,0){$\dots$}
\rput(3.1,0){$\dots$}
\rput(4,-0.6){$i$}
\rput(0,-0.6){$j$}
\rput(2,-0.6){$k$}
\psdots[{linewidth=0.7pt}](0,0)(2,0)(4,0)(5,0)
 \psset{linecolor=myc2}
\psbezier{-}(0,0)(0,2)(5,2)(5,0)
\psbezier{-}(2,0)(2,1.2)(4,1.2)(4,0)
\end{pspicture}\right)
  = \tilde i_N^d \hspace{-0.1cm}\left(
 \psset{unit=0.4}  
 \begin{pspicture}(-0.4,0)(5.4,0.8)
\rput(1.1,0){$\dots$}
\rput(3.1,0){$\dots$}
\rput(4,-0.6){$i$}
\rput(0,-0.6){$j$}
\rput(2,-0.6){$k$}
\psdots[{linewidth=0.7pt}](0,0)(2,0)(4,0)(5,0)
 \psset{linecolor=myc2}
\psarc{-}(4.5,0){0.5}{0}{180}
\psbezier{-}(0,0)(0,1.2)(2,1.2)(2,0)
  \end{pspicture}\right)$ \vspace{0.3cm}\\
    $
    \rightarrow \bar e_i \tilde T_{j,i+1} \tilde T_{k,i} Y(w) |0\rangle = \tilde T_{i,i+1} \tilde T_{j,k} Y(w) |0\rangle, 
$
\item[8)]
$
\tau\hspace{-0.1cm}\left(
\psset{unit=0.4}
\begin{pspicture}(-0.2,-1.0)(4.9,0.4)
\psset{linewidth=1pt}
\psset{linewidth=1pt}
\rput(0.6,0){$\dots$}
\rput(0.6,-1.5){$\dots$}
\rput(4.4,0){$\dots$}
\rput(4.4,-1.5){$\dots$}
\rput(2,-2.1){$i$}
\psdots[{linewidth=0.7pt}](2,0)(3,0)
\psdots[{linewidth=0.7pt}](2,-1.5)(3,-1.5)
\psset{linecolor=myc2}
\psarc{-}(2.5,0){0.5}{180}{360}
\psarc{-}(2.5,-1.5){0.5}{0}{180}
\end{pspicture} 
\right)
\tilde  i_N^d \hspace{-0.1cm}\left(
 \psset{unit=0.4}  
 \begin{pspicture}(-0.4,0)(5.4,0.8)
\rput(0.6,0){$\dots$}
\rput(4.4,0){$\dots$}
\rput(2,-0.6){$i$}
\psdots[{linewidth=0.7pt}](2,0)(3,0)
 \psset{linecolor=myc2}
\psbezier{-}(2,0)(2,0.8)(0.4,1)(0,1)
\psbezier{-}(3,0)(3,0.8)(4.6,1)(5,1)
\end{pspicture}\right)
  = (v^N + v^{-N})\tilde i_N^d \hspace{-0.1cm}\left(
 \psset{unit=0.4}  
 \begin{pspicture}(-0.4,0)(5.4,0.8)
\rput(0.6,0){$\dots$}
\rput(4.4,0){$\dots$}
\rput(2,-0.6){$i$}
\psdots[{linewidth=0.7pt}](2,0)(3,0)
 \psset{linecolor=myc2}
\psarc{-}(2.5,0){0.5}{0}{180}
  \end{pspicture}\right)$ \vspace{0.3cm}\\
    $\rightarrow \bar e_i \tilde T_{i+1,N+i} Y(w) |0\rangle = (v^{N}+v^{-N}) \tilde T_{i,i+1} Y(w) |0\rangle,
$ 
\item[9)]
$
\tau\hspace{-0.1cm}\left(
\psset{unit=0.4}
\begin{pspicture}(-0.5,-1.0)(5.5,0.4)
\psset{linewidth=1pt}
\psset{linewidth=1pt}
\rput(2.5,-0.75){$\dots$}
\psdots[{linewidth=0.7pt}](0,0)(1,0)(2,0)(3,0)(4,0)(5,0)
\psdots[{linewidth=0.7pt}](0,-1.5)(1,-1.5)(2,-1.5)(3,-1.5)(4,-1.5)(5,-1.5)
\psset{linecolor=myc2}
\psbezier{-}(0,0)(0,-0.5)(-0.3,-0.75)(-0.45,-0.75)
\psbezier{-}(5,-1.5)(5,-1)(5.3,-0.75)(5.45,-0.75)
\psbezier{-}(0,-1.5)(0,-0.75)(1,-0.75)(1,0)
\psbezier{-}(1,-1.5)(1,-0.75)(2,-0.75)(2,0)
\psbezier{-}(3,-1.5)(3,-0.75)(4,-0.75)(4,0)
\psbezier{-}(4,-1.5)(4,-0.75)(5,-0.75)(5,0)
\end{pspicture} 
\right)
\tilde  i_N^d (w)
  = \tilde i_N^d \hspace{-0.1cm}\left(\psset{unit=0.4}
\begin{pspicture}(-0.5,-1.0)(5.5,0.4)
\psset{linewidth=1pt}
\psset{linewidth=1pt}
\rput(2.5,-0.75){$\dots$}
\psdots[{linewidth=0.7pt}](0,0)(1,0)(2,0)(3,0)(4,0)(5,0)
\psdots[{linewidth=0.7pt}](0,-1.5)(1,-1.5)(2,-1.5)(3,-1.5)(4,-1.5)(5,-1.5)
\psset{linecolor=myc2}
\psbezier{-}(0,0)(0,-0.5)(-0.3,-0.75)(-0.45,-0.75)
\psbezier{-}(5,-1.5)(5,-1)(5.3,-0.75)(5.45,-0.75)
\psbezier{-}(0,-1.5)(0,-0.75)(1,-0.75)(1,0)
\psbezier{-}(1,-1.5)(1,-0.75)(2,-0.75)(2,0)
\psbezier{-}(3,-1.5)(3,-0.75)(4,-0.75)(4,0)
\psbezier{-}(4,-1.5)(4,-0.75)(5,-0.75)(5,0)
\end{pspicture} \, \, w \right)$ \vspace{0.3cm}\\
    $\rightarrow \bar \Omega^{\pm 1} \tilde i_N^d(w) = \tilde i_N^d(\Omega^{\pm 1}w)$. 
\end{itemize}
A few observations are useful. First, if the four indices of $\tilde T_{i,j}$ and $\tilde T_{k,l}$ are distinct, then these two linear maps commute. Second, because $Y(w)$ commutes with $\bar e_i$ and with those $\tilde T$s with indices in $\{i,i+1,j,k\}$, we can ignore it in our calculation. And third, the usual $\sigma_j^+\sigma_j^-|0\rangle=|0\rangle$ and $\sigma_j^+|0\rangle=0$ are keys in the computations to follow. Because of the latter identities, the relation 1) is trivially satisfied. Under the action of $\tau$, the number of defects is conserved, as it is in the representation $\omega_d$. The computations for the elements 2), 3) and 5) are
\begin{align*} \bar e_i \tilde T_{i,i+1} | 0 \rangle &=  \big( uv^{-1} \sigma^-_i + u^{3}v \sigma^-_{i+1} + u^{-1}v \sigma^-_{i+1} + u^{-3}v^{-1} \sigma^-_i\big) |0 \rangle \\
& = (u^2 + u^{-2}) (uv \sigma^-_{i+1} + (uv)^{-1} \sigma^-_{i}) |0 \rangle= (u^2 + u^{-2}) \tilde T_{i,i+1} |0 \rangle, \\
\bar e_i \tilde T_{i+1,j} |0\rangle &=  u^{-1} v^{-j+i+1} (v^{-2} \sigma_i^- + u^2 \sigma_{i+1}^-)| 0 \rangle=v^{i-j} \tilde T_{i,i+1} |0\rangle,\\
 \bar e_i \tilde T_{i,k} \tilde T_{i+1,j}| 0 \rangle &= \big((v^{k-i}u)(v^{-(j-i-1)}u^{-1}) (v^{-2} \sigma_i^-\sigma_k^- + u^{2} \sigma_k^-\sigma_{i+1}^-) \\ &\hspace{5 cm}+ (v^{-(k-i)}u^{-1})(v^{j-i-1}u (v^{2} \sigma_{i+1}^-\sigma_j^- + u^{-2} \sigma_i^-\sigma_{j}^-) \big) |0 \rangle \\
& = \big( uv \sigma_{i+1}^- (u v^{k-j} \sigma_k^- +u^{-1} v^{-(k-j)} \sigma_j^-) + (uv)^{-1} \sigma_{i}^- (u v^{k-j} \sigma_k^- +u^{-1} v^{-(k-j)} \sigma_j^-)\big) |0 \rangle  \\ &= \tilde T_{i,i+1}\tilde T_{j,k}|0 \rangle, \\
\intertext{and for 8), the link state $w$ has a boundary bubble connecting positions $i+1$ and $i+N$ and}
 \bar e_i \tilde T_{i+1,N+i}  |0\rangle &= (uv^{N-1}(v^2 \sigma_{i+1}^- + u^{-2}\sigma_i^-) +u^{-1}v^{1-N}(v^{-2}\sigma_i^- + u^2 \sigma_{i+1}^-))| 0 \rangle \\&= (v^{N}+v^{-N}) \tilde T_{i,i+1} |0\rangle.
\end{align*}
The proofs of 4), 6) and 7) are similar. For 9), from the definition of $\bar \Omega^{\pm 1}$, we have $\bar \Omega^{\pm 1} \tilde T_{i,j} = v^{\mp 2} \tilde T_{i\mp 1,j\mp 1} \bar \Omega^{\pm 1}$ and $v^{\pm 2 S^z} |0 \rangle = v^{\pm N} |0 \rangle$. In the subspace with $d$ defects, the number $|\psi(w)|$ of pairs $(i_m,j_m)$ in $\psi$ is $(N-d)/2$ and
\begin{align*}\bar \Omega^{\pm 1} \tilde i_N^d(w) &= \bar \Omega^{\pm 1}  \bigg( \prod_{(i,j) \in \psi(w)} \hspace{-0.3cm}\tilde{T}_{i,j} \bigg) | 0 \rangle  = v^{\mp 2 |\psi(w)|}\bigg( \prod_{(i,j) \in \psi(w)} \hspace{-0.3cm}\tilde{T}_{i \mp 1,j \mp 1} \bigg) \bar \Omega^{\pm 1}  | 0 \rangle \\
& = v^{\mp (N-d)} v^{\pm N} \bigg( \prod_{(i,j) \in \psi(w)} \hspace{-0.3cm}\tilde{T}_{i \mp 1,j \mp 1} \bigg) |0 \rangle = \tilde i_N^d(\Omega^{\pm 1} w),
 \end{align*} 
 as required. \hfill$\square$\\

The proofs of equations \eqref{eq:nontrivial} and \eqref{eq:conjuOmega} have been left out. To complete these, we will show that $\tilde i^d_N$ is an isomorphism between $\tilde V_N^d$ and the eigenspace where $S^z=d/2$, except for some critical values of  $u$ and $v$. Because the analogs of equations \eqref{eq:nontrivial} and \eqref{eq:conjuOmega} hold in the link state representation, then they will also hold in the spin eigenspace since the previous proposition showed that $\tilde i_N^d\circ\Omega^{\pm 1}=\bar\Omega^{\pm 1}\circ \tilde i_N^d$.

\subsection{The factorization of the Gram matrix in terms of the map $\tilde i_N^d$}
\label{sec:relGram}

Let $I^d_N$ be the matrix of the transformation $\tilde i^d_N$ expressed in the bases of link states for the domain and of spin states $\left\{\vec x  = |x_1x_2 \dots x_N\rangle, x_i\in\{+1,-1\}\text{\rm\ and\ }\sum_ix_i=d\right\}$ for $\left.(\mathbb C^2)^{\otimes N}\right|_{S^z=d/2}$, so that 
$\tilde i_N^d(w)  = \sum_{\vec x} | \vec x \rangle (I_N^d)_{\vec x, w}
$ for the basis vector $w\in\tilde B_N^d$. The matrix $I_N^d$ is square and has size $\left( \begin{smallmatrix} N \\ (N-d)/2 \end{smallmatrix} \right)$, and the linear map $\tilde i_N^d$ is an isomorphism if and only if $\det I_N^d \ne 0$.

Let $w_1$ and $w_2$ be link states $\in \tilde B_N$. The Gram diagram $D_\mathcal{G}(w_1,w_2)$ is obtained by taking the mirror image of $w_2$ by a horizontal axis and by connecting the entries of this state to those of $w_1$. 

\begin{Definition} 
The Gram product $\langle \cdot | \cdot \rangle_G : \tilde V_N\times\tilde V_N\rightarrow \mathbb C$ is a bilinear form defined on $w_1$ and $w_2\in\tilde B_N$ by 
\begin{equation*}
\langle w_1 | w_2 \rangle_G = \left\{ \begin{array}{l l} 0 & \quad  \textrm{if in $D_{\mathcal{G}}(w_1,w_2)$, two defects of $w_1$ (or $w_2$) are connected,}\\ \alpha^{n_\alpha}\beta^{n_\beta}v^{n_v}  & \quad  \textrm{otherwise,} \end{array} \right.
\end{equation*}
where $n_\beta$ and $n_\alpha$ are respectively the numbers of contractible and non-contractible closed loops in $D_\mathcal{G}(w_1,w_2)$, and $n_v = \sum_l \Delta_l$ where $\Delta_l$ is the displacement (to the left) of defect $l$ of $w_1$ connecting to another defect of $w_2$. Of course, only one of $n_\alpha$ or $n_v$ can be non-zero. 
\label{sec:dGram}\end{Definition}

Thus, if $w_1 \in \tilde B_N^{d_1}$ and $w_2 \in \tilde B_N^{d_2}$, $\langle w_1| w_2 \rangle_G = 0$ unless $d_1 = d_2$. The matrix of the Gram product restricted to $\tilde B_N^d$ is noted $\mathcal{\tilde G}_N^d$. Here are a few examples.
\begin{align*} & 1) \, \, w_1 =
\psset{unit=0.6}
\begin{pspicture}(5.3,-0.8)(10.3,-0.6)
\psdots(5.5,-1)(6,-1)(6.5,-1)(7,-1)(7.5,-1)(8,-1)(8.5,-1)(9,-1)(9.5,-1)(10,-1)
\psset{linewidth=1pt}
\psset{linecolor=myc2}
\psline{-}(5.5,-1)(5.5,0)
\psline{-}(10,-1)(10,0)
\psbezier{-}(6,-1)(6,0.5)(9.5,0.5)(9.5,-1)
\psbezier{-}(6.5,-1)(6.5,-0.25)(8.0,-0.25)(8,-1)
\psarc{-}(7.25,-1){0.25}{0}{180}
\psarc{-}(8.75,-1){0.25}{0}{180}
\end{pspicture},
 \,w_2 =
\begin{pspicture}(5.3,-0.8)(10.3,-0.6)
\psdots(5.5,-1)(6,-1)(6.5,-1)(7,-1)(7.5,-1)(8,-1)(8.5,-1)(9,-1)(9.5,-1)(10,-1)
\psset{linewidth=1pt}
\psset{linecolor=myc2}
\psline{-}(5.5,-1)(5.5,0)
\psline{-}(6,-1)(6,0)
\psbezier{-}(7.5,-1)(7.5,-0.25)(9.0,-0.25)(9,-1)
\psarc{-}(6.75,-1){0.25}{0}{180}
\psarc{-}(8.25,-1){0.25}{0}{180}
\psarc{-}(9.75,-1){0.25}{0}{180}
\end{pspicture}
, \, D_{\mathcal{G}}(w_1,w_2) = 
\begin{pspicture}(5.3,-1.2)(10.3,-0.6)
\psdots(5.5,-1)(6,-1)(6.5,-1)(7,-1)(7.5,-1)(8,-1)(8.5,-1)(9,-1)(9.5,-1)(10,-1)
\psset{linewidth=1pt}
\psset{linecolor=myc2}
\psline{-}(5.5,-1)(5.5,0)
\psline{-}(10,-1)(10,0)
\psbezier{-}(6,-1)(6,0.5)(9.5,0.5)(9.5,-1)
\psbezier{-}(6.5,-1)(6.5,-0.25)(8.0,-0.25)(8,-1)
\psarc{-}(7.25,-1){0.25}{0}{180}
\psarc{-}(8.75,-1){0.25}{0}{180}
\psline{-}(5.5,-1)(5.5,-2)
\psline{-}(6,-1)(6,-2)
\psbezier{-}(7.5,-1)(7.5,-1.75)(9,-1.75)(9,-1)
\psarc{-}(6.75,-1){0.25}{180}{360}
\psarc{-}(8.25,-1){0.25}{180}{360}
\psarc{-}(9.75,-1){0.25}{180}{360}
\end{pspicture}
, \, \langle w_1| w_2 \rangle = \beta v^{8},
%\end{equation*}
%
%
%\begin{equation*}
\\ \\ \\ & 2)  \, \, w_1 =
\psset{unit=0.6}
\begin{pspicture}(5.3,-0.8)(10.3,-0.6)
\psdots(5.5,-1)(6,-1)(6.5,-1)(7,-1)(7.5,-1)(8,-1)(8.5,-1)(9,-1)(9.5,-1)(10,-1)
\psset{linewidth=1pt}
\psset{linecolor=myc2}
\psline{-}(5.5,-1)(5.5,0)
\psline{-}(10,-1)(10,0)
\psbezier{-}(6,-1)(6,0.5)(9.5,0.5)(9.5,-1)
\psbezier{-}(6.5,-1)(6.5,-0.25)(8.0,-0.25)(8,-1)
\psarc{-}(7.25,-1){0.25}{0}{180}
\psarc{-}(8.75,-1){0.25}{0}{180}
\end{pspicture},
 \,w_2 =
\begin{pspicture}(5.3,-0.8)(10.3,-0.6)
\psdots(5.5,-1)(6,-1)(6.5,-1)(7,-1)(7.5,-1)(8,-1)(8.5,-1)(9,-1)(9.5,-1)(10,-1)
\psset{linewidth=1pt}
\psset{linecolor=myc2}
\psline{-}(8.5,-1)(8.5,0)
\psline{-}(9,-1)(9,0)
\psbezier{-}(6.0,-1)(6.0,-0.25)(7.5,-0.25)(7.5,-1)
\psbezier{-}(5.5,-1)(5.5,0.25)(8.0,0.25)(8.0,-1)
\psarc{-}(6.75,-1){0.25}{0}{180}
\psarc{-}(9.75,-1){0.25}{0}{180}
\end{pspicture}
, \, D_{\mathcal{G}}(w_1,w_2) = 
\begin{pspicture}(5.3,-1.2)(10.3,-0.6)
\psdots(5.5,-1)(6,-1)(6.5,-1)(7,-1)(7.5,-1)(8,-1)(8.5,-1)(9,-1)(9.5,-1)(10,-1)
\psset{linewidth=1pt}
\psset{linecolor=myc2}
\psline{-}(5.5,-1)(5.5,0)
\psline{-}(10,-1)(10,0)
\psbezier{-}(6,-1)(6,0.5)(9.5,0.5)(9.5,-1)
\psbezier{-}(6.5,-1)(6.5,-0.25)(8.0,-0.25)(8,-1)
\psarc{-}(7.25,-1){0.25}{0}{180}
\psarc{-}(8.75,-1){0.25}{0}{180}
\psline{-}(8.5,-1)(8.5,-2)
\psline{-}(9,-1)(9,-2)
\psbezier{-}(6.0,-1)(6.0,-1.75)(7.5,-1.75)(7.5,-1)
\psbezier{-}(5.5,-1)(5.5,-2.25)(8.0,-2.25)(8.0,-1)
\psarc{-}(6.75,-1){0.25}{180}{0}
\psarc{-}(9.75,-1){0.25}{180}{0}
\end{pspicture}
, \, \langle w_1| w_2 \rangle = 0,
%\end{equation*}
%
%
%\begin{equation*}
\\ \\ \\ & 3)  \, \, w_1 =
\psset{unit=0.6}
\begin{pspicture}(5.3,-0.8)(10.3,-0.6)
\psdots(5.5,-1)(6,-1)(6.5,-1)(7,-1)(7.5,-1)(8,-1)(8.5,-1)(9,-1)(9.5,-1)(10,-1)
\psset{linewidth=1pt}
\psset{linecolor=myc2}
%\psbezier{-}(8.5,-1)(8.5,-0.25)(10.0,-0.25)(10,-1)
\psbezier{-}(7,-1)(7,-0.25)(8.5,-0.25)(8.5,-1)
\psarc{-}(5.25,-1){0.25}{0}{90}
\psarc{-}(6.25,-1){0.25}{0}{180}
\psarc{-}(7.75,-1){0.25}{0}{180}
\psarc{-}(9.25,-1){0.25}{0}{180}
\psarc{-}(10.25,-1){0.25}{90}{180}
%\psarc{-}(8.75,-1){0.25}{0}{180}
\end{pspicture},
 \,w_2 =
\begin{pspicture}(5.3,-0.8)(10.3,-0.6)
\psdots(5.5,-1)(6,-1)(6.5,-1)(7,-1)(7.5,-1)(8,-1)(8.5,-1)(9,-1)(9.5,-1)(10,-1)
\psset{linewidth=1pt}
\psset{linecolor=myc2}
\psbezier{-}(7,-1)(7,-0.25)(5.5,-0)(5.25,-0)
\psbezier{-}(8.5,-1)(8.5,-0.25)(10.0,-0)(10.25,-0)
\psbezier{-}(7.5,-1)(7.5,-0.1)(5.5,0.35)(5.25,0.35)
\psbezier{-}(8,-1)(8,-0.1)(10.0,0.35)(10.25,0.35)
\psarc{-}(5.25,-1){0.25}{0}{90}
\psarc{-}(6.25,-1){0.25}{0}{180}
\psarc{-}(9.25,-1){0.25}{0}{180}
\psarc{-}(10.25,-1){0.25}{90}{180}
\end{pspicture}
, \, D_{\mathcal{G}}(w_1,w_2) = 
\begin{pspicture}(5.3,-1.2)(10.3,-0.6)
\psdots(5.5,-1)(6,-1)(6.5,-1)(7,-1)(7.5,-1)(8,-1)(8.5,-1)(9,-1)(9.5,-1)(10,-1)
\psset{linewidth=1pt}
\psset{linecolor=myc2}
\psbezier{-}(7,-1)(7,-0.25)(8.5,-0.25)(8.5,-1)
\psarc{-}(5.25,-1){0.25}{0}{90}
\psarc{-}(6.25,-1){0.25}{0}{180}
\psarc{-}(7.75,-1){0.25}{0}{180}
\psarc{-}(9.25,-1){0.25}{0}{180}
\psarc{-}(10.25,-1){0.25}{90}{180}
\psbezier{-}(7,-1)(7,-1.75)(5.5,-2)(5.25,-2)
\psbezier{-}(8.5,-1)(8.5,-1.75)(10.0,-2)(10.25,-2)
\psbezier{-}(7.5,-1)(7.5,-1.9)(5.5,-2.35)(5.25,-2.35)
\psbezier{-}(8,-1)(8,-1.9)(10.0,-2.35)(10.25,-2.35)
\psarc{-}(5.25,-1){0.25}{270}{0}
\psarc{-}(6.25,-1){0.25}{180}{0}
\psarc{-}(9.25,-1){0.25}{180}{0}
\psarc{-}(10.25,-1){0.25}{180}{270}
\end{pspicture}
, \, \langle w_1| w_2 \rangle = \alpha^2 \beta^3.
\end{align*}\ 
\\ 
\noindent For simplicity, we will sometimes denote $\langle w_1| w_2 \rangle_G$ by the corresponding Gram diagram. With this convention, example 3) becomes
\vspace{-0.2cm}

\begin{equation*}
\psset{unit=0.6}
%D_{\mathcal{G}}(v_1,v_2) = 
\begin{pspicture}(5.3,-1.2)(10.3,-0.6)
\psdots(5.5,-1)(6,-1)(6.5,-1)(7,-1)(7.5,-1)(8,-1)(8.5,-1)(9,-1)(9.5,-1)(10,-1)
\psset{linewidth=1pt}
\psset{linecolor=myc2}
\psbezier{-}(7,-1)(7,-0.25)(8.5,-0.25)(8.5,-1)
\psarc{-}(5.25,-1){0.25}{0}{90}
\psarc{-}(6.25,-1){0.25}{0}{180}
\psarc{-}(7.75,-1){0.25}{0}{180}
\psarc{-}(9.25,-1){0.25}{0}{180}
\psarc{-}(10.25,-1){0.25}{90}{180}
\psbezier{-}(7,-1)(7,-1.75)(5.5,-2)(5.25,-2)
\psbezier{-}(8.5,-1)(8.5,-1.75)(10.0,-2)(10.25,-2)
\psbezier{-}(7.5,-1)(7.5,-1.9)(5.5,-2.35)(5.25,-2.35)
\psbezier{-}(8,-1)(8,-1.9)(10.0,-2.35)(10.25,-2.35)
\psarc{-}(5.25,-1){0.25}{270}{0}
\psarc{-}(6.25,-1){0.25}{180}{0}
\psarc{-}(9.25,-1){0.25}{180}{0}
\psarc{-}(10.25,-1){0.25}{180}{270}
\end{pspicture}
 = \alpha^2 \beta^3.
\end{equation*}
\\ 
\\
\noindent Here are some remarks on this bilinear form. First the Gram product verifies $\langle w_1 | e_i w_2 \rangle_G = \langle e_i w_1 |  w_2 \rangle_G$ for $1 \le i\le N$ and $w_1 , w_2 \in \tilde B_N$, where the $e_i$s act as $\omega_d(v)$ on the first entry of the bilinear form and as $\omega_d(v^{-1})$ on the second. It is then natural to define the adjoint $u^\dagger$ of a word $u=e_{i_1}\dots e_{i_{k-1}}e_{i_k}$ in the generators as $e_{i_k}e_{i_{k-1}}\dots e_{i_1}$. With that definition, $\langle w_1 | u w_2 \rangle_G = \langle u^\dagger w_1 |  w_2 \rangle_G$ for all words $u$ and $w_1, w_2$. The second remark follows from the first: The radical of the Gram bilinear form $\textrm{Rad}_N^d=\{w\in \tilde V_N^d | \langle v | w\rangle_G=0, \textrm{for all }v\in\tilde V_N^d\}$ is a subspace of $\tilde V_N^d$ stable under the action of $\mathcal EPTL_N$. Third the matrix $\tilde{\mathcal G}_N^d$ is symmetric for $d=0$. For $d>0$, the Gram product is not symmetric but still satisfies $\langle w_1|w_2 \rangle_G = \left.\langle w_2|w_1 \rangle_G\right|_{v\rightarrow v^{-1}}$ for $w_1,w_2\in\tilde B_N$. Finally it will be useful to consider the Gram matrix restricted to link states in $B_N^d$. This matrix will be noted $\mathcal{G}_N^d$ without the ``$\tilde{\phantom A}$''. 

The relation between $I_N^d$ and the Gram product is given by the following proposition.
\begin{Theoreme} Let $Q_N^d = \big(I_N^d(u,v^{-1})\big)^TI_N^d(u,v)$. Then $Q_N^d = \mathcal{ \tilde G}_N^d $ with $\beta = u^2 + u^{-2}$, $\alpha = v^N + v^{-N}$ and twist parameter $v$.
\label{sec:homoGram}
\end{Theoreme}

\noindent Note that the $v^{-1}$ in $\big(I_N^d(u,v^{-1})\big)^T$ is consistent with the previous remark that $\tilde{ \mathcal G}_N^d =(\tilde{ \mathcal G}_N^d)^T\big|_{v \rightarrow v^{-1}} $. Here is a simple example of this remarkable factorisation for $N=4$ and $d=0$. The bases are ordered as $$\left\{\ 
| +-+- \rangle,\ \ 
| ++-- \rangle,\ \ 
| -++- \rangle,\ \ 
| -+-+ \rangle,\ \ 
| +--+ \rangle,\ \ 
| --++ \rangle
\ \right\}$$
for the spin basis and as (\ref{eq:linkbasis40}) for the link state basis. The matrices are
\begin{equation}\label{eq:exemplei42}
I_4^0(u,v)=
\left(
\begin{smallmatrix}
 u^2 v^2 & v^2 & v^{-2} & u^{-2}v^{-2} & v^{-2} & v^2 \\
 0 & u^2 v^4 & 0 & 1 & 0 & u^{-2}v^{-4} \\
 1 & 0 & u^2 v^4 & 0 & u^{-2}v^{-4} & 0 \\
u^{-2}v^{-2} & v^{-2} & v^2 & u^2 v^2 & v^2 & v^{-2} \\
 1 & 0 & u^{-2}v^{-4} & 0 & u^2 v^4 & 0 \\
 0 & u^{-2}v^{-4} & 0 & 1 & 0 & u^2 v^4
% u^{-2} v^{-4} & 1 & 0 & 0 & 0 & u^2 v^4 \\
% v^2 & u^{-2} v^{-2} & v^{-2} &
%   v^{-2} & u^2 v^2 & v^2 \\
% 0 & 0 & u^{-2} v^{-4} & u^2 v^4 & 1 & 0 \\
% 0 & 0 & u^2 v^4 & u^{-2} v^{-4} & 1 & 0 \\
% v^{-2} & u^2 v^2 & v^2 & v^2 & u^{-2}
%   v^{-2} & v^{-2} \\
% u^2 v^4 & 1 & 0 & 0 & 0 & u^{-2} v^{-4} \\
\end{smallmatrix}
\right)\quad\textrm{and}\quad
\tilde{\mathcal G}_4^0
=
\left(
\begin{smallmatrix}
 \beta ^2 & \beta  & \alpha  \beta  & \alpha  & \alpha  \beta  & \beta  \\
 \beta  & \beta ^2 & \alpha  & \alpha  \beta  & \alpha  & \alpha ^2 \\
 \alpha  \beta  & \alpha  & \beta ^2 & \beta  & \alpha ^2 & \alpha  \\
 \alpha  & \alpha  \beta  & \beta  & \beta ^2 & \beta  & \alpha  \beta  \\
 \alpha  \beta  & \alpha  & \alpha ^2 & \beta  & \beta ^2 & \alpha  \\
 \beta  & \alpha ^2 & \alpha  & \alpha  \beta  & \alpha  & \beta ^2
\end{smallmatrix}
\right).
%\left(
%\begin{smallmatrix}
% \beta ^2 & \alpha  \beta  & \alpha  & \alpha  & \beta
%    & \alpha ^2 \\
% \alpha  \beta  & \beta ^2 & \beta  & \beta  & \alpha 
%   & \alpha  \beta  \\
% \alpha  & \beta  & \beta ^2 & \alpha ^2 & \alpha 
%   \beta  & \alpha  \\
% \alpha  & \beta  & \alpha ^2 & \beta ^2 & \alpha 
%   \beta  & \alpha  \\
% \beta  & \alpha  & \alpha  \beta  & \alpha  \beta  &
%   \beta ^2 & \beta  \\
% \alpha ^2 & \alpha  \beta  & \alpha  & \alpha  &
%   \beta  & \beta ^2 \\
%\end{smallmatrix}
%\right)
\end{equation}
The equality can be checked by doing the product $\big(I_4^0(u,v^{-1})\big)^TI_4^0(u,v)$ and replacing in $\tilde{\mathcal G}_4^0$ the two variables $\alpha$ and $\beta$ by $v^4+v^{-4}$ and $u^2+u^{-2}$ respectively. Clearly this factorization is non-trivial.

\medskip

\noindent{\scshape Proof\ \ } We see $Q_N^d$ as an endomorphism of $\tilde V_N^d$ whose matrix elements in the link state basis are
\begin{equation*}
(Q_N^d)_{w_1,w_2} = \sum_{\vec x} \big(I_N^d(u,v^{-1})\big)^T_{w_1, \vec x}(I_N^d(u,v))_{\vec x, w_2}.
\end{equation*}
With the usual scalar product on spin states ($\langle x_1x_2\dots x_N|y_1y_2\dots y_N\rangle=\prod_i \delta_{x_i,y_i}$ for $x_i,y_i\in\{+1,-1\}$), matrix elements can be rewritten as 
\begin{equation*}
\big(I_N^d(u,v^{-1})\big)^T_{w_1, \vec x} = \big(I^d_N(u,v^{-1})\big)_{\vec x, w_1}= \langle \vec x |  \hspace{-0.1cm}\prod_{(i,j) \in \psi(w_1)} \hspace{-0.5cm}\tilde{T}_{i,j}(u,v^{-1}) \hspace{0.2cm} | 0 \rangle =  \langle  0 |  \hspace{-0.1cm}\prod_{(i,j) \in \psi(w_1)} \hspace{-0.5cm} \tilde{T}^T_{i,j}(u,v^{-1}) \hspace{0.2cm} | \vec{x} \rangle,
\end{equation*}
and then
\begin{align*}
(Q_N^d)_{w_1,w_2} &= \sum_{\vec x} \hspace{0.2cm}\langle  0 | \hspace{-0.1cm} \prod_{(i,j) \in \psi(w_1)} \hspace{-0.5cm} \tilde{T}^T_{i,j}(u,v^{-1}) \hspace{0.1cm} | \vec{x} \rangle \hspace{0.1cm} \langle \vec x | \hspace{-0.1cm} \prod_{(k,l) \in \psi(w_2)} \hspace{-0.5cm} \tilde{T}_{k,l}(u,v) \hspace{0.1cm} | 0 \rangle  \\
& =  \langle  0 | \hspace{-0.1cm} \prod_{(i,j) \in \psi(w_1)} \hspace{-0.5cm}\tilde{T}^T_{i,j}(u,v^{-1}) \hspace{-0.2cm}   \prod_{(k,l) \in \psi(w_2)} \hspace{-0.5cm}\tilde{T}_{k,l}(u,v) \hspace{0.2cm} | 0 \rangle.
\end{align*}
Notice that all the elements of the second product commute among each other, and the same goes for those of the first. However, elements of the first product do not commute with some elements of the second. More precisely, an element $\tilde T_{i,j}^T$ of the first product does not commute with a $\tilde T_{k,l}$ of the second if $\{i,j\}\cap\{k,l\}\neq \emptyset$. This suggests to break down the above products into clusters, namely subsets of indices corresponding to connected components of the Gram diagram $D_{\mathcal G}(w_1,w_2)$. Such a cluster is the set of labels visited by one loop or one defect in the diagram. Then $(Q_N^d)_{w_1,w_2}$ factors as a product over clusters:
\begin{equation*}
(Q_N^d)_{w_1,w_2} = \langle  0 |   \prod_{m=1}^{n_c}  \bigg( \prod_{(i,j) \in \psi(w_1) \cap c_m} \hspace{-0.5cm}\tilde T^T_{i,j}(u,v^{-1})\hspace{-0.3cm}\prod_{(k,l) \in \psi(w_2) \cap c_m} \hspace{-0.5cm}\tilde T_{k,l}(u,v) \bigg)| 0 \rangle,
\end{equation*}
where the $c_m$s are the clusters and $n_c$ is their number in $D_{\mathcal G}(w_1,w_2)$. We want to show that 
\begin{equation}\label{eq:QNd}
(Q_N^d)_{w_1,w_2}=\left\{ \begin{array}{l l} 0 & \quad  \textrm{if two defects of $w_1$ or $w_2$ are connected,}\\ (v^N + v^{-N})^{n_\alpha} (u^2 + u^{-2})^{n_\beta} v^{n_v}& \quad  \textrm{otherwise,} \end{array} \right.
\end{equation}
where $n_\alpha$, $n_\beta$ and $n_v$ have been defined in definition \ref{sec:dGram}.
To show this, we can concentrate on a single cluster, say the one containing the point $i$, and simplify it by removing pairs of indices recursively. More precisely, we proceed as in proposition \ref{sec:fermeture}: We identify local relations between the $\tilde T$s and $\tilde{T}^T$s that, if true, would show that each cluster gives rise to its proper contribution in the final result of \eqref{eq:QNd}. Here are these local relations where the $u$ and $v$ dependence of $\tilde T_{i,j}(u,v)$ and $\tilde T^T_{i,j}(u,v^{-1})$ has been removed and therefore $T_{i,j}$ and $T_{i,j}^T$ stand for $\tilde T_{i,j}(u,v)$ and $\tilde T_{i,j}^T(u,v^{-1}) $ respectively.
\begin{itemize}
\item[1)]
\begin{equation*} 
\psset{unit=0.6}
\begin{pspicture}(-4.2,-1.2)(1.2,-0.6)
\psset{linewidth=1pt}
\rput(0,-1.5){$i$}
\rput(1,-1.5){$j$}
\psdots(0,-1)(1,-1)
\psset{linecolor=myc2}
%\psline{-}(0,1)(0,0)
%\psline{-}(1,1)(1,0)
\psarc{-}(0.5,-1){0.5}{0}{360}
\end{pspicture} 
 =u^2 + u^{-2}   \qquad \rightarrow \qquad  T_{i,j}^T  T_{i,j} |0\rangle = (u^2+u^{-2})|0\rangle, \end{equation*}
\item[2)]
\begin{equation*} 
\psset{unit=0.6}
\begin{pspicture}(-4.63,-1.2)(1.5,-0.6)
\psset{linewidth=1pt}
\rput(0,-1.5){$i$}
\rput(1,-1.5){$j$}
\psdots(0,-1)(1,-1)
\psset{linecolor=myc2}
\psarc{-}(0.5,-1){0.5}{180}{360}
\psarc{-}(1.5,-1){0.5}{90}{180}
\psarc{-}(-0.5,-1){0.5}{0}{90}
\end{pspicture} 
 = v^N + v^{-N}   \qquad \rightarrow \qquad  T_{i,j}^T  T_{j,i+N} |0\rangle = (v^N + v^{-N}) |0\rangle, \end{equation*}
\item[3)]
\begin{equation*} 
\psset{unit=0.6}
\begin{pspicture}(-2.32,-1.2)(1.2,-0.6)
\psset{linewidth=1pt}
\rput(0,-1.5){$i$}
\rput(1,-1.5){$j$}
\psdots(0,-1)(1,-1)
\psset{linecolor=myc2}
\psline{-}(0,-1)(0,0)
\psline{-}(1,-1)(1,0)
\psarc{-}(0.5,-1){0.5}{180}{360}
\end{pspicture} 
 = 0 =  \begin{pspicture}(-0.2,-1.2)(1.2,-0.6)
\psset{linewidth=1pt}
\rput(0,-0.5){$i$}
\rput(1,-0.5){$j$}
\psdots(0,-1)(1,-1)
\psset{linecolor=myc2}
\psline{-}(0,-1)(0,-2)
\psline{-}(1,-1)(1,-2)
\psarc{-}(0.5,-1){0.5}{0}{180}
\end{pspicture}   \qquad \rightarrow \qquad  T_{i,j}^T  |0\rangle = 0 = \langle 0 |  T_{i,j}, \end{equation*}
\item[4)]
\begin{equation*} 
\psset{unit=0.6}
\begin{pspicture}(-0.14,-1.2)(3.2,-0.6)
\psset{linewidth=1pt}
\rput(0,-1.5){$i$}
\rput(1,-1.5){$j$}
\rput(2,-1.5){$k$}
\rput(3,-1.5){$l$}
\psdots(0,-1)(1,-1)(2,-1)(3,-1)
\psset{linecolor=myc2}
%\psline{-}(0,-1)(0,0)
%\psline{-}(1,-1)(1,0)
\psarc{-}(0.5,-1){0.5}{00}{180}
\psarc{-}(2.5,-1){0.5}{00}{180}
\psarc{-}(1.5,-1){0.5}{180}{360}
\end{pspicture} 
 = \begin{pspicture}(-0.2,-1.2)(3.2,-0.6)
\psset{linewidth=1pt}
\rput(0,-1.5){$i$}
\rput(1,-1.5){$j$}
\rput(2,-1.5){$k$}
\rput(3,-1.5){$l$}
\psdots(0,-1)(1,-1)(2,-1)(3,-1)
\psset{linecolor=myc2}
%\psline{-}(0,-1)(0,0)
%\psline{-}(1,-1)(1,0)
\psbezier{-}(0,-1)(0,0)(3,0)(3,-1)
\end{pspicture}    \qquad \rightarrow \qquad  T_{j,k}^T  T_{i,j}  T_{k,l}  |0\rangle \simeq    T_{i,l} |0\rangle, \end{equation*}
 \item[5)]
\begin{equation*} 
\psset{unit=0.6}
\begin{pspicture}(-0.14,-1.2)(3.2,-0.6)
\psset{linewidth=1pt}
\rput(0,-1.5){$i$}
\rput(1,-1.5){$j$}
\rput(2,-1.5){$k$}
\rput(3,-1.5){$l$}
\psdots(0,-1)(1,-1)(2,-1)(3,-1)
\psset{linecolor=myc2}
\psbezier{-}(0,-1)(0,0)(3,0)(3,-1)
\psarc{-}(1.5,-1){0.5}{00}{180}
\psarc{-}(0.5,-1){0.5}{180}{360}
\end{pspicture} 
 = \begin{pspicture}(-0.2,-1.2)(3.2,-0.6)
\psset{linewidth=1pt}
\rput(0,-1.5){$i$}
\rput(1,-1.5){$j$}
\rput(2,-1.5){$k$}
\rput(3,-1.5){$l$}
\psdots(0,-1)(1,-1)(2,-1)(3,-1)
\psset{linecolor=myc2}
\psarc{-}(2.5,-1){0.5}{0}{180}
\end{pspicture}    \qquad \rightarrow \qquad  T_{i,j}^T  T_{i,l}  T_{j,k}  |0\rangle \simeq    T_{k,l} |0\rangle, \end{equation*}
\item[6)]
\begin{equation*} 
\psset{unit=0.6}
\begin{pspicture}(-0.14,-1.2)(3.2,-0.6)
\psset{linewidth=1pt}
\rput(0,-1.5){$i$}
\rput(1,-1.5){$j$}
\rput(2,-1.5){$k$}
\rput(3,-1.5){$l$}
\psdots(0,-1)(1,-1)(2,-1)(3,-1)
\psset{linecolor=myc2}
\psbezier{-}(0,-1)(0,0)(3,0)(3,-1)
\psarc{-}(1.5,-1){0.5}{00}{180}
\psarc{-}(2.5,-1){0.5}{180}{360}
\end{pspicture} 
 = \begin{pspicture}(-0.2,-1.2)(3.2,-0.6)
\psset{linewidth=1pt}
\rput(0,-1.5){$i$}
\rput(1,-1.5){$j$}
\rput(2,-1.5){$k$}
\rput(3,-1.5){$l$}
\psdots(0,-1)(1,-1)(2,-1)(3,-1)
\psset{linecolor=myc2}
%\psline{-}(0,-1)(0,0)
%\psline{-}(1,-1)(1,0)
\psarc{-}(0.5,-1){0.5}{0}{180}
\end{pspicture}    \qquad \rightarrow \qquad  T_{k,l}^T  T_{i,l}  T_{j,k}  |0\rangle \simeq    T_{i,j} |0\rangle, \end{equation*}
\item[7)]
\begin{equation*} 
\psset{unit=0.6}
\begin{pspicture}(-0.38,-1.2)(2.2,-0.6)
\psset{linewidth=1pt}
\rput(0,-1.5){$i$}
\rput(1,-1.5){$j$}
\rput(2,-1.5){$k$}
\psdots(0,-1)(1,-1)(2,-1)
\psset{linecolor=myc2}
\psarc{-}(1.5,-1){0.5}{00}{180}
\psarc{-}(0.5,-1){0.5}{180}{360}
\psline{-}(0,-1)(0,0)
\end{pspicture} 
 = v^{i-k}\begin{pspicture}(-0.2,-1.2)(2.2,-0.6)
\psset{linewidth=1pt}
\rput(0,-1.5){$i$}
\rput(1,-1.5){$j$}
\rput(2,-1.5){$k$}
\psdots(0,-1)(1,-1)(2,-1)
\psset{linecolor=myc2}
\psline{-}(2,-1)(2,0)
\end{pspicture}    \qquad \rightarrow \qquad  T_{i,j}^T  T_{j,k}  |0\rangle =  v^{i-k}  |0\rangle, \end{equation*}

\item[8)]
\begin{equation*} 
\psset{unit=0.6}
\begin{pspicture}(-0.38,-1.2)(2.2,-0.6)
\psset{linewidth=1pt}
\rput(0,-1.5){$i$}
\rput(1,-1.5){$j$}
\rput(2,-1.5){$k$}
\psdots(0,-1)(1,-1)(2,-1)
\psset{linecolor=myc2}
\psarc{-}(0.5,-1){0.5}{00}{180}
\psarc{-}(1.5,-1){0.5}{180}{360}
\psline{-}(2,-1)(2,0)
\end{pspicture} 
 = v^{k-i} \begin{pspicture}(-0.2,-1.2)(2.2,-0.6)
\psset{linewidth=1pt}
\rput(0,-1.5){$i$}
\rput(1,-1.5){$j$}
\rput(2,-1.5){$k$}
\psdots(0,-1)(1,-1)(2,-1)
\psset{linecolor=myc2}
\psline{-}(0,-1)(0,0)
\end{pspicture}    \qquad \rightarrow \qquad  T_{j,k}^T  T_{i,j}  |0\rangle = v^{k-i}   |0\rangle, \end{equation*}

\item[9)]
\begin{equation*} 
\psset{unit=0.6}
\begin{pspicture}(-0.24,-1.2)(2.7,-0.6)
\psset{linewidth=1pt}
\psdots(0,-1)(1,-1)(2.5,-1)
\psset{linecolor=myc2}
\rput(1.85,-1){$\dots$}
\psline{-}(2.5,-1)(2.5,0)
\psline{-}(1,-1)(1,0)
\psline{-}(0,-1)(0,0)
\psline{-}(2.5,-1)(2.5,-2)
\psline{-}(1,-1)(1,-2)
\psline{-}(0,-1)(0,-2)
\end{pspicture} 
 = 1    \qquad \rightarrow \qquad \langle 0 |0\rangle =    1, \end{equation*}
\end{itemize}\vspace{0.2cm}
\noindent where $\simeq$ means the equality holds modulo terms that will be $0$ when the product with $\langle 0 |$ is computed. For instance, relation 4) is checked as follows
\begin{align*}  T_{j,k}^T  T_{i,j}  T_{k,l} |0 \rangle &= (u v^{j-k} \sigma_k^+ + u^{-1}v^{k-j} \sigma_j^+) (uv^{j-i} \sigma_j^-  + u^{-1}v^{i-j} \sigma_i^-) (uv^{l-k} \sigma_l^-  + u^{-1}v^{k-l} \sigma_k^-) |0 \rangle \\ &= (uv^{2j-i-l} \sigma_j^-  + u^{-1}v^{i-l} \sigma_i^- + uv^{l-i} \sigma_l^-  + u^{-1} v^{2k-i-l}\sigma_k^-) |0 \rangle
  \\& = ( T_{i,l} +  uv^{2j-i-l} \sigma_j^- + u^{-1} v^{2k-i-l}\sigma_k^-) |0 \rangle \simeq  T_{i,l}  |0 \rangle
 \end{align*}
where, in the last step, we used the fact that in what remains of $\prod  T^T \prod  T$, everything commutes with $\sigma_k^-$ and $\sigma_j^-$ and $\langle 0|  \sigma_j^- = \langle 0|  \sigma_k^- = 0$. Relation 9) is trivial and the seven others are proved in a similar fashion.
Finally, we note that the case \\
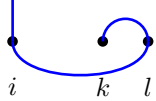
\begin{equation*}
\psset{unit=0.6}
\begin{pspicture}(-0.2,-1.2)(3.2,-0.6)
\psset{linewidth=1pt}
\rput(0,-2.0){$i$}
%\rput(1,-2.0){$j$}
\rput(2,-2.0){$k$}
\rput(3,-2.0){$l$}
\psdots(0,-1)(2,-1)(3,-1)
\psset{linecolor=myc2}
\psline{-}(0,-1)(0,0)
%\psline{-}(1,-1)(1,0)
\psbezier{-}(0,-1)(0,-2)(3,-2)(3,-1)
\psarc{-}(2.5,-1){0.5}{0}{180}
\end{pspicture}\,   
\end{equation*}
\\
does not need to be studied because there has to be a $j$ (with $i<j<k$) for which the diagram reads
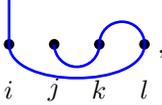
\begin{equation*}
\psset{unit=0.6}
\begin{pspicture}(-0.2,-1.2)(3.2,-0.6)
\psset{linewidth=1pt}
\rput(0,-2.0){$i$}
\rput(1,-2.0){$j$}
\rput(2,-2.0){$k$}
\rput(3,-2.0){$l$}
\psdots(0,-1)(1,-1)(2,-1)(3,-1)
\psset{linecolor=myc2}
\psline{-}(0,-1)(0,0)
%\psline{-}(1,-1)(1,0)
\psbezier{-}(0,-1)(0,-2)(3,-2)(3,-1)
\psarc{-}(2.5,-1){0.5}{0}{180}
\psarc{-}(1.5,-1){0.5}{180}{360}
\end{pspicture}\,,  
\end{equation*}
\\
and then relation 6) (or rather its transpose) may be used first. In conclusion, in calculating $(Q^d_N)_{w_1,w_2}$, each cluster gives rise to a factor through relations 1), 2), 7) and 8) that is equal to the factor obtained from the corresponding closed loop in the Gram diagram and the proof is complete.
\hfill$\square$

\medskip

From the previous proposition, $\det I_N^d (u,v)\det I_N^d (u,v^{-1}) = \det \mathcal {\tilde G}_N^d$. The computation of these determinants will be done in section \ref{sec:Gram}, that of $\mathcal{\tilde G}_N^d$ leading naturally to that of $I_N^d$. The latter will reveal under what conditions the map $\tilde i_N^d$ is actually an isomorphism.

%%%%%%%%%%%%%%%%
%
\section{The critical curves of the intertwiner $\tilde i_N^d$}\label{sec:Gram}
%
%%%%%%%%%%%%%%%%

{\it All equalities involving $\det\tilde{\mathcal G}_N^d$ or $\det I_N^d$ are valid up to a sign.}
\\

One goal of this section is to finish the proof that the map $\tilde i_N^d$ is a $\mathcal EPTL_N$-homomorphism, therefore proving that it intertwines loop and spin representations (theorem \ref{thm:enfin}). But the crucial result of the section is theorem \ref{prop:detInd} which gives the {\em critical curves} in the complex plane $(u,v)$ where $\tilde i_N^d$ stops being an isomorphism. 
The identification of these critical curves amounts to computing the determinant of $\tilde i_N^d$ in some appropriate bases. We shall recover, along the way, the determinant of the Gram matrix.

The computation of $\det \tilde{ \mathcal G}_N^d$ and $\det I_N^d$ is technical. Fortunately it is significantly simplified by a change of bases. In this new basis, the determinant for the periodic Gram matrix $\tilde{\mathcal G}_N^d$ can be seen to be related to that of the Gram matrix $\mathcal G_N^d$ for the open boundary case, that is the restriction of $\tilde{\mathcal G}_N^d$ to $B_N^d$. For $v=1$, the latter determinant is already known (see for example \cite{PDF} and \cite{Westbury}):
\begin{equation}
\det  \mathcal{G}_N^d =  \prod_{k=1}^{(N-d)/2} \left({S_{d+k+1}}/{S_k}\right)^{\dim V_N^{d+2k}} 
\label{eq:detG0}\end{equation}
where the following notation is used:
\begin{equation*}\qquad S_k = \sin(k \Lambda), \qquad C_k = \cos(k \Lambda),\qquad \Lambda = \pi - \lambda,  \qquad \beta = -2C_1\end{equation*}
and, as usual, $\dim V_N^d = \left(\begin{smallmatrix}N\\  (N-d)/2  \end{smallmatrix}\right) - \left(\begin{smallmatrix}N\\  (N-d-2)/2  \end{smallmatrix}\right)$. Section \ref{sec:Gv} is devoted to generalizing this result for arbitrary values of $v$. In section \ref{sec:relationtoopen}, we show that an appropriate change of bases allows for the factorization of the Gram determinant in the periodic case in terms of Gram determinants on the strip and factors $K_{d,r}$. The computation of these new factors $K_{d,r}$ will be the following step. It will require treating the subspaces with $d=0$ and $d>0$ separately, as non-contractible loops yield factors of $\alpha$ in the former, while in the latter defects are twisted, yielding powers of the twist parameter $v$. We shall then be able to prove the following theorem (theorem \ref{thm:detGram}) and characterize the values of $u$ and $v$ for which $\tilde i_N^d$ is singular (theorem \ref{prop:detInd}).

\medskip

\noindent{\scshape Theorem}\ \ {\em The determinant of the Gram matrices $\tilde{\mathcal G}_N^d$ is
\begin{equation}
\det  \mathcal{ \tilde G}_N^0 =  \prod_{k=1}^{N/2} (\alpha^2 - 4 C_{k}^2)^{ \left(\begin{smallmatrix}N\\  N/2-k  \end{smallmatrix}\right)}, \label{eq:detGN0}\end{equation} 
\begin{equation}\text{\rm and}\qquad
\det  \mathcal{ \tilde G}_N^d =  \prod_{k=1}^{(N-d)/2} (4 \cos^2(\mu N)-4C_{k+d/2}^2)^{ \left(\begin{smallmatrix}N\\  (N-d)/2-k  \end{smallmatrix}\right)},\qquad\text{\rm if\ }d>0,
\label{eq:detGtilde0}
\end{equation}
where $v = e^{i\mu}$. 
Trivially, for $d = N$, $\det  \mathcal{ \tilde G}_N^N = 1$. 
}

\medskip

\noindent Formulas \eqref{eq:detGN0} and \eqref{eq:detGtilde0} have been proved in various contexts before. To our knowledge, they first appeared in Martin and Saleur's work \cite{MartinSaleur} on the representation theory of the full Temperley-Lieb algebra on a graph $G$. (The periodic Temperley-Lieb corresponds to choosing the graph $G$ to be the affine $\hat A_N$.) Graham and Lehrer \cite{Lehrer} gave a full description of the representation theory of the periodic Temperley-Lieb algebra (affine Temperley-Lieb algebra), including a proof of the above determinant, within the context of category theory. Their affine cell representation $W_{t,z}$ is introduced as a functor between two categories. For a given $n$ (our $N$), their modules $W_{t,z}(n)$ are related to our $\omega_d$, with their $t$ fixing our $d$. Their parameter $z$ is tied to our parameter $v$, though the exact relationship depends on $d$. Their proof is (very) different from ours. Chen and Przytycki \cite{ChenPRZ} recovered recently the case $d=0$ in still another way. Clearly this result is crucial and different proofs will bring into light different properties of the problem. Ours is based mainly on diagrammatic arguments, rests upon the Wenzl-Jones projectors and underlines a remarkable property of the Gram determinant for the (original) Temperley-Lieb algebra to be proved next.

\subsection{The Gram determinant on the strip}\label{sec:Gv}

The goal of this section is to show that equation (\ref{eq:detG0}), which gives the Gram determinant on the geometry of the strip for $v=1$, actually holds for any value of the twist parameter. In fact, we will pursue an even more ambitious goal. Let $\langle \cdot  | \cdot \rangle_G^{\bf v} : V_N^d \times V_N^d \rightarrow \mathbb C$, with ${\bf v} = (v_1, v_2, ..., v_d)$, be the bilinear form defined for $w_1, w_2 \in B_N^d$ as
\begin{equation*}
\langle w_1 | w_2 \rangle_G^{\bf v} = \left\{ \begin{array}{l l} 0 & \quad  \textrm{if, in $D_{\mathcal{G}}(w_1,w_2)$, two defects of $w_1$ (or $w_2$) are connected,}\\ \alpha^{n_\alpha}\beta^{n_\beta} \prod_i v_i^{\Delta_i}  & \quad  \textrm{otherwise,} \end{array} \right.
\end{equation*}
where now $\Delta_i$ denotes the displacement, towards the left, of defect $i$ with the index $i$ labelling defects from left to right. Note that this new bilinear form coincides with the usual one on $V_N^d$ for $d=0$. Because $w_1$ and $w_2$ are in $B_N^d$, $\langle w_1 | w_2 \rangle_G^{\bf v}$ is non-zero only if defect $i$ of $w_1$ is connected to defect $i$ of $w_2$. In this section we show that the determinant of the matrix $\mathcal G_N^{d,{\bf v}}$ of this bilinear form is independent of the $v_i$s and therefore
\begin{equation}
\det  \mathcal{G}_N^{d,\bf v} = \det  \mathcal{G}_N^{d}.
\label{eq:detGv}\end{equation}
The main tool will be the Wenzl-Jones projector (see \cite{GoodmanWenzl,Wenzl}).

\begin{Definition}\label{def:WJ} For each $1 \le n\le N$,  the Wenzl-Jones projector $\wj{n} \in \mathcal E PTL_N(\beta, \alpha)$ is defined as $\wj{1}=id$ and for $n> 1$ as
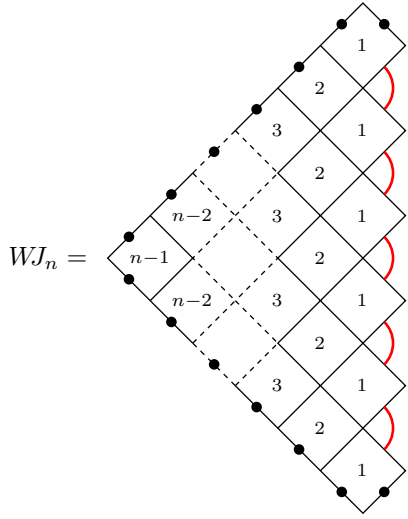
\begin{equation*}
\psset{unit=0.565685}\begin{pspicture}(0,-6)(7,6)
\rput(-1.4,0){$\wj{n} =$}
\lw
\psline{-}(2,2)(0,0)(2,-2)
\psline{-}(3,3)(6,6)(7,5)(6,4)(7,3)(6,2)(7,1)(6,0)(7,-1)(6,-2)(7,-3)(6,-4)(7,-5)(6,-6)(4,-4)(5,-3)(4,-2)(5,-1)(4,0)(5,1)(4,2)(5,3)(4,4)
\psline{-}(3,-3)(4,-4)
\psline{-}(5,5)(6,4)(5,3)(6,2)(5,1)(6,0)(5,-1)(6,-2)(5,-3)(6,-4)(5,-5)
\psline{-}(1,1)(2,0)(1,-1)
\psline[linestyle=dashed,dash=2pt 2pt]{-}(2,2)(3,3)
\psline[linestyle=dashed,dash=2pt 2pt]{-}(2,-2)(3,-3)
\psline[linestyle=dashed,dash=2pt 2pt]{-}(3,-3)(4,-2)(3,-1)(4,0)(3,1)(4,2)(3,3)
\psline[linestyle=dashed,dash=2pt 2pt]{-}(2,2)(3,1)(2,0)(3,-1)(2,-2)
\rput(6,5){$_1$}
\rput(6,3){$_1$}
\rput(6,1){$_1$}
\rput(6,-1){$_1$}
\rput(6,-3){$_1$}
\rput(6,-5){$_1$}
\rput(5,4){$_2$}
\rput(5,2){$_2$}
\rput(5,0){$_2$}
\rput(5,-2){$_2$}
\rput(5,-4){$_2$}
\rput(4,3){$_3$}
\rput(4,1){$_3$}
\rput(4,-1){$_3$}
\rput(4,-3){$_3$}
\rput(2,1){$_{n-2}$}
\rput(2,-1){$_{n-2}$}
\rput(1,0){$_{n-1}$}\unlw
\psdots(0.5,0.5)(1.5,1.5)(2.5,2.5)(3.5,3.5)(4.5,4.5)(5.5,5.5)(6.5,5.5)
\psdots(0.5,-0.5)(1.5,-1.5)(2.5,-2.5)(3.5,-3.5)(4.5,-4.5)(5.5,-5.5)(6.5,-5.5)

\psset{linecolor=myc}
\psarc{-}(6,-4){0.7071}{-45}{45}
\psarc{-}(6,-2){0.7071}{-45}{45}
\psarc{-}(6,0){0.7071}{-45}{45}
\psarc{-}(6,2){0.7071}{-45}{45}
\psarc{-}(6,4){0.7071}{-45}{45} 
\end{pspicture}
\end{equation*}
where each box is
\begin{equation}
\psset{linewidth=1pt}
\begin{pspicture}(-0.5,-0.5)(0.4,0.5)
\lw
\psline{-}(-0.5,0)(0,-0.5)(0.5,0)(0,0.5)(-0.5,0)
\rput(0,0){$_{k}$}
\end{pspicture}\ \ 
\begin{pspicture}(-0.2,-0.5)(0.2,0.5)
\rput(-0.1,-0.030){$_{=}$}\lw
\end{pspicture}
\begin{pspicture}(-0.5,-0.5)(0.5,0.5)\lw
%\rput(-0.8,-0.03){$_{=}$}
\psline{-}(-0.5,0)(0,-0.5)(0.5,0)(0,0.5)(-0.5,0)
\psset{linecolor=myc}\unlw
\psarc{-}(-0.5,0){0.3536}{-45}{45}
\psarc{-}(0.5,0){0.3536}{135}{225}
\end{pspicture}
\begin{pspicture}(-0.9,-0.5)(0.6,0.5)
\rput(-0.1,-0.03){$_{+\ \displaystyle{\frac{S_k}{S_{k+1}}} }$}
\end{pspicture}
\begin{pspicture}(-0.5,-0.5)(0.5,0.5)\lw
%\rput(-1.0,-0.03){$_{+\ \frac{S_k}{S_{k+1}}\ \ }$}
\psline{-}(-0.5,0)(0,-0.5)(0.5,0)(0,0.5)(-0.5,0)
\psset{linecolor=myc}\unlw
\psarc{-}(0,0.5){0.3536}{225}{315}
\psarc{-}(0,-0.5){0.3536}{45}{135}
\end{pspicture}
\begin{pspicture}(-1.3,-0.5)(0.6,0.5)
\rput(-0.3,-0.03){$_{=\  \displaystyle{\frac{(-1)^k}{S_{k+1}}} }$}\lw
\end{pspicture}
\begin{pspicture}(-0.5,-0.5)(0.5,0.5)\lw
%\rput(-1.2,-0.03){$_{=  \frac1{S_{k+1}}\ \ }$}
\psline{-}(-0.5,0)(0,-0.5)(0.5,0)(0,0.5)(-0.5,0)
\rput(0,0){$_{-k \lambda}$}
\psarc[linewidth=0.5pt]{-}(0,-0.5){0.176777}{45}{135}
%\rput(0.65,-0.2){.}
\end{pspicture}.
\label{eq:kbox}
\end{equation}
\label{sec:WJdefinition}
\end{Definition}
The projector $\wj{n}$ acts on $n$ of the $N$ positions of $\mathcal EPTL_N$. If these positions are chosen to be $1$ to $n$, then $\wj{n}$ satisfies the four properties \cite{GoodmanWenzl, Wenzl, KauffmanLins}:
\begin{itemize}
\item[{(i)}]  $\wj{n} e_i = e_i \wj{n} =0$ for $n \ge 2$ and $i=1,2,...,n-1$.
\item[{(ii)}]  $(\wj{n})^2 = \wj{n}$.
\item[{(iii)}] The reflection through a horizontal mirror of the diagram defining $\wj{n}$ is equal to $\wj{n}$. Equivalently $\wj n=(\wj n)^\dagger$. (See the definition of ``$\dagger$'' before theorem \ref{sec:homoGram}.)
\item[{(iv)}] $\wj n$ is also unchanged if the diagram defining the projector is reflected through a vertical mirror. This follows from the unicity of the projector defined by (i) and (ii), as proved for example in \cite{KauffmanLins}.
\end{itemize}
If $\wj{n}$ is chosen to act on positions $k$ to $k+n-1$ or, even, on a subset of $\{1, 2, \dots, N\}$ of $n$ elements, then {(ii)}, {(iii)} and {(iv)} still stand, but {(i)} then needs to be replaced by the statement that any bubble joining neighboring positions of $\wj{n}$ gives zero \cite{PRZ, AMDSA}. The property {(iii)} implies that $\langle w_1|\wj nw_2\rangle_G=\langle \wj nw_1|w_2\rangle_G$ for $w_1,w_2\in\tilde V_N$.  Note that the projector $\wj{}$ is defined only if none of the sine functions in the denominators of equation \eqref{eq:kbox} vanish; this requirement is clearly a condition on $\Lambda$.

To show equation (\ref{eq:detGv}), we partition $B_N^d$ as $\mathcal W_1 \cup \mathcal W_2 $: $\mathcal W_1$ contains all link states in $B_N^d$ with a defect in first position, while $\mathcal W_2$ contains link states with a bubble starting in first position. For $N=5, d=1$,

\begin{equation*}
\psset{unit=0.6}
 \qquad B_5^1 = \big\{
 \underbrace{
 \begin{pspicture}(-0.8,0)(1.8,0.4)
\psdots(-0.5,0)(0,0)(0.5,0)(1,0)(1.5,0)
\psset{linewidth=1pt}
\psset{linecolor=myc2}
\psarc{-}(0.25,0){0.25}{0}{180}
\psarc{-}(1.25,0){0.25}{0}{180}
\psline{-}(-0.5,0)(-0.5,0.75)
\end{pspicture},
 \begin{pspicture}(-0.8,0)(1.8,0.64)
\psdots(-0.5,0)(0,0)(0.5,0)(1,0)(1.5,0)
\psset{linewidth=1pt}
\psset{linecolor=myc2}
\psarc{-}(0.75,0){0.25}{0}{180}
\psline{-}(-0.5,0)(-0.5,0.75)
\psbezier{-}(0,0)(0,0.75)(1.5,0.75)(1.5,0)
\end{pspicture}}_{\mathcal W_1},
\underbrace{
 \begin{pspicture}(-0.8,0)(1.8,0.64)
\psdots(-0.5,0)(0,0)(0.5,0)(1,0)(1.5,0)
\psset{linewidth=1pt}
\psset{linecolor=myc2}
\psarc{-}(-0.25,0){0.25}{0}{180}
\psarc{-}(1.25,0){0.25}{0}{180}
\psline{-}(0.5,0)(0.5,0.75)
\end{pspicture},
 \begin{pspicture}(-0.8,0)(1.8,0.64)
\psdots(-0.5,0)(0,0)(0.5,0)(1,0)(1.5,0)
\psset{linewidth=1pt}
\psset{linecolor=myc2}
\psarc{-}(-0.25,0){0.25}{0}{180}
\psarc{-}(0.75,0){0.25}{0}{180}
\psline{-}(1.5,0)(1.5,0.75)
\end{pspicture},
 \begin{pspicture}(-0.8,0)(1.8,0.64)
\psdots(-0.5,0)(0,0)(0.5,0)(1,0)(1.5,0)
\psset{linewidth=1pt}
\psset{linecolor=myc2}
\psarc{-}(0.25,0){0.25}{0}{180}
\psline{-}(1.5,0)(1.5,0.75)
\psbezier{-}(-0.5,0)(-0.5,0.75)(1.0,0.75)(1.0,0)
\end{pspicture}}_{\mathcal W_2}
 \big\}.
\end{equation*}
\noindent The next step is to replace the elements of $\mathcal W_2$. To each $w \in \mathcal W_2$, we do the following manipulations: Remove all arcs except the one in first position, apply a projector $\wj{d+1}$ on the $d$ defects and on the right part of the arc in first position and, finally, restore the $(N-d-2)/2$ arcs into their original positions. We note by $\mathcal V(w)\in V_N^d$ the resulting vector and by $\mathcal{VW}_2$ the new set. 

Note that the action of the projector $\wj{}$ on the link states with $d$ defects is that of definition \ref{sec:omegarep}, but with $d$ different twist parameters, one for each defect. Such an action is a representation of the (usual) Temperley-Lieb algebra (generated by $e_i$s, with $i = 1, ..., N-1$, satisfying equation (\ref{eq:TLPN})). The proof of this claim is identical to that of proposition \ref{prop:omegaRep}. Again one can check that the relations (\ref{eq:TLPN}) are always satisfied (each involves at most four entries of the link state) and that a given defect $i$ is always associated with the same twist parameter $v_i$, a property that is lost for the periodic case. 
 
For the above example, we get
\vspace{0.2cm}\\ 

\begin{equation*}
\psset{unit=0.6}
 \qquad \mathcal {VW}_2 = \big\{
 \begin{pspicture}(-0.8,0)(1.8,0.64)
\psline{-}(0.25,0.5)(0.75,1)(0.25,1.5)(-0.25,1)(0.25,0.5)
\rput(0.25,1){$_1$}
\psdots(-0.5,0)(0,0)(0.5,0)(1,0)(1.5,0)
\psset{linewidth=1pt}
\psset{linecolor=myc2}
\psarc{-}(-0.25,1.25){0.25}{0}{180}
\psarc{-}(1.25,0){0.25}{0}{180}
\psline{-}(0.5,0)(0.5,0.75)
\psline{-}(0.,0)(0.,0.75)
\psline{-}(0.5,1.25)(0.5,2)
\psline{-}(-0.5,0)(-0.5,1.25)
\end{pspicture},
 \begin{pspicture}(-0.8,0)(1.8,0.64)
\psdots(-0.5,0)(0,0)(0.5,0)(1,0)(1.5,0)
\psline{-}(1.25,0.5)(1.75,1)(1.25,1.5)(0.75,1)(1.25,0.5)
\rput(1.25,1){$_1$}
\psset{linewidth=1pt}
\psset{linecolor=myc2}
\psarc{-}(0.75,1.25){0.25}{0}{180}
\psarc{-}(0.75,0){0.25}{0}{180}
\psline{-}(1.5,1.25)(1.5,2)
\psline{-}(0.5,1.25)(0.5,1)
\psbezier{-}(-0.5,0)(-0.5,0.5)(0.5,0.5)(0.5,1)
\psbezier{-}(0,0)(0,0.5)(1,0.5)(1,0.75)
\psline{-}(1.5,0)(1.5,0.75)
\end{pspicture},
 \begin{pspicture}(-0.8,0)(1.8,0.64)
 \psline{-}(1.25,0.5)(1.75,1)(1.25,1.5)(0.75,1)(1.25,0.5)
\rput(1.25,1){$_1$}
\psdots(-0.5,0)(0,0)(0.5,0)(1,0)(1.5,0)
\psset{linewidth=1pt}
\psset{linecolor=myc2}
\psarc{-}(0.25,0){0.25}{0}{180}
\psarc{-}(0.75,1.25){0.25}{0}{180}
\psline{-}(1,0)(1,0.75)
\psline{-}(0.5,1)(0.5,1.25)
\psline{-}(1.5,0)(1.5,0.75)
\psline{-}(1.5,1.25)(1.5,2)
\psbezier{-}(0.5,1)(0.5,0.5)(-0.5,0.5)(-0.5,0)
\end{pspicture}
 \big\}.
\end{equation*}
To compute the twist factor $\prod_i v_i^{\Delta_i}$ in a Gram diagram, one has to be cautious to identify correctly the position of the defects. For example, the second element of $\mathcal{VW}_2$ above will be a sum of two link states. The (only) defect is at position $1$ in one of the two link states (with an added factor $v_1^4$) and at position $5$ in the other (with no factor added).

By construction, if $\Lambda$ is chosen such that $\wj{d+1}$ exists, the vector $\mathcal V(w)$ for  $w\in\mathcal W_2$ has the form $w + w'$ for some $w'$ in $\mathcal W_1$. Therefore $\mathcal W_1 \cup \mathcal {VW}_2$ is a basis of $V_N^d$ and the determinant of the change of bases is $1$. 

Our interest in this new basis is that for $w_1 \in \mathcal W_1$ and $w_2 \in \mathcal W_2$, $\langle w_1| \mathcal V(w_2) \rangle_G^{\bf v} = 0$. Indeed, the entries at positions $2$ to $N-1$ in $\mathcal V(w_2)$ are connected to $d+1$ entries of a $\wj{d+1}$ and $(N-d-2)/2$ bubbles, while those of $w_1$ are connected to $d-1$ defects and $(N-d)/2$ bubbles. In $D_{\mathcal G}(w_1,\mathcal V(w_2))$, two entries of the projector are tied by bubbles and the result is $0$. The new Gram product $\langle u_1 | w_1 \rangle_G^{\bf v}$ between two states $u_1$ and $w_1$ in $\mathcal W_1$ is just the Gram product between states in $V_{N-1}^{d-1}$ obtained from $u_1$ and $w_1$ by removing the defect at the first position, and with ${\bf v'} = (v_2, ..., v_d)$. 

The product $\langle \mathcal V(u_2)|\mathcal  V(w_2) \rangle_G^{\bf v}$ for $u_2, w_2 \in \mathcal {W}_2$ can be seen to factor into a constant $L_{d}$ independent of ${\bf v}$ times $\langle \mathcal Y(u_2)| \mathcal Y(w_2)\rangle_G^{\bf v''}$ where $\mathcal Y(w)$ stands for the link states in $B_{N-1}^{d+1}$ obtained by removing the bubble connecting positions $1$ and $x$, putting a defect at $x$ and removing position $1$ altogether.  The added defect is considered to have $i=0$, and we must impose $v_0 = 1$ for the twist factor to be evaluated correctly, so ${\bf v''} = (1, v_1, ..., v_d)$. The $v_i$ dependence in the original $\langle \mathcal V(u_2)| \mathcal V(w_2) \rangle_G^{\bf v}$ is contained in $\langle \mathcal Y(u_2)| \mathcal Y(w_2)\rangle_G^{\bf v''}$. Finally, the constant $L_d$ is given by

\begin{equation*}
\psset{unit=0.6}
\begin{pspicture}(-0.2,-6.5)(4.5,4.2)
\rput(-1.25,-1.25){$L_d=$}
\psdots(0,-1.25)(0.5,-1.25)(1,-1.25)(1.5,-1.25)(2,-1.25)(2.5,-1.25)(3,-1.25)
\psset{linewidth=1pt}
\psline{-}(2.75,-1)(0.25,1.5)(2.75,4)(3.25,3.5)(2.75,3)(3.25,2.5)(2.75,2)(3.25,1.5)(2.75,1)(3.25,0.5)(2.75,0)(3.25,-0.5)(2.75,-1)
\psline{-}(2.25,-0.5)(2.75,-0)(2.25,0.5)(2.75,1)(2.25,1.5)(2.75,2)(2.25,2.5)(2.75,3)(2.25,3.5)
\psline{-}(1.75,0)(2.25,0.5)(1.75,1)(2.25,1.5)(1.75,2)(2.25,2.5)(1.75,3)
\psline{-}(1.25,0.5)(1.75,1)(1.25,1.5)(1.75,2)(1.25,2.5)
\psline{-}(0.75,1)(1.25,1.5)(0.75,2)
\rput(0.75,1.5){$_{d}$}
\rput(0.75,-4.5){$_{d}$}
\psline{-}(2.75,-2)(0.25,-4.5)(2.75,-7)(3.25,-6.5)(2.75,-6)(3.25,-5.5)(2.75,-5)(3.25,-4.5)(2.75,-4)(3.25,-3.5)(2.75,-3)(3.25,-2.5)(2.75,-2)
\psline{-}(2.25,-2.5)(2.75,-3)(2.25,-3.5)(2.75,-4)(2.25,-4.5)(2.75,-5)(2.25,-5.5)(2.75,-6)(2.25,-6.5)
\psline{-}(1.75,-3)(2.25,-3.5)(1.75,-4)(2.25,-4.5)(1.75,-5)(2.25,-5.5)(1.75,-6)
\psline{-}(1.25,-3.5)(1.75,-4)(1.25,-4.5)(1.75,-5)(1.25,-5.5)
\psline{-}(0.75,-4)(1.25,-4.5)(0.75,-5)
\psset{linecolor=myc2}
\psline{-}(3,-1.25)(3,-0.75)
\psline{-}(2.5,-1.25)(2.5,-0.75)
\psline{-}(2,-1.25)(2,-0.25)
\psline{-}(1.5,-1.25)(1.5,0.25)
\psline{-}(1,-1.25)(1,0.75)
\psline{-}(0.5,-1.25)(0.5,1.25)
\psline{-}(0.5,1.75)(0.5,3.5)
\psline{-}(0,-1.25)(0,3.5)
\psline{-}(1,2.25)(1,4.25)
\psline{-}(1.5,2.75)(1.5,4.25)
\psline{-}(2,3.25)(2,4.25)
\psline{-}(2.5,3.75)(2.5,4.25)
\psline{-}(3,3.75)(3,4.25)
\psarc{-}(0.25,3.5){0.25}{0}{180}
\psline{-}(3,-1.25)(3,-2.25)
\psline{-}(2.5,-1.25)(2.5,-2.25)
\psline{-}(2,-1.25)(2,-2.75)
\psline{-}(1.5,-1.25)(1.5,-3.25)
\psline{-}(1,-1.25)(1,-3.75)
\psline{-}(0.5,-1.25)(0.5,-4.25)
\psline{-}(0,-1.25)(0,-6.5)
\psline{-}(0.5,-4.75)(0.5,-6.5)
\psline{-}(1,-5.25)(1,-7.25)
\psline{-}(1.5,-5.75)(1.5,-7.25)
\psline{-}(2,-6.25)(2,-7.25)
\psline{-}(2.5,-6.75)(2.5,-7.25)
\psline{-}(3,-6.75)(3,-7.25)
\psarc{-}(0.25,-6.5){0.25}{180}{360}
\psset{linecolor=myc}
\psarc{-}(2.75,3){0.353553}{-45}{45}
\psarc{-}(2.75,2){0.353553}{-45}{45}
\psarc{-}(2.75,1){0.353553}{-45}{45}
\psarc{-}(2.75,0){0.353553}{-45}{45}
\psarc{-}(2.75,-6){0.353553}{-45}{45}
\psarc{-}(2.75,-5){0.353553}{-45}{45}
\psarc{-}(2.75,-4){0.353553}{-45}{45}
\psarc{-}(2.75,-3){0.353553}{-45}{45}
\end{pspicture} 
\begin{pspicture}(-0.2,-3.8)(4.5,4.2)
\psset{linewidth=1pt}
\psline{-}(2.75,-1)(0.25,1.5)(2.75,4)(3.25,3.5)(2.75,3)(3.25,2.5)(2.75,2)(3.25,1.5)(2.75,1)(3.25,0.5)(2.75,0)(3.25,-0.5)(2.75,-1)
\psline{-}(2.25,-0.5)(2.75,-0)(2.25,0.5)(2.75,1)(2.25,1.5)(2.75,2)(2.25,2.5)(2.75,3)(2.25,3.5)
\psline{-}(1.75,0)(2.25,0.5)(1.75,1)(2.25,1.5)(1.75,2)(2.25,2.5)(1.75,3)
\psline{-}(1.25,0.5)(1.75,1)(1.25,1.5)(1.75,2)(1.25,2.5)
\psline{-}(0.75,1)(1.25,1.5)(0.75,2)
\rput(0.75,1.5){$_{d}$}
\rput(-0.75,1.375){$=$}
\psset{linecolor=myc2}
\psline{-}(3,-1.25)(3,-0.75)
\psline{-}(2.5,-1.25)(2.5,-0.75)
\psline{-}(2,-1.25)(2,-0.25)
\psline{-}(1.5,-1.25)(1.5,0.25)
\psline{-}(1,-1.25)(1,0.75)
\psline{-}(1,2.25)(1,4.25)
\psline{-}(1.5,2.75)(1.5,4.25)
\psline{-}(2,3.25)(2,4.25)
\psline{-}(2.5,3.75)(2.5,4.25)
\psline{-}(3,3.75)(3,4.25)
\psarc{-}(0.25,1.5){0.353553}{45}{-45}

\psset{linecolor=myc}
\psarc{-}(2.75,3){0.353553}{-45}{45}
\psarc{-}(2.75,2){0.353553}{-45}{45}
\psarc{-}(2.75,1){0.353553}{-45}{45}
\psarc{-}(2.75,0){0.353553}{-45}{45}
\end{pspicture}
\begin{pspicture}(-1.2,-3.8)(4.5,4.2)
\psset{linewidth=1pt}
\psline{-}(2.75,-1)(0.75,1)
\psline{-}(0.75,2)(2.75,4)(3.25,3.5)(2.75,3)(3.25,2.5)(2.75,2)(3.25,1.5)(2.75,1)(3.25,0.5)(2.75,0)(3.25,-0.5)(2.75,-1)
\psline{-}(2.25,-0.5)(2.75,-0)(2.25,0.5)(2.75,1)(2.25,1.5)(2.75,2)(2.25,2.5)(2.75,3)(2.25,3.5)
\psline{-}(1.75,0)(2.25,0.5)(1.75,1)(2.25,1.5)(1.75,2)(2.25,2.5)(1.75,3)
\psline{-}(1.25,0.5)(1.75,1)(1.25,1.5)(1.75,2)(1.25,2.5)
\psline{-}(0.75,1)(1.25,1.5)(0.75,2)
\psbezier[linewidth=0.5pt](1,1.0)(0.5,1.0)(0.25,0.75)(0,0)
\rput(0,-0.25){$_{d-1}$}
\rput(-0.75,1.375){$= - \frac{S_{d+2}}{S_{d+1}}$}
\rput(4.75,1.375){$= - \frac{S_{d+2}}{S_{d+1}}$,}
\psset{linecolor=myc2}
\psline{-}(3,-1.25)(3,-0.75)
\psline{-}(2.5,-1.25)(2.5,-0.75)
\psline{-}(2,-1.25)(2,-0.25)
\psline{-}(1.5,-1.25)(1.5,0.25)
\psline{-}(1,-1.25)(1,0.75)
\psline{-}(1,2.25)(1,4.25)
\psline{-}(1.5,2.75)(1.5,4.25)
\psline{-}(2,3.25)(2,4.25)
\psline{-}(2.5,3.75)(2.5,4.25)
\psline{-}(3,3.75)(3,4.25)
\psarc{-}(1.25,1.5){0.353553}{135}{-135}
\psset{linecolor=myc}
\psarc{-}(2.75,3){0.353553}{-45}{45}
\psarc{-}(2.75,2){0.353553}{-45}{45}
\psarc{-}(2.75,1){0.353553}{-45}{45}
\psarc{-}(2.75,0){0.353553}{-45}{45}
\end{pspicture}
\end{equation*}
\\
\noindent where, at the last step, the only configuration contributing has all tiles set to
$\psset{unit=0.4}
\begin{pspicture}(-0.5,-0.3)(0.5,0.5)\lw
%\rput(-1.0,-0.03){$_{+\ \frac{S_k}{S_{k+1}}\ \ }$}
\psline{-}(-0.5,0)(0,-0.5)(0.5,0)(0,0.5)(-0.5,0)
\psset{linecolor=myc}\unlw
\psarc{-}(0.5,0){0.3536}{135}{-135}
\psarc{-}(-0.5,0){0.3536}{-45}{45}
\end{pspicture}$. From the previous remarks, up to a sign, the Gram determinants obey the recursion relation
\begin{equation}
\det(\mathcal G_N^{d, \bf v}) =\det(\mathcal G_{N-1}^{d-1, \bf v'}) \times\left( \frac{S_{d+2}}{S_{d+1}}\right)^{\dim V_{N-1}^{d+1}}\det(\mathcal G_{N-1}^{d+1, \bf v''}).
\label{eq:Gramrec1}
\end{equation}
%The case $d=0$ is particular ($\mathcal W_1$ is empty and there is no $v$ dependence) and instead,
%\begin{equation}
%\det(\mathcal G_N^0) = \left( \frac{S_{2}}{S_{1}}\right)^{\dim V_{N-1}^1}\det(\mathcal G_{N-1}^{1, \bf v}) \qquad \textrm{with} \quad {\bf v} = (1).
%\label{eq:Gramrec2}
%\end{equation}

\noindent Two limiting cases are known:
\begin{equation}
\det \mathcal G_N^{0,\bf v}=\det \mathcal G_N^0\qquad \textrm{and} \qquad
\det \mathcal G_N^{N,\bf v}=1.
\label{eq:Gramrec2}
\end{equation}
The use of \eqref{eq:Gramrec1} lowers the bottom index $N$. Its repeated use will lead either to an upper index $d=0$ through the first term of the right side or to equal upper and lower indices through its second term. Thus equations (\ref{eq:Gramrec1}) and (\ref{eq:Gramrec2}) determine $\det(\mathcal G_N^{d, \bf v})$ completely. The expression given in equation (\ref{eq:detG0}) satisfies all these and must thus be the solution. 
This ends the proof of \eqref{eq:detGv}. 

Even though the matrices $\mathcal G_N^d$ explicitly depend upon the twist parameters $v_i$, their determinants do not! Again, for the previous example, the Gram matrix with ${\bf v}=\{v_1\}$ is  
$$\mathcal G^{1,\bf v}_5 = \begin{pmatrix} 
\beta^2 & \beta & \beta v_1^{-2} & v_1^{-4} & \beta v_1^{-4}\\
\beta & \beta^2 & v_1^{-2} & \beta v_1^{-4} & v_1^{-4}\\
\beta v_1^{2} & v_1^{2} & \beta^2 & \beta v_1^{-2} & v_1^{-2}\\
v_1^{4} & \beta v_1^{4} & \beta v_1^{2} & \beta^2 & \beta\\
\beta v_1^{4} & v_1^{4} & v_1^{2} & \beta & \beta^2\\
\end{pmatrix}$$
and a direct computation gives
 $$\det(\mathcal G_5^{1, \bf v}) =  (\beta^2 - 1)^4 (\beta^2 - 2) = (S_3 / S_1)^{\dim V_5^3} (S_4 / S_2)^{\dim V_5^5}.$$

\subsection{The relation between the open and periodic cases}\label{sec:relationtoopen}

\begin{Definition} The subset $\tilde B_N^{d,r}\subset \tilde B_N^d$ contains the link patterns that have precisely $r$ bubbles crossing the imaginary boundaries at $x=\frac12$ and $x=N+\frac12$. Then $\tilde B_N^d=\cup_{0\le r\le (N-d)/2} \tilde B_N^{d,r}$ is a partition of $\tilde B_N^d$ and $\tilde V_N^{d,r}\subset \tilde V_N^d$ is the subspace spanned by $\tilde B_N^{d,r}$.

A bijection $\mathcal C$ between $\tilde B_N^{d,r}$ and $B_N^{d+2r}$ is defined by identifying $w\in\tilde B_N^{d,r}$ to $\mathcal C(w)\in B_N^{d+2r}$ obtained by replacing the $r$ boundary bubbles of $w$ by defects, and leaving the rest of $w$ unchanged. If $r=0$, $\mathcal C(w)=w$.
\end{Definition}
From now on, the basis $\tilde B_N^d$ will be (partially) ordered in ascending order of $r$. Here are examples of pairs $w\leftrightarrow \mathcal C(w)$:
\begin{equation*} \mathcal C \big(
\psset{unit=0.6}
\begin{pspicture}(5.5,-1)(10,-0.6)
\psdots(6,-1)(6.5,-1)(7,-1)(7.5,-1)(8,-1)(8.5,-1)(9,-1)(9.5,-1)
\psset{linewidth=1pt}
\psset{linecolor=myc2}
\psbezier{-}(6.5,-1)(6.5,-0.5)(6,-0.4)(5.75,-0.4)
\psbezier{-}(9,-1)(9,-0.5)(9.6,-0.4)(9.75,-0.4)
\psbezier{-}(7,-1)(7,-0.25)(8.5,-0.25)(8.5,-1)
\psarc{-}(7.75,-1){0.25}{0}{180}
\psarc{-}(5.75,-1){0.25}{0}{90}
\psarc{-}(9.75,-1){0.25}{90}{180}
\end{pspicture} \big) = 
\begin{pspicture}(5.8,-1)(9.8,-0.6)
\psdots(6,-1)(6.5,-1)(7,-1)(7.5,-1)(8,-1)(8.5,-1)(9,-1)(9.5,-1)
\psset{linewidth=1pt}
\psset{linecolor=myc2}
\psline{-}(6,-1)(6,0)
\psline{-}(6.5,-1)(6.5,0)
\psline{-}(9,-1)(9,0)
\psline{-}(9.5,-1)(9.5,0)
\psbezier{-}(7,-1)(7,-0.25)(8.5,-0.25)(8.5,-1)
\psarc{-}(7.75,-1){0.25}{0}{180}
\end{pspicture},
 \qquad \mathcal C \big(
 \begin{pspicture}(5.5,-1)(10,-0.6)
\psdots(6,-1)(6.5,-1)(7,-1)(7.5,-1)(8,-1)(8.5,-1)(9,-1)(9.5,-1)
\psset{linewidth=1pt}
\psset{linecolor=myc2}
\psline{-}(6.5,-1)(6.5,0)
\psline{-}(7,-1)(7,0)
\psline{-}(7.5,-1)(7.5,0)
\psline{-}(9.0,-1)(9.0,0)
\psarc{-}(8.25,-1){0.25}{0}{180}
\psarc{-}(5.75,-1){0.25}{0}{90}
\psarc{-}(9.75,-1){0.25}{90}{180}
\end{pspicture} \big) = 
\begin{pspicture}(5.8,-1)(9.8,-0.6)
\psdots(6,-1)(6.5,-1)(7,-1)(7.5,-1)(8,-1)(8.5,-1)(9,-1)(9.5,-1)
\psset{linewidth=1pt}
\psset{linecolor=myc2}
\psline{-}(6.0,-1)(6.0,0)
\psline{-}(6.5,-1)(6.5,0)
\psline{-}(7,-1)(7,0)
\psline{-}(7.5,-1)(7.5,0)
\psline{-}(9.0,-1)(9.0,0)
\psline{-}(9.5,-1)(9.5,0)
\psarc{-}(8.25,-1){0.25}{0}{180}
\end{pspicture}.
\end{equation*}

We now introduce a linear transformation $\mathcal U$ such that, in the new basis $\{\mathcal U(w), w\in\tilde B_N^d\}$, the Gram matrix is block-diagonal. 

\begin{Definition}\label{def:linearU} The linear transformation $\mathcal U: \tilde V_N^d \rightarrow \tilde V_N^d$ is defined by its action on elements $w$ of each $\tilde B_N^{d,r}$. If $r>0$, $\mathcal U(w)$ is obtained by first removing all arcs of $w$ that do not cross the boundary, acting with $\wj{d+2r}$ on the $r$ remaining bubbles and the $d$ defects, and then inserting back the bubbles that were first removed at their original positions. If $r=0$, $\mathcal U(w) = w$.
\end{Definition}
Here is an example for $w\in\tilde B_8^{4,1}$: \vspace{0.6cm}

\begin{equation*}
\psset{unit=0.6}
 \qquad  \mathcal U \big(
 \begin{pspicture}(5.5,-0.5)(10,-1)
\psdots(6,-1)(6.5,-1)(7,-1)(7.5,-1)(8,-1)(8.5,-1)(9,-1)(9.5,-1)
\psset{linewidth=1pt}
\psset{linecolor=myc2}
\psline{-}(6.5,-1)(6.5,0)
\psline{-}(7,-1)(7,0)
\psline{-}(7.5,-1)(7.5,0)
\psline{-}(9.0,-1)(9.0,0)
\psarc{-}(8.25,-1){0.25}{0}{180}
\psarc{-}(5.75,-1){0.25}{0}{90}
\psarc{-}(9.75,-1){0.25}{90}{180}
\end{pspicture} \big) = 
\begin{pspicture}(-0.2,0.2)(4.0,4)
\psdots(0,-2)(0.5,-2)(1,-2)(1.5,-2)(2,-2)(2.5,-2)(3,-2)(3.5,-2)
\psdots(0,4.5)(0.5,4.5)(1,4.5)(1.5,4.5)(2,4.5)(2.5,4.5)(3,4.5)(3.5,4.5)
\psset{linewidth=1pt}
\psline{-}(2.75,-1)(0.25,1.5)(2.75,4)(3.25,3.5)(2.75,3)(3.25,2.5)(2.75,2)(3.25,1.5)(2.75,1)(3.25,0.5)(2.75,0)(3.25,-0.5)(2.75,-1)
\psline{-}(2.25,-0.5)(2.75,-0)(2.25,0.5)(2.75,1)(2.25,1.5)(2.75,2)(2.25,2.5)(2.75,3)(2.25,3.5)
\psline{-}(1.75,0)(2.25,0.5)(1.75,1)(2.25,1.5)(1.75,2)(2.25,2.5)(1.75,3)
\psline{-}(1.25,0.5)(1.75,1)(1.25,1.5)(1.75,2)(1.25,2.5)
\psline{-}(0.75,1)(1.25,1.5)(0.75,2)
\rput(0.75,1.5){$_5$}
\psset{linecolor=myc2}
\psbezier{-}(3.5,-2)(3.5,-1.375)(3,-1.375)(3,-0.75)
\psbezier{-}(3.0,-2)(3.0,-1.375)(2.5,-1.375)(2.5,-0.75)
\psbezier{-}(1.5,-2)(1.5,-1.125)(2.0,-1.125)(2.0,-0.25)
\psbezier{-}(1.0,-2)(1.0,-0.875)(1.5,-0.875)(1.5,0.25)
\psbezier{-}(0.5,-2)(0.5,-0.625)(1.0,-0.625)(1.0,0.75)
\psbezier{-}(0.,-2)(0,-0.375)(0.5,-0.375)(0.5,1.25)
\psarc{-}(2.25,-2){0.25}{0}{180}
\psline{-}(0.5,4.5)(0.5,5.25)
\psline{-}(1,4.5)(1,5.25)
\psline{-}(1.5,4.5)(1.5,5.25)
\psline{-}(3,4.5)(3,5.25)
\psbezier{-}(3.0,3.75)(3.0,4.125)(3.5,4.125)(3.5,4.5)
\psbezier{-}(2.5,3.75)(2.5,4.125)(3.0,4.125)(3.0,4.5)
\psbezier{-}(2.0,3.25)(2.0,3.875)(1.5,3.875)(1.5,4.5)
\psbezier{-}(1.5,2.75)(1.5,3.625)(1.0,3.625)(1.0,4.5)
\psbezier{-}(1.0,2.25)(1.0,3.375)(0.5,3.375)(0.5,4.5)
\psbezier{-}(0.5,1.75)(0.5,3.125)(0.0,3.125)(0,4.5)
\psarc{-}(-0.25,4.5){0.25}{0}{90}
\psarc{-}(3.75,4.5){0.25}{90}{180}
\psset{linecolor=myc}
\psarc{-}(2.75,3){0.353553}{-45}{45}
\psarc{-}(2.75,2){0.353553}{-45}{45}
\psarc{-}(2.75,1){0.353553}{-45}{45}
\psarc{-}(2.75,0){0.353553}{-45}{45}
\end{pspicture}.
\end{equation*}
\\
\\
\\
The Wenzl-Jones projector $\wj{n}$ is a product of terms of the form $(id+ e_i S_k/S_{k+1})$ and therefore its expansion contains words $e_{i_1}e_{i_2}\dots e_{i_\ell}$ in the $(n-1)$ first generators of $\mathcal EPTL_N$. The identity $id$, that is the empty word, occurs with a factor one. The removal of the bubbles that do not cross the boundary in $w\in\tilde B_N^{d,r}$ gives a state in $\tilde B_{d+2r}^{d,r}$. But the non-empty words $e_{i_1}e_{i_2}\dots e_{i_\ell}$ cannot create new bubbles, they can only move them around. So these words either act as zero or give, up to a constant, a vector in $\tilde B_{d+2r}^{d,r'}$ with $r'<r$. Thus $\mathcal U(w)=w+w'$ where $w'$ is a linear combination of vectors in the $\tilde B_N^{d,r'}$s with $r'< r$. The matrix $U_N^d$ representing the linear transformation $\mathcal U$ in the basis $\tilde B_N^d$, ordered with increasing $r$s, is therefore upper block triangular with identity matrices along the diagonal. For $\Lambda$s where the projectors $\wj{}$ exist,  the matrix $\mathcal U$ exists, is invertible and $\mathcal U (\tilde B_N^d) = \{ \mathcal U(w) | w \in \tilde B_N^d \}$ is a basis of $\tilde V_N^d$. The matrix elements of $U_N^d$ depend on $\beta$, through $\Lambda$, and on $v$. If $N$ is even and $d=0$, some words of $\wj{}$ may close non-contractible loops and $\alpha$ may also appear.

Here are the new bases of the three $\tilde V_N^d$ for $N=4$:
\vspace{2 cm}
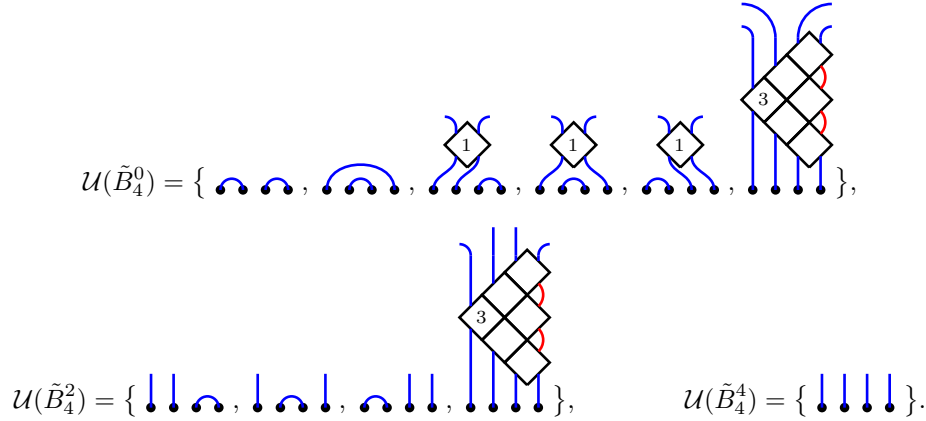
\begin{equation*}
\psset{unit=0.6}
 \qquad  \mathcal U(\tilde B_4^0) = \big\{
 \begin{pspicture}(-0.3,0)(1.8,0.4)
\psdots(0,0)(0.5,0)(1,0)(1.5,0)
\psset{linewidth=1pt}
\psset{linecolor=myc2}
\psarc{-}(0.25,0){0.25}{0}{180}
\psarc{-}(1.25,0){0.25}{0}{180}
\end{pspicture},
 \begin{pspicture}(-0.3,0)(1.8,0.64)
\psdots(0,0)(0.5,0)(1,0)(1.5,0)
\psset{linewidth=1pt}
\psset{linecolor=myc2}
\psarc{-}(0.75,0){0.25}{0}{180}
\psbezier{-}(0,0)(0,0.75)(1.5,0.75)(1.5,0)
\end{pspicture},
 \begin{pspicture}(-0.3,0)(1.8,0.4)
\psdots(0,0)(0.5,0)(1,0)(1.5,0)
\psset{linewidth=1pt}
\psline{-}(0.75,0.5)(1.25,1)(0.75,1.5)(0.25,1)(0.75,0.5)
\rput(0.75,1){$_1$}
\psset{linecolor=myc2}
\psbezier{-}(0,0)(0,0.375)(0.5,0.375)(0.5,0.75)
\psbezier{-}(0.5,0)(0.5,0.375)(1,0.375)(1,0.75)
\psarc{-}(1.25,0){0.25}{0}{180}
\psline{-}(0.5,1.25)(0.5,1.375)
\psline{-}(1,1.25)(1,1.375)
\psarc{-}(0.25,1.375){0.25}{0}{90}
\psarc{-}(1.25,1.375){0.25}{90}{180}
\end{pspicture},
  \begin{pspicture}(-0.3,0)(1.8,0.4)
\psdots(0,0)(0.5,0)(1,0)(1.5,0)
\psset{linewidth=1pt}
\psline{-}(0.75,0.5)(1.25,1)(0.75,1.5)(0.25,1)(0.75,0.5)
\rput(0.75,1){$_1$}
\psset{linecolor=myc2}
\psbezier{-}(0,0)(0,0.375)(0.5,0.375)(0.5,0.75)
\psbezier{-}(1.5,0)(1.5,0.375)(1,0.375)(1,0.75)
\psarc{-}(0.75,0){0.25}{0}{180}
\psline{-}(0.5,1.25)(0.5,1.375)
\psline{-}(1,1.25)(1,1.375)
\psarc{-}(0.25,1.375){0.25}{0}{90}
\psarc{-}(1.25,1.375){0.25}{90}{180}\end{pspicture},
 \begin{pspicture}(-0.3,0)(1.8,0.4)
\psdots(0,0)(0.5,0)(1,0)(1.5,0)
\psset{linewidth=1pt}
\psline{-}(0.75,0.5)(1.25,1)(0.75,1.5)(0.25,1)(0.75,0.5)
\rput(0.75,1){$_1$}
\psset{linecolor=myc2}
\psbezier{-}(1,0)(1,0.375)(0.5,0.375)(0.5,0.75)
\psbezier{-}(1.5,0)(1.5,0.375)(1,0.375)(1,0.75)
\psarc{-}(0.25,0){0.25}{0}{180}
\psline{-}(0.5,1.25)(0.5,1.375)
\psline{-}(1,1.25)(1,1.375)
\psarc{-}(0.25,1.375){0.25}{0}{90}
\psarc{-}(1.25,1.375){0.25}{90}{180}
\end{pspicture},
 \begin{pspicture}(-0.3,0)(1.8,0.4)
\psdots(0,0)(0.5,0)(1,0)(1.5,0)
\psset{linewidth=1pt}
\psline{-}(1.25,0.5)(1.75,1)(1.25,1.5)(0.75,1)(1.25,0.5)
\psline{-}(1.25,1.5)(1.75,2)(1.25,2.5)(0.75,2)(1.25,1.5)
\psline{-}(1.25,2.5)(1.75,3)(1.25,3.5)(0.75,3)(1.25,2.5)
\psline{-}(0.75,1)(1.25,1.5)(0.75,2)(0.25,1.5)(0.75,1)
\psline{-}(0.75,2)(1.25,2.5)(0.75,3)(0.25,2.5)(0.75,2)
\psline{-}(0.25,1.5)(0.75,2)(0.25,2.5)(-0.25,2)(0.25,1.5)
\rput(0.25,2){$_3$}
\psset{linecolor=myc2}
\psline{-}(0,0)(0,1.75)
\psline{-}(0.5,0)(0.5,1.25)
\psline{-}(1,0)(1,0.75)
\psline{-}(1.5,0)(1.5,0.75)
\psline{-}(1.0,3.25)(1.0,3.375)
\psline{-}(1.5,3.25)(1.5,3.375)
\psline{-}(0.0,2.25)(0.0,3.375)
\psline{-}(0.5,2.75)(0.5,3.375)
\psarc{-}(-0.25,3.375){0.25}{0}{90}
\psarc{-}(-0.25,3.375){0.75}{0}{90}
\psarc{-}(1.75,3.375){0.25}{90}{180}
\psarc{-}(1.75,3.375){0.75}{90}{180}
\psset{linecolor=myc}
\psarc{-}(1.25,1.5){0.355}{-45}{45}
\psarc{-}(1.25,2.5){0.355}{-45}{45}
\end{pspicture}
 \big\} ,
\end{equation*}
\vspace{1.7 cm}
\begin{equation*}
\psset{unit=0.6}
 \qquad  \mathcal U(\tilde B_4^2) = \big\{
 \begin{pspicture}(-0.3,0)(1.8,0.4)
\psdots(0,0)(0.5,0)(1,0)(1.5,0)
\psset{linewidth=1pt}
\psset{linecolor=myc2}
\psarc{-}(1.25,0){0.25}{0}{180}
\psline{-}(0,0)(0,0.75)
\psline{-}(0.5,0)(0.5,0.75)
\end{pspicture},
 \begin{pspicture}(-0.3,0)(1.8,0.4)
\psdots(0,0)(0.5,0)(1,0)(1.5,0)
\psset{linewidth=1pt}
\psset{linecolor=myc2}
\psarc{-}(0.75,0){0.25}{0}{180}
\psline{-}(0,0)(0,0.75)
\psline{-}(1.5,0)(1.5,0.75)
\end{pspicture},
 \begin{pspicture}(-0.3,0)(1.8,0.4)
\psdots(0,0)(0.5,0)(1,0)(1.5,0)
\psset{linewidth=1pt}
\psset{linecolor=myc2}
\psarc{-}(0.25,0){0.25}{0}{180}
\psline{-}(1,0)(1,0.75)
\psline{-}(1.5,0)(1.5,0.75)
\end{pspicture},
 \begin{pspicture}(-0.3,0)(1.8,0.4)
\psdots(0,0)(0.5,0)(1,0)(1.5,0)
\psset{linewidth=1pt}
\psline{-}(1.25,0.5)(1.75,1)(1.25,1.5)(0.75,1)(1.25,0.5)
\psline{-}(1.25,1.5)(1.75,2)(1.25,2.5)(0.75,2)(1.25,1.5)
\psline{-}(1.25,2.5)(1.75,3)(1.25,3.5)(0.75,3)(1.25,2.5)
\psline{-}(0.75,1)(1.25,1.5)(0.75,2)(0.25,1.5)(0.75,1)
\psline{-}(0.75,2)(1.25,2.5)(0.75,3)(0.25,2.5)(0.75,2)
\psline{-}(0.25,1.5)(0.75,2)(0.25,2.5)(-0.25,2)(0.25,1.5)
\rput(0.25,2){$_3$}
\psset{linecolor=myc2}
\psline{-}(0,0)(0,1.75)
\psline{-}(0.5,0)(0.5,1.25)
\psline{-}(1,0)(1,0.75)
\psline{-}(1.5,0)(1.5,0.75)
\psline{-}(0.0,2.25)(0.0,3.375)
\psline{-}(0.5,2.75)(0.5,3.375)\psline{-}(0.5,3.375)(0.5,4)
\psline{-}(1.0,3.25)(1.0,3.375)\psline{-}(1,3.375)(1,4)
\psline{-}(1.5,3.25)(1.5,3.375)
\psarc{-}(-0.25,3.375){0.25}{0}{90}
\psarc{-}(1.75,3.375){0.25}{90}{180}
\psset{linecolor=myc}
\psarc{-}(1.25,1.5){0.355}{-45}{45}
\psarc{-}(1.25,2.5){0.355}{-45}{45}
\end{pspicture}
 \big\}, \qquad \qquad 
  \mathcal U(\tilde B_4^4) = \big\{
  \begin{pspicture}(-0.3,0)(1.8,0.4)
\psdots(0,0)(0.5,0)(1,0)(1.5,0)
\psset{linewidth=1pt}
\psset{linecolor=myc2}
\psline{-}(0,0)(0,0.75)
\psline{-}(0.5,0)(0.5,0.75)
\psline{-}(1,0)(1,0.75)
\psline{-}(1.5,0)(1.5,0.75)
\end{pspicture}
 \big\}.
\end{equation*}
The Gram matrix is much simpler in these new bases for the $\tilde V_N^d$s.

\begin{Proposition} For $w_1 \in \tilde B_N^{d,r_1}, w_2 \in \tilde B_N^{d,r_2}$
\begin{equation}
\langle  \mathcal U(w_2) |  \mathcal U(w_1) \rangle_G =   \delta_{r_1,r_2} \, K_{d,r_1}   \,\langle  \mathcal C(w_2) |  \mathcal C(w_1) \rangle_G^{\bf v} ,\quad \textrm{with} \quad K_{d,r} = \langle   w^{d,r} |  \wj{d+2r}w^{d,r} \rangle_G,
\end{equation}
where $w^{d,r}$ is the (unique) link state $\in \tilde B_{d+2r}^{d,r}$ and ${\bf v} = (\underbrace{1, ..., 1}_r, \underbrace{v, ..., v}_d, \underbrace{1, ..., 1}_r)$.
\end{Proposition}

\noindent{\scshape Proof\ \ } The only states $w_3$ that can potentially satisfy $\langle  w_3 |  \mathcal U(w_1) \rangle_G \neq 0$ are those in $V_N^d$. This is why $w_1$ and $w_2$ have been taken with the same number of defects. %Il faudra porter une attention particuli\`ere \`a la restriction sur $d$ dans la d\'efinition de $\mathcal U$ en calculant $\langle  \mathcal U(w_2) |  \mathcal U(w_1) \rangle_G$, s'assurant que les deux membres aient toujours $d$ d\'efauts.

What happens if we calculate explicitly $\langle  \mathcal U(w_2) |  \mathcal U(w_1) \rangle_G$? If $r_1 > r_2$, in the Gram diagram, $\wj{d+2r_1}$ has more entries than $\wj{d+2r_2}$ and some entries of $\wj{d+2r_1}$ are necessarily connected pairwise by some (non-boundary) bubbles of the original $w_2$. From the property {(i)} of the $\wj{}$ projector, the result is zero if $r_1 > r_2$, so we restrict our study to $r_1 = r_2 \equiv r$. Two scenarios may occur for the diagram $D_{\mathcal G}(\mathcal U(w_2), \mathcal U(w_1))$.

In the first, the bottom $d+2r$ points of the top projector are not all connected to entries of the bottom projector. If this happens, the top $\wj{}$ has some of its $N$ points connected to non-boundary bubbles and the result is $0$, by the same argument used above for $r_1 \neq r_2$. Under these conditions, $\langle  \mathcal C(w_2) |  \mathcal C(w_1) \rangle_G^{\bf v}$ vanishes. Indeed, the bubbles connecting entry points of the Wenzl-Jones projector, say in $\mathcal U(w_1)$, now connect two defects of $\mathcal C(w_1)$ in $D_{\mathcal G}(\mathcal C(w_2), \mathcal C(w_1))$ and the result is zero.

In the second scenario, the $d+2r$ entries of the top Wenzl-Jones projector coincide with those of the bottom one. Then the pattern of contractible bubbles is the same in $D_{\mathcal G}(\mathcal U(w_2), \mathcal U(w_1))$ and $D_{\mathcal G}(\mathcal C(w_2), \mathcal C(w_1))$ and the corresponding factors of $\beta$ are equal. Let us then concentrate on the $d$ defects and $r$ boundary bubbles of each diagram. The $d+2r$ corresponding entries of the diagram $D_{\mathcal G}(\mathcal U(w_2), \mathcal U(w_1))$ start, from the top, as a state with $d$ defects and $r$ boundary bubbles, that is $w^{d,r}$, go through two copies of the Wenzl-Jones projector $\wj{d+2r}$ and then connect with a second $w^{d,r}$, as in the following example,

\begin{equation*}
\langle \mathcal U(
\psset{unit=0.6}
 \begin{pspicture}(-0.3,0)(2.8,0.6)
\psdots(0,0)(0.5,0)(1,0)(1.5,0)(2,0)(2.5,0)
\psset{linewidth=1pt}
\psset{linecolor=myc2}
\psarc{-}(-0.25,0){0.25}{0}{90}
\psarc{-}(2.75,0){0.25}{90}{180}
\psarc{-}(0.75,0){0.25}{0}{180}
\psline{-}(2,0)(2,0.65)
\psline{-}(1.5,0)(1.5,0.65)
\end{pspicture}
)
| \mathcal U(
 \begin{pspicture}(-0.3,0)(2.8,0.6)
\psdots(0,0)(0.5,0)(1,0)(1.5,0)(2,0)(2.5,0)
\psset{linewidth=1pt}
\psset{linecolor=myc2}
\psarc{-}(-0.25,0){0.25}{0}{90}
\psarc{-}(2.75,0){0.25}{90}{180}
\psarc{-}(1.25,0){0.25}{0}{180}
\psline{-}(2,0)(2,0.65)
\psline{-}(0.5,0)(0.5,0.65)
\end{pspicture}
 )\rangle_G = \, \, \, \, \,
%
% \psset{unit=0.6}
 \begin{pspicture}(-0.3,-0.15)(1.8,4.5)
\psdots(-0.5,0)(0,0)(0.5,0)(1,0)(1.5,0)(2,0)
\psdots(-0.5,4)(0,4)(0.5,4)(1,4)(1.5,4)(2,4)
\psdots(-0.5,-4)(0,-4)(0.5,-4)(1,-4)(1.5,-4)(2,-4)
\psset{linewidth=1pt}
\psline{-}(1.25,0.5)(1.75,1)(1.25,1.5)(0.75,1)(1.25,0.5)
\psline{-}(1.25,1.5)(1.75,2)(1.25,2.5)(0.75,2)(1.25,1.5)
\psline{-}(1.25,2.5)(1.75,3)(1.25,3.5)(0.75,3)(1.25,2.5)
\psline{-}(0.75,1)(1.25,1.5)(0.75,2)(0.25,1.5)(0.75,1)
\psline{-}(0.75,2)(1.25,2.5)(0.75,3)(0.25,2.5)(0.75,2)
\psline{-}(0.25,1.5)(0.75,2)(0.25,2.5)(-0.25,2)(0.25,1.5)
\rput(0.25,2){$_3$}
%\psline[linestyle=dashed,dash=2pt 2pt]{-}(-0.25,0.25)(1.75,0.25)
%\rput(2.375,0.25){$_{d=2}$}
\psset{linecolor=myc2}
\psarc{-}(0.25,0){0.25}{0}{180}
\psbezier{-}(0.0,1.75)(0.0,0.875)(-0.5,0.875)(-0.5,0)
\psbezier{-}(0.5,1.25)(0.5,0.625)(1.0,0.625)(1.0,0)
\psbezier{-}(1.0,0.75)(1.0,0.375)(1.5,0.375)(1.5,0)
\psbezier{-}(1.5,0.75)(1.5,0.375)(2,0.375)(2,0)
\psbezier{-}(0.0,2.25)(0.0,3.125)(-0.5,3.125)(-0.5,4)
\psbezier{-}(0.5,2.75)(0.5,3.375)(1.0,3.375)(1.0,4)
\psbezier{-}(1.0,3.25)(1.0,3.625)(1.5,3.625)(1.5,4)
\psbezier{-}(1.5,3.25)(1.5,3.625)(2,3.625)(2,4)
\psarc{-}(-0.75,4){0.25}{0}{90}
\psarc{-}(2.25,4){0.25}{90}{180}
\psline{-}(1,4)(1,4.7)
\psline{-}(1.5,4)(1.5,4.7)
\psset{linecolor=myc}
\psarc{-}(1.25,1.5){0.355}{-45}{45}
\psarc{-}(1.25,2.5){0.355}{-45}{45}
%\psarc{-}(0.75,2.00){0.355}{-45}{45}\psarc{-}(1.75,2.00){0.355}{135}{-135}
%\psarc{-}(0.75,3.00){0.355}{-45}{45}\psarc{-}(1.75,3.00){0.355}{135}{-135}
%\psarc{-}(0.25,2.50){0.355}{-45}{45}\psarc{-}(1.25,2.50){0.355}{135}{-135}
%\psarc{-}(0.25,1.50){0.355}{45}{135}\psarc{-}(0.25,2.50){0.355}{-135}{-45}
%\psarc{-}(0.75,1.00){0.355}{45}{135}\psarc{-}(0.75,2.00){0.355}{-135}{-45}
%\psarc{-}(1.25,0.50){0.355}{45}{135}\psarc{-}(1.25,1.50){0.355}{-135}{-45}
\psset{linecolor=black}
\psline{-}(1.25,-0.5)(1.75,-1)(1.25,-1.5)(0.75,-1)(1.25,-0.5)
\psline{-}(1.25,-1.5)(1.75,-2)(1.25,-2.5)(0.75,-2)(1.25,-1.5)
\psline{-}(1.25,-2.5)(1.75,-3)(1.25,-3.5)(0.75,-3)(1.25,-2.5)
\psline{-}(0.75,-1)(1.25,-1.5)(0.75,-2)(0.25,-1.5)(0.75,-1)
\psline{-}(0.75,-2)(1.25,-2.5)(0.75,-3)(0.25,-2.5)(0.75,-2)
\psline{-}(0.25,-1.5)(0.75,-2)(0.25,-2.5)(-0.25,-2)(0.25,-1.5)
\rput(0.25,-2){$_3$}
\psset{linecolor=myc2}
\psarc{-}(0.75,0){-0.25}{0}{180}
\psbezier{-}(0.0,-1.75)(0.0,-0.875)(-0.5,-0.875)(-0.5,0)
\psbezier{-}(0.5,-1.25)(0.5,-0.625)(0.0,-0.625)(0.0,0)
\psbezier{-}(1.0,-0.75)(1.0,-0.375)(1.5,-0.375)(1.5,0)
\psbezier{-}(1.5,-0.75)(1.5,-0.375)(2,-0.375)(2,0)
\psbezier{-}(0.0,-2.25)(0.0,-3.125)(-0.5,-3.125)(-0.5,-4)
\psbezier{-}(0.5,-2.75)(0.5,-3.375)(0.0,-3.375)(0.0,-4)
\psbezier{-}(1.0,-3.25)(1.0,-3.625)(1.5,-3.625)(1.5,-4)
\psbezier{-}(1.5,-3.25)(1.5,-3.625)(2,-3.625)(2,-4)
\psarc{-}(-0.75,-4){0.25}{-90}{0}
\psarc{-}(2.25,-4){0.25}{180}{270}
\psline{-}(0,-4)(0,-4.7)
\psline{-}(1.5,-4)(1.5,-4.7)
\psset{linecolor=myc}
\psarc{-}(1.25,-1.5){0.355}{-45}{45}
\psarc{-}(1.25,-2.5){0.355}{-45}{45}
%\psarc{-}(0.75,-3.00){0.355}{-45}{45}\psarc{-}(1.75,-3.00){0.355}{135}{-135}
%\psarc{-}(0.75,-2.00){0.355}{-45}{45}\psarc{-}(1.75,-2.00){0.355}{135}{-135}
%\psarc{-}(0.75,-1.00){0.355}{-45}{45}\psarc{-}(1.75,-1.00){0.355}{135}{-135}
%\psarc{-}(0.25,-2.50){0.355}{-45}{45}\psarc{-}(1.25,-2.50){0.355}{135}{-135}
%\psarc{-}(0.25,-1.50){0.355}{-45}{45}\psarc{-}(1.25,-1.50){0.355}{135}{-135}
%\psarc{-}(-0.25,-2.00){0.355}{-45}{45}\psarc{-}(0.75,-2.00){0.355}{135}{-135}
\end{pspicture} \, \, \, \, \,= \, \, \,
 \begin{pspicture}(-0.3,1.85)(1.8,4.5)
\psdots(-0.5,0)(0,0)(0.5,0)(1,0)(1.5,0)(2,0)
\psdots(-0.5,4)(0,4)(0.5,4)(1,4)(1.5,4)(2,4)
%\psdots(-0.5,-4)(0,-4)(0.5,-4)(1,-4)(1.5,-4)(2,-4)
\psset{linewidth=1pt}
\psline{-}(1.25,0.5)(1.75,1)(1.25,1.5)(0.75,1)(1.25,0.5)
\psline{-}(1.25,1.5)(1.75,2)(1.25,2.5)(0.75,2)(1.25,1.5)
\psline{-}(1.25,2.5)(1.75,3)(1.25,3.5)(0.75,3)(1.25,2.5)
\psline{-}(0.75,1)(1.25,1.5)(0.75,2)(0.25,1.5)(0.75,1)
\psline{-}(0.75,2)(1.25,2.5)(0.75,3)(0.25,2.5)(0.75,2)
\psline{-}(0.25,1.5)(0.75,2)(0.25,2.5)(-0.25,2)(0.25,1.5)
\rput(0.25,2){$_3$}
\psset{linecolor=myc2}
\psarc{-}(0.25,0){0.25}{0}{180}
\psbezier{-}(0.0,1.75)(0.0,0.875)(-0.5,0.875)(-0.5,0)
\psbezier{-}(0.5,1.25)(0.5,0.625)(1.0,0.625)(1.0,0)
\psbezier{-}(1.0,0.75)(1.0,0.375)(1.5,0.375)(1.5,0)
\psbezier{-}(1.5,0.75)(1.5,0.375)(2,0.375)(2,0)
\psbezier{-}(0.0,2.25)(0.0,3.125)(-0.5,3.125)(-0.5,4)
\psbezier{-}(0.5,2.75)(0.5,3.375)(1.0,3.375)(1.0,4)
\psbezier{-}(1.0,3.25)(1.0,3.625)(1.5,3.625)(1.5,4)
\psbezier{-}(1.5,3.25)(1.5,3.625)(2,3.625)(2,4)
\psarc{-}(-0.75,4){0.25}{0}{90}
\psarc{-}(2.25,4){0.25}{90}{180}
\psline{-}(1,4)(1,4.7)
\psline{-}(1.5,4)(1.5,4.7)
\psarc{-}(-0.75,-0){0.25}{-90}{0}
\psarc{-}(2.25,-0){0.25}{180}{270}
\psline{-}(0,-0)(0,-0.7)
\psline{-}(1.5,-0)(1.5,-0.7)
\psarc{-}(0.75,0){-0.25}{0}{180}
\psset{linecolor=myc}
\psarc{-}(1.25,1.5){0.355}{-45}{45}
\psarc{-}(1.25,2.5){0.355}{-45}{45}
\end{pspicture}\, \, \, \, \,= v^2 \, \, \,
 \begin{pspicture}(-0.3,1.85)(1.8,4.5)
\psdots(0,0)(0.5,0)(1,0)(1.5,0)
\psdots(0,4)(0.5,4)(1,4)(1.5,4)
\psset{linewidth=1pt}
\psline{-}(1.25,0.5)(1.75,1)(1.25,1.5)(0.75,1)(1.25,0.5)
\psline{-}(1.25,1.5)(1.75,2)(1.25,2.5)(0.75,2)(1.25,1.5)
\psline{-}(1.25,2.5)(1.75,3)(1.25,3.5)(0.75,3)(1.25,2.5)
\psline{-}(0.75,1)(1.25,1.5)(0.75,2)(0.25,1.5)(0.75,1)
\psline{-}(0.75,2)(1.25,2.5)(0.75,3)(0.25,2.5)(0.75,2)
\psline{-}(0.25,1.5)(0.75,2)(0.25,2.5)(-0.25,2)(0.25,1.5)
\rput(0.25,2){$_3$}
\psset{linecolor=myc2}
\psline{-}(0.0,1.75)(0,0)
\psline{-}(0.5,1.25)(0.5,0)
\psline{-}(1.0,0.75)(1.0,0)
\psline{-}(1.5,0.75)(1.5,0)
\psline{-}(0.0,2.25)(0,4)
\psline{-}(0.5,2.75)(0.5,4)
\psline{-}(1.0,3.25)(1.0,4)
\psline{-}(1.5,3.25)(1.5,4)
\psarc{-}(-0.25,4){0.25}{0}{90}
\psarc{-}(1.75,4){0.25}{90}{180}
\psline{-}(1,4)(1,4.7)
\psline{-}(0.5,4)(0.5,4.7)
\psarc{-}(-0.25,-0){0.25}{-90}{0}
\psarc{-}(1.75,-0){0.25}{180}{270}
\psline{-}(0.5,-0)(0.5,-0.7)
\psline{-}(1.0,-0)(1.0,-0.7)
\psset{linecolor=myc}
\psarc{-}(1.25,1.5){0.355}{-45}{45}
\psarc{-}(1.25,2.5){0.355}{-45}{45}
\end{pspicture}
\, = v^2 K_{2,1},
\end{equation*}
\\ \\ \\ \\ \\ \\
\noindent where properties (ii) and (iii) of the $WJ$ projector were used at the second equality. The product $\langle \mathcal U(w_2)| \mathcal U(w_1) \rangle_G$ is thus  $\beta^{n_{\beta}}v^{\tilde \Delta} K_{d,r}$ for some $\tilde \Delta$, and with $K_{d,r}$  given by
$$K_{d,r} = \langle  w^{d,r} | \wj{d+2r} \, w^{d,r} \rangle_G.$$
For a given configuration of the $\wj{}$ projector, each defect $i$ will contribute a factor $v^{\tilde \Delta_i + \Delta'_i}$ to the total weight,
%$\langle  \mathcal U(w_2) |  \mathcal U(w_1) \rangle_G $
where $\tilde \Delta_i$ is the displacement of the defect $i+r$ in the diagram $D_{\mathcal G}(\mathcal C(w_2), \mathcal C(w_1))$ and $\Delta'_i$ depends upon the configuration chosen for the $WJ$ projector. Defects $1$ to $r$ and $d+r+1$ to $d+2r$ in $D_{\mathcal G}(\mathcal C(w_2), \mathcal C(w_1))$ correspond to boundary bubbles of the original diagram $D_{\mathcal G}(\mathcal U(w_2), \mathcal U(w_1))$ and must contribute $v^0$, which justifies our choice of ${\bf v}$ in $\langle \mathcal C(w_2)| \mathcal C(w_1)\rangle_G^{\bf v}$. For example, the diagram above has a twist factor of $v^2$ which is exactly the twist one finds in computing 
$$
\langle \mathcal C(
\psset{unit=0.6}
 \begin{pspicture}(-0.3,0)(2.8,0.6)
\psdots(0,0)(0.5,0)(1,0)(1.5,0)(2,0)(2.5,0)
\psset{linewidth=1pt}
\psset{linecolor=myc2}
\psarc{-}(-0.25,0){0.25}{0}{90}
\psarc{-}(2.75,0){0.25}{90}{180}
\psarc{-}(0.75,0){0.25}{0}{180}
\psline{-}(2,0)(2,0.65)
\psline{-}(1.5,0)(1.5,0.65)
\end{pspicture}
)
| \mathcal C(
 \begin{pspicture}(-0.3,0)(2.8,0.6)
\psdots(0,0)(0.5,0)(1,0)(1.5,0)(2,0)(2.5,0)
\psset{linewidth=1pt}
\psset{linecolor=myc2}
\psarc{-}(-0.25,0){0.25}{0}{90}
\psarc{-}(2.75,0){0.25}{90}{180}
\psarc{-}(1.25,0){0.25}{0}{180}
\psline{-}(2,0)(2,0.65)
\psline{-}(0.5,0)(0.5,0.65)
\end{pspicture})
 \rangle_G^{\bf v} = 
 \langle 
 \begin{pspicture}(-0.2,0)(2.7,0.6)
\psdots(0,0)(0.5,0)(1,0)(1.5,0)(2,0)(2.5,0)
\psset{linewidth=1pt}
\psset{linecolor=myc2}
\psarc{-}(0.75,0){0.25}{0}{180}
\psline{-}(2.5,0)(2.5,0.65)
\psline{-}(2,0)(2,0.65)
\psline{-}(1.5,0)(1.5,0.65)
\psline{-}(0.0,0)(0.0,0.65)
\end{pspicture}
| 
 \begin{pspicture}(-0.2,0)(2.7,0.6)
\psdots(0,0)(0.5,0)(1,0)(1.5,0)(2,0)(2.5,0)
\psset{linewidth=1pt}
\psset{linecolor=myc2}
\psarc{-}(1.25,0){0.25}{0}{180}
\psline{-}(2,0)(2,0.65)
\psline{-}(2.5,0)(2.5,0.65)
\psline{-}(0,0)(0,0.65)
\psline{-}(0.5,0)(0.5,0.65)
\end{pspicture}
 \rangle_G^{\bf v} = v^2
$$
with ${\bf v} = (1,v,v,1)$ and where $\tilde \Delta_1=2$ and $\tilde \Delta_2=0$. The factor $v^{\sum_i\Delta_i'}$ will depend on the choice of configuration of the projector $\wj{}$ and will be accounted for in the computation of $K_{d,r}$. This will be apparent in section \ref{sec:kdr}. The product $\langle \mathcal U(w_2)| \mathcal U(w_1) \rangle_G$ is thus given by $K_{d,r} \langle \mathcal C(w_2)| \mathcal C(w_1) \rangle_G^{\bf v} $, as given in the proposition. \hfill$\square$ 

\medskip

For $d=0$, the dependence on $\alpha$ is hidden in the constant $K_{d,r}$, while for $d>0$ both $K_{d,r}$ and $\langle \mathcal C(w_2)| \mathcal C(w_1)\rangle_G^{\bf v}$ have a $v$ dependence. The calculation of $K_{d,r}$ will be done in section \ref{sec:kdr}. But we can already sum up the simplification afforded by the new basis.

\begin{Corollaire}
\begin{equation}
\det \tilde {\mathcal G}_N^d = \prod_{r=0}^{{(N-d)}/2} \det( \mathcal G_N^{d+2r}) K_{d,r}^{\dim V_N^{d+2r}}\label{eq:Gint}
\end{equation}
\end{Corollaire}
\noindent{\scshape Proof\ \ } For the proof, we calculate the matrix $\Gamma_N^d = (U_N^d)^T  \tilde {\mathcal G}_N^d U_N^d $, whose matrix elements are given by
\begin{equation*}
(\Gamma_N^d)_{w_1,w_2} = \langle  \mathcal U(w_2) |  \mathcal U(w_1) \rangle_G. 
\end{equation*}
Because $U_N^d$ is upper triangular and has only $1$s on the diagonal, $\det \tilde {\mathcal G}_N^d = \det  \Gamma_N^d$. In the previous proposition, we have found $\Gamma_N^d|_{\tilde V_N^{d,r}} = K_{d,r} \mathcal G_N^{d+2r,\bf v}$. But $\det \mathcal G_N^{d+2r,\bf v}$ is independent of $v$ and given by \eqref{eq:detG0}, and this completes the proof. \hfill$\square$ \\

Again examples are useful:
\vspace{-0cm}
\begin{align*}
\psset{unit=0.55}
 \Gamma_4^0= 
 .
\end{equation*}
\\
\\ \vspace{-0.2cm}

\noindent The next section is devoted to the computation of $K_{d,r}$. 

\subsection{The factor $K_{d,r}$}\label{sec:kdr}

We construct recurrence relations for $K_{d,r}$ and use them to find their values. For $d=0$, $K_{0,r}$ satisfies the relation
\begin{equation} K_{0,r} = K_{0,r-1} (\alpha^2 - 4 C_r^2) \frac{S_r^2}{S_{2r}S_{2r-1}}.\label{eq:recK0r}
\end{equation}
To show this, we sum over all configurations of the top diagonal row. Many of these have weight $0$. All configurations with
$\psset{unit=0.3}
%
$
give $0$ from property (i) of the $\wj{}$ projector. Also, the top entries of the projector $\wj{2r}$ are of two types: those connected to the boundary at $x=N+1/2$ and those connected to the boundary at $x=1/2$. Configurations with 
$\psset{unit=0.3}
%
$
connecting two entries of the same type also give $0$. In the end, only two configurations have non-zero contribution:

\begin{equation*}
\psset{unit=0.4}
\quad K_{0,r} = \, \,
%
\end{equation*} \vspace{1.4cm}

\noindent In the first term, a non-contractible loop is closed and a factor of $\alpha$ is added. Summing over the lower diagonal row and using the same arguments as before gives a unique contribution, and the result is
\begin{equation*}
\alpha \frac{S_r^2}{S_{2r}S_{2r-1}}
\psset{unit=0.4}
%
 = \alpha^2 \frac{S_r^2}{S_{2r}S_{2r-1}} K_{0,r-1}.
\end{equation*}
\\ \\ \\ \vspace{-0.2cm}

\noindent In the second term, we use property (iv) of the $WJ$ projector and find
\begin{equation*}
\psset{unit=0.4}
%
  \, = - \frac{S_{2r}}{S_{2r-1}} K_{0,r-1}
\end{equation*}
\\ \\ \\ \\
where we used

\begin{equation*}
\psset{unit=0.6}
%
 
\end{equation*}
to obtain the second equality. This concludes the proof of equation (\ref{eq:recK0r}). One can compute the initial condition $K_{0,1} = (\alpha^2 - 4 C_1^2) \frac{S_1}{S_2}$ and find, finally, 
\begin{equation}
K_{0,r} = \prod_{k=1}^r  \frac{S_k^2}{S_{2k}S_{2k-1}} (\alpha^2 - 4 C_k^2) = \prod_{k=1}^r  \frac{S_k}{S_{r+k}} (\alpha^2 - 4 C_k^2). \label{eq:finK0r}
\end{equation}

The case $d>0$ depends on the twist parameter $v=e^{i \mu}$ of which we must keep track when writing the recurrence relation
\begin{equation}
K_{d,r} = K_{d,r-1} ( 4 \cos^2 \mu N - 4 C^2_{r+d/2} ) \frac{S_{r}S_{r+d}}{S_{2r+d}S_{2r+d-1}}. \label{eq:recKdr}
\end{equation}

\noindent The steps are otherwise similar to those of the case $d=0$:

\begin{equation}
\psset{unit=0.4}
\quad K_{d,r} = \, \,
\begin{pspicture}(-0.2,-0.25)(4,4.5)
\psset{linewidth=1pt}
\psline{-}(0,0)(0.5,0.5)(1,0)(0.5,-0.5)(0,0)
\psline{-}(1,0)(1.5,0.5)(2,0)(1.5,-0.5)(1,0)
\psline{-}(1,1)(1.5,1.5)(2,1)(1.5,0.5)(1,1)
\psline{-}(1,-1)(1.5,-0.5)(2,-1)(1.5,-1.5)(1,-1)%
\psline{-}(2,0)(2.5,0.5)(3,0)(2.5,-0.5)(2,0)
\psline{-}(2,1)(2.5,1.5)(3,1)(2.5,0.5)(2,1)
\psline{-}(2,2)(2.5,2.5)(3,2)(2.5,1.5)(2,2)
\psline{-}(2,-1)(2.5,-0.5)(3,-1)(2.5,-1.5)(2,-1)
\psline{-}(2,-2)(2.5,-1.5)(3,-2)(2.5,-2.5)(2,-2)%
\psline{-}(3,0)(3.5,0.5)(4,0)(3.5,-0.5)(3,0)
\psline{-}(3,1)(3.5,1.5)(4,1)(3.5,0.5)(3,1)
\psline{-}(3,2)(3.5,2.5)(4,2)(3.5,1.5)(3,2)
\psline{-}(3,3)(3.5,2.5)(4,3)(3.5,3.5)(3,3)
\psline{-}(3,-1)(3.5,-0.5)(4,-1)(3.5,-1.5)(3,-1)
\psline{-}(3,-2)(3.5,-1.5)(4,-2)(3.5,-2.5)(3,-2)
\psline{-}(3,-3)(3.5,-2.5)(4,-3)(3.5,-3.5)(3,-3)%
\psline{-}(0.5,0.5)(1,1)(1.5,0.5)(1,0)(0.5,0.5)
\psline{-}(0.5,-0.5)(1,0)(1.5,-0.5)(1,-1)(0.5,-0.5)
\psline{-}(1.5,0.5)(2,1)(2.5,0.5)(2,0)(1.5,0.5)
\psline{-}(1.5,1.5)(2,2)(2.5,1.5)(2,1)(1.5,1.5)
\psline{-}(1.5,-0.5)(2,0)(2.5,-0.5)(2,-1)(1.5,-0.5)
\psline{-}(1.5,-1.5)(2,-1)(2.5,-1.5)(2,-2)(1.5,-1.5)
\psline{-}(2.5,2.5)(3,3)(3.5,2.5)(3,2)(2.5,2.5)
\psline{-}(2.5,1.5)(3,2)(3.5,1.5)(3,1)(2.5,1.5)
\psline{-}(2.5,0.5)(3,1)(3.5,0.5)(3,0)(2.5,0.5)
\psline{-}(2.5,-0.5)(3,0)(3.5,-0.5)(3,-1)(2.5,-0.5)
\psline{-}(2.5,-1.5)(3,-1)(3.5,-1.5)(3,-2)(2.5,-1.5)
\psline{-}(2.5,-2.5)(3,-2)(3.5,-2.5)(3,-3)(2.5,-2.5)
\psbezier[linewidth=0.5pt](0.5,-0)(0,-0)(-0.25,-0.25)(-0.5,-1.)
\rput(-1.15,-1.25){$_{2r+d-1}$}
\psset{linecolor=myc2}
\psline{-}(3.75,3.25)(3.75,3.75)
\psline{-}(3.25,3.25)(3.25,3.75)
\psline{-}(2.75,2.75)(2.75,4.75)
\psline{-}(2.25,2.25)(2.25,4.75)
\psline{-}(1.75,1.75)(1.75,4.75)
\psline{-}(1.25,1.25)(1.25,4.75)
\psline{-}(0.75,0.75)(0.75,3.75)
\psline{-}(0.25,0.25)(0.25,3.75)%
\psline{-}(3.75,-3.25)(3.75,-3.75)
\psline{-}(3.25,-3.25)(3.25,-3.75)
\psline{-}(2.75,-2.75)(2.75,-4.75)
\psline{-}(2.25,-2.25)(2.25,-4.75)
\psline{-}(1.75,-1.75)(1.75,-4.75)
\psline{-}(1.25,-1.25)(1.25,-4.75)
\psline{-}(0.75,-0.75)(0.75,-3.75)
\psline{-}(0.25,-0.25)(0.25,-3.75)
\psarc{-}(0,3.75){0.25}{0}{90}
\psbezier{-}(0.75,3.75)(0.75,4.125)(0.5,4.25)(0,4.25)
\psbezier{-}(3.25,3.75)(3.25,4.125)(3.5,4.25)(4,4.25)
\psarc{-}(4,3.75){0.25}{90}{180}
\psarc{-}(0,-3.75){0.25}{-90}{0}
\psbezier{-}(0.75,-3.75)(0.75,-4.125)(0.5,-4.25)(0,-4.25)
\psbezier{-}(3.25,-3.75)(3.25,-4.125)(3.5,-4.25)(4,-4.25)
\psarc{-}(4,-3.75){0.25}{180}{-90}
%\psbezier{-}(0.25,3.75)()
\psset{linecolor=myc}
\psarc{-}(3.5,2.5){0.353553}{-45}{45}
\psarc{-}(3.5,1.5){0.353553}{-45}{45}
\psarc{-}(3.5,0.5){0.353553}{-45}{45}
\psarc{-}(3.5,-0.5){0.353553}{-45}{45}
\psarc{-}(3.5,-1.5){0.353553}{-45}{45}
\psarc{-}(3.5,-2.5){0.353553}{-45}{45}
\end{pspicture} 
\, = \frac{S_{r}}{S_{2r+d}}
\begin{pspicture}(-0.2,-0.25)(4,4.5)
\psset{linewidth=1pt}
\psline{-}(0,0)(0.5,0.5)(1,0)(0.5,-0.5)(0,0)
\psline{-}(1,0)(1.5,0.5)(2,0)(1.5,-0.5)(1,0)
\psline{-}(1,1)(1.5,1.5)(2,1)(1.5,0.5)(1,1)
\psline{-}(1,-1)(1.5,-0.5)(2,-1)(1.5,-1.5)(1,-1)%
\psline{-}(2,0)(2.5,0.5)(3,0)(2.5,-0.5)(2,0)
\psline{-}(2,1)(2.5,1.5)(3,1)(2.5,0.5)(2,1)
\psline{-}(2,2)(2.5,2.5)(3,2)(2.5,1.5)(2,2)
\psline{-}(2,-1)(2.5,-0.5)(3,-1)(2.5,-1.5)(2,-1)
\psline{-}(2,-2)(2.5,-1.5)(3,-2)(2.5,-2.5)(2,-2)%
\psline{-}(3,0)(3.5,0.5)(4,0)(3.5,-0.5)(3,0)
\psline{-}(3,1)(3.5,1.5)(4,1)(3.5,0.5)(3,1)
\psline{-}(3,2)(3.5,2.5)(4,2)(3.5,1.5)(3,2)
\psline{-}(3,3)(3.5,2.5)(4,3)(3.5,3.5)(3,3)
\psline{-}(3,-1)(3.5,-0.5)(4,-1)(3.5,-1.5)(3,-1)
\psline{-}(3,-2)(3.5,-1.5)(4,-2)(3.5,-2.5)(3,-2)
\psline{-}(3,-3)(3.5,-2.5)(4,-3)(3.5,-3.5)(3,-3)%
\psline{-}(0.5,0.5)(1,1)(1.5,0.5)(1,0)(0.5,0.5)
\psline{-}(0.5,-0.5)(1,0)(1.5,-0.5)(1,-1)(0.5,-0.5)
\psline{-}(1.5,0.5)(2,1)(2.5,0.5)(2,0)(1.5,0.5)
\psline{-}(1.5,1.5)(2,2)(2.5,1.5)(2,1)(1.5,1.5)
\psline{-}(1.5,-0.5)(2,0)(2.5,-0.5)(2,-1)(1.5,-0.5)
\psline{-}(1.5,-1.5)(2,-1)(2.5,-1.5)(2,-2)(1.5,-1.5)
\psline{-}(2.5,2.5)(3,3)(3.5,2.5)(3,2)(2.5,2.5)
\psline{-}(2.5,1.5)(3,2)(3.5,1.5)(3,1)(2.5,1.5)
\psline{-}(2.5,0.5)(3,1)(3.5,0.5)(3,0)(2.5,0.5)
\psline{-}(2.5,-0.5)(3,0)(3.5,-0.5)(3,-1)(2.5,-0.5)
\psline{-}(2.5,-1.5)(3,-1)(3.5,-1.5)(3,-2)(2.5,-1.5)
\psline{-}(2.5,-2.5)(3,-2)(3.5,-2.5)(3,-3)(2.5,-2.5)
%\psbezier[linewidth=0.5pt](0.5,-0)(0,-0)(-0.25,-0.25)(-0.5,-1.)
%\rput(-0.75,-1.25){$_{2r-1}$}
%\rput(0.5,0){$_d$}
\psset{linecolor=myc2}
\psline{-}(3.75,3.25)(3.75,3.75)
\psline{-}(3.25,3.25)(3.25,3.75)
\psline{-}(2.75,2.75)(2.75,4.75)
\psline{-}(2.25,2.25)(2.25,4.75)
\psline{-}(1.75,1.75)(1.75,4.75)
\psline{-}(1.25,1.25)(1.25,4.75)
\psline{-}(0.75,0.75)(0.75,3.75)
\psline{-}(0.25,0.25)(0.25,3.75)%
\psline{-}(3.75,-3.25)(3.75,-3.75)
\psline{-}(3.25,-3.25)(3.25,-3.75)
\psline{-}(2.75,-2.75)(2.75,-4.75)
\psline{-}(2.25,-2.25)(2.25,-4.75)
\psline{-}(1.75,-1.75)(1.75,-4.75)
\psline{-}(1.25,-1.25)(1.25,-4.75)
\psline{-}(0.75,-0.75)(0.75,-3.75)
\psline{-}(0.25,-0.25)(0.25,-3.75)
\psarc{-}(0,3.75){0.25}{0}{90}
\psbezier{-}(0.75,3.75)(0.75,4.125)(0.5,4.25)(0,4.25)
\psbezier{-}(3.25,3.75)(3.25,4.125)(3.5,4.25)(4,4.25)
\psarc{-}(4,3.75){0.25}{90}{180}
\psarc{-}(0,-3.75){0.25}{-90}{0}
\psbezier{-}(0.75,-3.75)(0.75,-4.125)(0.5,-4.25)(0,-4.25)
\psbezier{-}(3.25,-3.75)(3.25,-4.125)(3.5,-4.25)(4,-4.25)
\psarc{-}(4,-3.75){0.25}{180}{-90}
\psset{linecolor=myc}
\psarc{-}(3.5,2.5){0.353553}{-45}{45}
\psarc{-}(3.5,1.5){0.353553}{-45}{45}
\psarc{-}(3.5,0.5){0.353553}{-45}{45}
\psarc{-}(3.5,-0.5){0.353553}{-45}{45}
\psarc{-}(3.5,-1.5){0.353553}{-45}{45}
\psarc{-}(3.5,-2.5){0.353553}{-45}{45}
\psarc{-}(0.5,0.5){0.353553}{-135}{-45}\psarc{-}(0.5,-0.5){0.353553}{45}{135}
\psarc{-}(1.0,1.0){0.353553}{-135}{-45}\psarc{-}(1.0,-0.0){0.353553}{45}{135}
\psarc{-}(1.5,1.5){0.353553}{-135}{-45}\psarc{-}(1.5,0.5){0.353553}{45}{135}
\psarc{-}(2.0,2.0){0.353553}{-135}{-45}\psarc{-}(2.0,1.0){0.353553}{45}{135}%
\psarc{-}(2.5,2.5){0.353553}{-135}{-45}\psarc{-}(2.5,1.5){0.353553}{45}{135}
\psarc{-}(3.0,3.0){0.353553}{-135}{-45}\psarc{-}(3.0,2.0){0.353553}{45}{135}
\psarc{-}(3.0,3.0){0.353553}{-45}{45}\psarc{-}(4.0,3.0){0.353553}{135}{-135}
\end{pspicture}  \,+ \frac{S_{r+d}}{S_{2r+d}}
\begin{pspicture}(-0.2,-0.25)(4,4.5)
\psset{linewidth=1pt}
\psline{-}(0,0)(0.5,0.5)(1,0)(0.5,-0.5)(0,0)
\psline{-}(1,0)(1.5,0.5)(2,0)(1.5,-0.5)(1,0)
\psline{-}(1,1)(1.5,1.5)(2,1)(1.5,0.5)(1,1)
\psline{-}(1,-1)(1.5,-0.5)(2,-1)(1.5,-1.5)(1,-1)%
\psline{-}(2,0)(2.5,0.5)(3,0)(2.5,-0.5)(2,0)
\psline{-}(2,1)(2.5,1.5)(3,1)(2.5,0.5)(2,1)
\psline{-}(2,2)(2.5,2.5)(3,2)(2.5,1.5)(2,2)
\psline{-}(2,-1)(2.5,-0.5)(3,-1)(2.5,-1.5)(2,-1)
\psline{-}(2,-2)(2.5,-1.5)(3,-2)(2.5,-2.5)(2,-2)%
\psline{-}(3,0)(3.5,0.5)(4,0)(3.5,-0.5)(3,0)
\psline{-}(3,1)(3.5,1.5)(4,1)(3.5,0.5)(3,1)
\psline{-}(3,2)(3.5,2.5)(4,2)(3.5,1.5)(3,2)
\psline{-}(3,3)(3.5,2.5)(4,3)(3.5,3.5)(3,3)
\psline{-}(3,-1)(3.5,-0.5)(4,-1)(3.5,-1.5)(3,-1)
\psline{-}(3,-2)(3.5,-1.5)(4,-2)(3.5,-2.5)(3,-2)
\psline{-}(3,-3)(3.5,-2.5)(4,-3)(3.5,-3.5)(3,-3)%
\psline{-}(0.5,0.5)(1,1)(1.5,0.5)(1,0)(0.5,0.5)
\psline{-}(0.5,-0.5)(1,0)(1.5,-0.5)(1,-1)(0.5,-0.5)
\psline{-}(1.5,0.5)(2,1)(2.5,0.5)(2,0)(1.5,0.5)
\psline{-}(1.5,1.5)(2,2)(2.5,1.5)(2,1)(1.5,1.5)
\psline{-}(1.5,-0.5)(2,0)(2.5,-0.5)(2,-1)(1.5,-0.5)
\psline{-}(1.5,-1.5)(2,-1)(2.5,-1.5)(2,-2)(1.5,-1.5)
\psline{-}(2.5,2.5)(3,3)(3.5,2.5)(3,2)(2.5,2.5)
\psline{-}(2.5,1.5)(3,2)(3.5,1.5)(3,1)(2.5,1.5)
\psline{-}(2.5,0.5)(3,1)(3.5,0.5)(3,0)(2.5,0.5)
\psline{-}(2.5,-0.5)(3,0)(3.5,-0.5)(3,-1)(2.5,-0.5)
\psline{-}(2.5,-1.5)(3,-1)(3.5,-1.5)(3,-2)(2.5,-1.5)
\psline{-}(2.5,-2.5)(3,-2)(3.5,-2.5)(3,-3)(2.5,-2.5)
%\psbezier[linewidth=0.5pt](0.5,-0)(0,-0)(-0.25,-0.25)(-0.5,-1.)
%\rput(-0.75,-1.25){$_{2r-1}$}
%\rput(0.5,0){$_d$}
\psset{linecolor=myc2}
\psline{-}(3.75,3.25)(3.75,3.75)
\psline{-}(3.25,3.25)(3.25,3.75)
\psline{-}(2.75,2.75)(2.75,4.75)
\psline{-}(2.25,2.25)(2.25,4.75)
\psline{-}(1.75,1.75)(1.75,4.75)
\psline{-}(1.25,1.25)(1.25,4.75)
\psline{-}(0.75,0.75)(0.75,3.75)
\psline{-}(0.25,0.25)(0.25,3.75)%
\psline{-}(3.75,-3.25)(3.75,-3.75)
\psline{-}(3.25,-3.25)(3.25,-3.75)
\psline{-}(2.75,-2.75)(2.75,-4.75)
\psline{-}(2.25,-2.25)(2.25,-4.75)
\psline{-}(1.75,-1.75)(1.75,-4.75)
\psline{-}(1.25,-1.25)(1.25,-4.75)
\psline{-}(0.75,-0.75)(0.75,-3.75)
\psline{-}(0.25,-0.25)(0.25,-3.75)
\psarc{-}(0,3.75){0.25}{0}{90}
\psbezier{-}(0.75,3.75)(0.75,4.125)(0.5,4.25)(0,4.25)
\psbezier{-}(3.25,3.75)(3.25,4.125)(3.5,4.25)(4,4.25)
\psarc{-}(4,3.75){0.25}{90}{180}
\psarc{-}(0,-3.75){0.25}{-90}{0}
\psbezier{-}(0.75,-3.75)(0.75,-4.125)(0.5,-4.25)(0,-4.25)
\psbezier{-}(3.25,-3.75)(3.25,-4.125)(3.5,-4.25)(4,-4.25)
\psarc{-}(4,-3.75){0.25}{180}{-90}
\psset{linecolor=myc}
\psarc{-}(3.5,2.5){0.353553}{-45}{45}
\psarc{-}(3.5,1.5){0.353553}{-45}{45}
\psarc{-}(3.5,0.5){0.353553}{-45}{45}
\psarc{-}(3.5,-0.5){0.353553}{-45}{45}
\psarc{-}(3.5,-1.5){0.353553}{-45}{45}
\psarc{-}(3.5,-2.5){0.353553}{-45}{45}
\psarc{-}(0.5,0.5){0.353553}{-135}{-45}\psarc{-}(0.5,-0.5){0.353553}{45}{135}
\psarc{-}(1.0,1.0){0.353553}{-135}{-45}\psarc{-}(1.0,-0.0){0.353553}{45}{135}
\psarc{-}(1.0,1.0){0.353553}{-45}{45}\psarc{-}(2.0,1.0){0.353553}{135}{-135}
\psarc{-}(1.5,1.5){0.353553}{-45}{45}\psarc{-}(2.5,1.5){0.353553}{135}{-135}
\psarc{-}(2.0,2.0){0.353553}{-45}{45}\psarc{-}(3.0,2.0){0.353553}{135}{-135}
\psarc{-}(2.5,2.5){0.353553}{-45}{45}\psarc{-}(3.5,2.5){0.353553}{135}{-135}
\psarc{-}(3.0,3.0){0.353553}{-45}{45}\psarc{-}(4.0,3.0){0.353553}{135}{-135}
\end{pspicture}
\, +1 \, 
\begin{pspicture}(-0.2,-0.25)(4,4.5)
\psset{linewidth=1pt}
\psline{-}(0,0)(0.5,0.5)(1,0)(0.5,-0.5)(0,0)
\psline{-}(1,0)(1.5,0.5)(2,0)(1.5,-0.5)(1,0)
\psline{-}(1,1)(1.5,1.5)(2,1)(1.5,0.5)(1,1)
\psline{-}(1,-1)(1.5,-0.5)(2,-1)(1.5,-1.5)(1,-1)%
\psline{-}(2,0)(2.5,0.5)(3,0)(2.5,-0.5)(2,0)
\psline{-}(2,1)(2.5,1.5)(3,1)(2.5,0.5)(2,1)
\psline{-}(2,2)(2.5,2.5)(3,2)(2.5,1.5)(2,2)
\psline{-}(2,-1)(2.5,-0.5)(3,-1)(2.5,-1.5)(2,-1)
\psline{-}(2,-2)(2.5,-1.5)(3,-2)(2.5,-2.5)(2,-2)%
\psline{-}(3,0)(3.5,0.5)(4,0)(3.5,-0.5)(3,0)
\psline{-}(3,1)(3.5,1.5)(4,1)(3.5,0.5)(3,1)
\psline{-}(3,2)(3.5,2.5)(4,2)(3.5,1.5)(3,2)
\psline{-}(3,3)(3.5,2.5)(4,3)(3.5,3.5)(3,3)
\psline{-}(3,-1)(3.5,-0.5)(4,-1)(3.5,-1.5)(3,-1)
\psline{-}(3,-2)(3.5,-1.5)(4,-2)(3.5,-2.5)(3,-2)
\psline{-}(3,-3)(3.5,-2.5)(4,-3)(3.5,-3.5)(3,-3)%
\psline{-}(0.5,0.5)(1,1)(1.5,0.5)(1,0)(0.5,0.5)
\psline{-}(0.5,-0.5)(1,0)(1.5,-0.5)(1,-1)(0.5,-0.5)
\psline{-}(1.5,0.5)(2,1)(2.5,0.5)(2,0)(1.5,0.5)
\psline{-}(1.5,1.5)(2,2)(2.5,1.5)(2,1)(1.5,1.5)
\psline{-}(1.5,-0.5)(2,0)(2.5,-0.5)(2,-1)(1.5,-0.5)
\psline{-}(1.5,-1.5)(2,-1)(2.5,-1.5)(2,-2)(1.5,-1.5)
\psline{-}(2.5,2.5)(3,3)(3.5,2.5)(3,2)(2.5,2.5)
\psline{-}(2.5,1.5)(3,2)(3.5,1.5)(3,1)(2.5,1.5)
\psline{-}(2.5,0.5)(3,1)(3.5,0.5)(3,0)(2.5,0.5)
\psline{-}(2.5,-0.5)(3,0)(3.5,-0.5)(3,-1)(2.5,-0.5)
\psline{-}(2.5,-1.5)(3,-1)(3.5,-1.5)(3,-2)(2.5,-1.5)
\psline{-}(2.5,-2.5)(3,-2)(3.5,-2.5)(3,-3)(2.5,-2.5)
%\psbezier[linewidth=0.5pt](0.5,-0)(0,-0)(-0.25,-0.25)(-0.5,-1.)
%\rput(-0.75,-1.25){$_{2r-1}$}
%\rput(0.5,0){$_d$}
\psset{linecolor=myc2}
\psline{-}(3.75,3.25)(3.75,3.75)
\psline{-}(3.25,3.25)(3.25,3.75)
\psline{-}(2.75,2.75)(2.75,4.75)
\psline{-}(2.25,2.25)(2.25,4.75)
\psline{-}(1.75,1.75)(1.75,4.75)
\psline{-}(1.25,1.25)(1.25,4.75)
\psline{-}(0.75,0.75)(0.75,3.75)
\psline{-}(0.25,0.25)(0.25,3.75)%
\psline{-}(3.75,-3.25)(3.75,-3.75)
\psline{-}(3.25,-3.25)(3.25,-3.75)
\psline{-}(2.75,-2.75)(2.75,-4.75)
\psline{-}(2.25,-2.25)(2.25,-4.75)
\psline{-}(1.75,-1.75)(1.75,-4.75)
\psline{-}(1.25,-1.25)(1.25,-4.75)
\psline{-}(0.75,-0.75)(0.75,-3.75)
\psline{-}(0.25,-0.25)(0.25,-3.75)
\psarc{-}(0,3.75){0.25}{0}{90}
\psbezier{-}(0.75,3.75)(0.75,4.125)(0.5,4.25)(0,4.25)
%\psbezier{-}(1.25,3.75)(1.25,4.3125)(0.8333,4.5)(0,4.5)
%\psbezier{-}(1.75,3.75)(1.75,4.5)(1.166,4.75)(0,4.75)
\psbezier{-}(3.25,3.75)(3.25,4.125)(3.5,4.25)(4,4.25)
%\psbezier{-}(2.75,3.75)(2.75,4.3125)(3.1666,4.5)(4,4.5)
%\psbezier{-}(2.25,3.75)(2.25,4.5)(2.8333,4.75)(4,4.75)
\psarc{-}(4,3.75){0.25}{90}{180}
\psarc{-}(0,-3.75){0.25}{-90}{0}
\psbezier{-}(0.75,-3.75)(0.75,-4.125)(0.5,-4.25)(0,-4.25)
%\psbezier{-}(1.25,-3.75)(1.25,-4.3125)(0.8333,-4.5)(0,-4.5)
%\psbezier{-}(1.75,-3.75)(1.75,-4.5)(1.166,-4.75)(0,-4.75)
\psbezier{-}(3.25,-3.75)(3.25,-4.125)(3.5,-4.25)(4,-4.25)
%\psbezier{-}(2.75,-3.75)(2.75,-4.3125)(3.1666,-4.5)(4,-4.5)
%\psbezier{-}(2.25,-3.75)(2.25,-4.5)(2.8333,-4.75)(4,-4.75)
\psarc{-}(4,-3.75){0.25}{180}{-90}
%\psbezier{-}(0.25,3.75)()
\psset{linecolor=myc}
\psarc{-}(3.5,2.5){0.353553}{-45}{45}
\psarc{-}(3.5,1.5){0.353553}{-45}{45}
\psarc{-}(3.5,0.5){0.353553}{-45}{45}
\psarc{-}(3.5,-0.5){0.353553}{-45}{45}
\psarc{-}(3.5,-1.5){0.353553}{-45}{45}
\psarc{-}(3.5,-2.5){0.353553}{-45}{45}
\psarc{-}(0.0,0.0){0.353553}{-45}{45}\psarc{-}(1.0,0.0){0.353553}{135}{-135}
\psarc{-}(0.5,0.5){0.353553}{-45}{45}\psarc{-}(1.5,0.5){0.353553}{135}{-135}
\psarc{-}(1.0,1.0){0.353553}{-45}{45}\psarc{-}(2.0,1.0){0.353553}{135}{-135}
\psarc{-}(1.5,1.5){0.353553}{-45}{45}\psarc{-}(2.5,1.5){0.353553}{135}{-135}
\psarc{-}(2.0,2.0){0.353553}{-45}{45}\psarc{-}(3.0,2.0){0.353553}{135}{-135}
\psarc{-}(2.5,2.5){0.353553}{-45}{45}\psarc{-}(3.5,2.5){0.353553}{135}{-135}
\psarc{-}(3.0,3.0){0.353553}{-45}{45}\psarc{-}(4.0,3.0){0.353553}{135}{-135}
\end{pspicture}\, .
\label{eq:dessinrecKdr}\end{equation}
\\ \\ \\ \\
Top entries of the projector are of three types: Besides the left and right boundary bubbles encountered before, they can also be connected to defects. Whenever 
$\psset{unit=0.3}
\begin{pspicture}(-1,-0.6)(0.5,0.5)\lw
\psline{-}(-0.5,0)(0,-0.5)(0.5,0)(0,0.5)(-0.5,0)
\psline{-}(-0.5,0)(0,-0.5)(-0.5,-1)(-1,-0.5)(-0.5,0)
\psset{linecolor=myc}\unlw
\psarc{-}(-0.5,0){0.3536}{225}{315}
\psarc{-}(-0.5,-1){0.3536}{45}{135}
\psarc{-}(-0.5,0){0.3536}{-45}{45}
\psarc{-}(0.5,0){0.3536}{135}{-135}
\end{pspicture}$
connects two entries of the same type, the result is $0$ as before. The connections of the first and second term of (\ref{eq:dessinrecKdr}) are identical, but the weight due to the twist in the defects is not and remains to be computed.  When computing twist factors, we must not forget that the original diagram has $N$ positions and that contractible loops can be present between the entries of the projector $\wj{}$.
The first diagram provides a good example. Entries of the projector $\wj{}$ are labeled by integers $i = 1, \dots, 2r+d$ (defects occupy positions $r+1$ to $r+d$) and correspond to some positions $p_i$ in the original diagram. In (\ref{eq:dessinrecKdr}), the $(d-1)$ leftmost defects entering the projector $\wj{}$ all connect two positions to the right of their entry point, so their contribution to the twist factor is $v^{p_i-p_{i+2}}$, for $i = r+1, ..., r+d-1$. The rightmost defect enters from the top at position $r+d$, moves right across the imaginary boundary, and then connects at position $r+2$. Because the original diagram has $N$ positions, this last defect contributes $v^{p_{r+d} - (N+p_{r+2})}$. The total twist weight of the $d$ defects of this first diagram sums to $v^{-(N+\delta)}$, with $\delta = p_{r+d+1}-p_{r+1}$. Each defect of the second diagram has the same entry and exit points, except for the leftmost defect that wraps around in the left direction. This defect gives the only contribution to the second diagram, namely $v^{(N-\delta)}$. With these twist weights, contributions of the first and second diagrams combine and give
\begin{align*}
\underbrace{v^{-\delta} \left( v^N \frac{S_{r+d}}{S_{2r+d}} + v^{-N} \frac{S_{r}}{S_{2r+d}} \right)}_{A(r,d,N)}
&
\psset{unit=0.4}
\begin{pspicture}(-0.7,-0.25)(4,4.5)
\psset{linewidth=1pt}
\psline{-}(0,0)(0.5,0.5)(1,0)(0.5,-0.5)(0,0)
\psline{-}(1,0)(1.5,0.5)(2,0)(1.5,-0.5)(1,0)
\psline{-}(1,1)(1.5,1.5)(2,1)(1.5,0.5)(1,1)
\psline{-}(1,-1)(1.5,-0.5)(2,-1)(1.5,-1.5)(1,-1)%
\psline{-}(2,0)(2.5,0.5)(3,0)(2.5,-0.5)(2,0)
\psline{-}(2,1)(2.5,1.5)(3,1)(2.5,0.5)(2,1)
\psline{-}(2,2)(2.5,2.5)(3,2)(2.5,1.5)(2,2)
\psline{-}(2,-1)(2.5,-0.5)(3,-1)(2.5,-1.5)(2,-1)
\psline{-}(2,-2)(2.5,-1.5)(3,-2)(2.5,-2.5)(2,-2)%
%\psline{-}(3,0)(3.5,0.5)(4,0)(3.5,-0.5)(3,0)
%\psline{-}(3,1)(3.5,1.5)(4,1)(3.5,0.5)(3,1)
%\psline{-}(3,2)(3.5,2.5)(4,2)(3.5,1.5)(3,2)
%\psline{-}(3,3)(3.5,2.5)(4,3)(3.5,3.5)(3,3)
%\psline{-}(3,-1)(3.5,-0.5)(4,-1)(3.5,-1.5)(3,-1)
%\psline{-}(3,-2)(3.5,-1.5)(4,-2)(3.5,-2.5)(3,-2)
%\psline{-}(3,-3)(3.5,-2.5)(4,-3)(3.5,-3.5)(3,-3)%
\psline{-}(0.5,0.5)(1,1)(1.5,0.5)(1,0)(0.5,0.5)
\psline{-}(0.5,-0.5)(1,0)(1.5,-0.5)(1,-1)(0.5,-0.5)
\psline{-}(1.5,0.5)(2,1)(2.5,0.5)(2,0)(1.5,0.5)
\psline{-}(1.5,1.5)(2,2)(2.5,1.5)(2,1)(1.5,1.5)
\psline{-}(1.5,-0.5)(2,0)(2.5,-0.5)(2,-1)(1.5,-0.5)
\psline{-}(1.5,-1.5)(2,-1)(2.5,-1.5)(2,-2)(1.5,-1.5)
\psline{-}(2.5,2.5)(3,3)(3.5,2.5)(3,2)(2.5,2.5)
\psline{-}(2.5,1.5)(3,2)(3.5,1.5)(3,1)(2.5,1.5)
\psline{-}(2.5,0.5)(3,1)(3.5,0.5)(3,0)(2.5,0.5)
\psline{-}(2.5,-0.5)(3,0)(3.5,-0.5)(3,-1)(2.5,-0.5)
\psline{-}(2.5,-1.5)(3,-1)(3.5,-1.5)(3,-2)(2.5,-1.5)
\psline{-}(2.5,-2.5)(3,-2)(3.5,-2.5)(3,-3)(2.5,-2.5)
\psbezier[linewidth=0.5pt](0.5,0.0)(0.0,0.0)(-0.25,0.25)(-0.5,1.25)
\rput(-0.75,1.55){$_{2r+d-2}$}
\psset{linecolor=myc2}
%\psline{-}(3.75,3.25)(3.75,3.75)
\psline{-}(3.25,2.75)(3.25,3.25)
\psline{-}(2.75,2.75)(2.75,4.25)
\psline{-}(2.25,2.25)(2.25,4.25)
\psline{-}(1.75,1.75)(1.75,4.25)
\psline{-}(1.25,1.25)(1.25,4.25)
\psline{-}(0.75,0.75)(0.75,3.25)
\psline{-}(0.25,0.25)(0.25,0.75)%
\psarc{-}(0,0.75){0.25}{0}{180}
%\psline{-}(3.75,-3.25)(3.75,-3.75)
\psline{-}(3.25,-2.75)(3.25,-3.25)
\psline{-}(2.75,-2.75)(2.75,-3.25)
\psline{-}(2.25,-2.25)(2.25,-4.25)
\psline{-}(1.75,-1.75)(1.75,-4.25)
\psline{-}(1.25,-1.25)(1.25,-4.25)
\psline{-}(0.75,-0.75)(0.75,-4.25)
\psline{-}(0.25,-0.25)(0.25,-3.25)
\psline{-}(-0.25,0.75)(-0.25,-3.25)
%\psarc{-}(0,3.75){0.25}{0}{90}
\psbezier{-}(0.75,3.25)(0.75,3.625)(0.5,3.75)(0,3.75)
%\psbezier{-}(1.25,3.25)(1.25,3.7125)(0.8333,4.0)(0,4.0)
%\psbezier{-}(1.75,3.25)(1.75,4.0)(1.166,4.25)(0,4.25)
\psbezier{-}(3.25,3.25)(3.25,3.625)(3.5,3.75)(4,3.75)
%\psbezier{-}(2.75,3.25)(2.75,3.8125)(3.1666,4.0)(4,4.0)
%\psbezier{-}(2.25,3.25)(2.25,4.0)(2.8333,4.25)(4,4.25)
%\psarc{-}(4,3.75){0.25}{90}{180}
%
\psarc{-}(-0.5,-3.25){0.25}{-90}{0}
\psbezier{-}(0.25,-3.25)(0.25,-3.625)(0.0,-3.75)(-0.5,-3.75)
\psbezier{-}(2.75,-3.25)(2.75,-3.625)(3.0,-3.75)(3.5,-3.75)
\psarc{-}(3.5,-3.25){0.25}{180}{-90}
%\psbezier{-}(0.25,3.75)()
\psset{linecolor=myc}
\psarc{-}(3.0,2.0){0.353553}{-45}{45}
\psarc{-}(3.0,1.0){0.353553}{-45}{45}
\psarc{-}(3.0,0.0){0.353553}{-45}{45}
\psarc{-}(3.0,-1.0){0.353553}{-45}{45}
\psarc{-}(3.0,-2.0){0.353553}{-45}{45}
\end{pspicture} = A(r,d,N) \left( \frac{S_{r+d}}{S_{2r+d-1}}
\begin{pspicture}(-0.7,-0.25)(4,4.5)
\psset{linewidth=1pt}
\psline{-}(0,0)(0.5,0.5)(1,0)(0.5,-0.5)(0,0)
\psline{-}(1,0)(1.5,0.5)(2,0)(1.5,-0.5)(1,0)
\psline{-}(1,1)(1.5,1.5)(2,1)(1.5,0.5)(1,1)
\psline{-}(1,-1)(1.5,-0.5)(2,-1)(1.5,-1.5)(1,-1)%
\psline{-}(2,0)(2.5,0.5)(3,0)(2.5,-0.5)(2,0)
\psline{-}(2,1)(2.5,1.5)(3,1)(2.5,0.5)(2,1)
\psline{-}(2,2)(2.5,2.5)(3,2)(2.5,1.5)(2,2)
\psline{-}(2,-1)(2.5,-0.5)(3,-1)(2.5,-1.5)(2,-1)
\psline{-}(2,-2)(2.5,-1.5)(3,-2)(2.5,-2.5)(2,-2)%
%\psline{-}(3,0)(3.5,0.5)(4,0)(3.5,-0.5)(3,0)
%\psline{-}(3,1)(3.5,1.5)(4,1)(3.5,0.5)(3,1)
%\psline{-}(3,2)(3.5,2.5)(4,2)(3.5,1.5)(3,2)
%\psline{-}(3,3)(3.5,2.5)(4,3)(3.5,3.5)(3,3)
%\psline{-}(3,-1)(3.5,-0.5)(4,-1)(3.5,-1.5)(3,-1)
%\psline{-}(3,-2)(3.5,-1.5)(4,-2)(3.5,-2.5)(3,-2)
%\psline{-}(3,-3)(3.5,-2.5)(4,-3)(3.5,-3.5)(3,-3)%
\psline{-}(0.5,0.5)(1,1)(1.5,0.5)(1,0)(0.5,0.5)
\psline{-}(0.5,-0.5)(1,0)(1.5,-0.5)(1,-1)(0.5,-0.5)
\psline{-}(1.5,0.5)(2,1)(2.5,0.5)(2,0)(1.5,0.5)
\psline{-}(1.5,1.5)(2,2)(2.5,1.5)(2,1)(1.5,1.5)
\psline{-}(1.5,-0.5)(2,0)(2.5,-0.5)(2,-1)(1.5,-0.5)
\psline{-}(1.5,-1.5)(2,-1)(2.5,-1.5)(2,-2)(1.5,-1.5)
\psline{-}(2.5,2.5)(3,3)(3.5,2.5)(3,2)(2.5,2.5)
\psline{-}(2.5,1.5)(3,2)(3.5,1.5)(3,1)(2.5,1.5)
\psline{-}(2.5,0.5)(3,1)(3.5,0.5)(3,0)(2.5,0.5)
\psline{-}(2.5,-0.5)(3,0)(3.5,-0.5)(3,-1)(2.5,-0.5)
\psline{-}(2.5,-1.5)(3,-1)(3.5,-1.5)(3,-2)(2.5,-1.5)
\psline{-}(2.5,-2.5)(3,-2)(3.5,-2.5)(3,-3)(2.5,-2.5)
\psbezier[linewidth=0.5pt](1,0.5)(0.5,0.5)(0.25,0.75)(0.,1.5)
\rput(-0.65,1.75){$_{2r+d-3}$}
\psset{linecolor=myc2}
%\psline{-}(3.75,3.25)(3.75,3.75)
\psline{-}(3.25,2.75)(3.25,3.25)
\psline{-}(2.75,2.75)(2.75,4.25)
\psline{-}(2.25,2.25)(2.25,4.25)
\psline{-}(1.75,1.75)(1.75,4.25)
\psline{-}(1.25,1.25)(1.25,4.25)
\psline{-}(0.75,0.75)(0.75,3.25)
\psline{-}(0.25,0.25)(0.25,0.75)%
\psarc{-}(0,0.75){0.25}{0}{180}
%\psline{-}(3.75,-3.25)(3.75,-3.75)
\psline{-}(3.25,-2.75)(3.25,-3.25)
\psline{-}(2.75,-2.75)(2.75,-3.25)
\psline{-}(2.25,-2.25)(2.25,-4.25)
\psline{-}(1.75,-1.75)(1.75,-4.25)
\psline{-}(1.25,-1.25)(1.25,-4.25)
\psline{-}(0.75,-0.75)(0.75,-4.25)
\psline{-}(0.25,-0.25)(0.25,-3.25)
\psline{-}(-0.25,0.75)(-0.25,-3.25)
%\psarc{-}(0,3.75){0.25}{0}{90}
\psbezier{-}(0.75,3.25)(0.75,3.625)(0.5,3.75)(0,3.75)
%\psbezier{-}(1.25,3.25)(1.25,3.7125)(0.8333,4.0)(0,4.0)
%\psbezier{-}(1.75,3.25)(1.75,4.0)(1.166,4.25)(0,4.25)
\psbezier{-}(3.25,3.25)(3.25,3.625)(3.5,3.75)(4,3.75)
%\psbezier{-}(2.75,3.25)(2.75,3.8125)(3.1666,4.0)(4,4.0)
%\psbezier{-}(2.25,3.25)(2.25,4.0)(2.8333,4.25)(4,4.25)
%\psarc{-}(4,3.75){0.25}{90}{180}
%
\psarc{-}(-0.5,-3.25){0.25}{-90}{0}
\psbezier{-}(0.25,-3.25)(0.25,-3.625)(0.0,-3.75)(-0.5,-3.75)
\psbezier{-}(2.75,-3.25)(2.75,-3.625)(3.0,-3.75)(3.5,-3.75)
\psarc{-}(3.5,-3.25){0.25}{180}{-90}
%\psbezier{-}(0.25,3.75)()
\psset{linecolor=myc}
\psarc{-}(3.0,2.0){0.353553}{-45}{45}
\psarc{-}(3.0,1.0){0.353553}{-45}{45}
\psarc{-}(3.0,0.0){0.353553}{-45}{45}
\psarc{-}(3.0,-1.0){0.353553}{-45}{45}
\psarc{-}(3.0,-2.0){0.353553}{-45}{45}
\psarc{-}(0.5,0.5){0.353553}{-135}{-45}\psarc{-}(0.5,-0.5){0.353553}{45}{135}
\psarc{-}(0.5,-0.5){0.353553}{-45}{45}\psarc{-}(1.5,-0.5){0.353553}{135}{-135}
\psarc{-}(1.0,-1.0){0.353553}{-45}{45}\psarc{-}(2.0,-1.0){0.353553}{135}{-135}
\psarc{-}(1.5,-1.5){0.353553}{-45}{45}\psarc{-}(2.5,-1.5){0.353553}{135}{-135}
\psarc{-}(2.0,-2.0){0.353553}{-45}{45}\psarc{-}(3.0,-2.0){0.353553}{135}{-135}
\psarc{-}(2.5,-2.5){0.353553}{-45}{45}\psarc{-}(3.5,-2.5){0.353553}{135}{-135}
\end{pspicture} \, + \frac{S_r}{S_{2r+d-1}}
\begin{pspicture}(-0.7,-0.25)(4,4.5)
\psset{linewidth=1pt}
\psline{-}(0,0)(0.5,0.5)(1,0)(0.5,-0.5)(0,0)
\psline{-}(1,0)(1.5,0.5)(2,0)(1.5,-0.5)(1,0)
\psline{-}(1,1)(1.5,1.5)(2,1)(1.5,0.5)(1,1)
\psline{-}(1,-1)(1.5,-0.5)(2,-1)(1.5,-1.5)(1,-1)%
\psline{-}(2,0)(2.5,0.5)(3,0)(2.5,-0.5)(2,0)
\psline{-}(2,1)(2.5,1.5)(3,1)(2.5,0.5)(2,1)
\psline{-}(2,2)(2.5,2.5)(3,2)(2.5,1.5)(2,2)
\psline{-}(2,-1)(2.5,-0.5)(3,-1)(2.5,-1.5)(2,-1)
\psline{-}(2,-2)(2.5,-1.5)(3,-2)(2.5,-2.5)(2,-2)%
%\psline{-}(3,0)(3.5,0.5)(4,0)(3.5,-0.5)(3,0)
%\psline{-}(3,1)(3.5,1.5)(4,1)(3.5,0.5)(3,1)
%\psline{-}(3,2)(3.5,2.5)(4,2)(3.5,1.5)(3,2)
%\psline{-}(3,3)(3.5,2.5)(4,3)(3.5,3.5)(3,3)
%\psline{-}(3,-1)(3.5,-0.5)(4,-1)(3.5,-1.5)(3,-1)
%\psline{-}(3,-2)(3.5,-1.5)(4,-2)(3.5,-2.5)(3,-2)
%\psline{-}(3,-3)(3.5,-2.5)(4,-3)(3.5,-3.5)(3,-3)%
\psline{-}(0.5,0.5)(1,1)(1.5,0.5)(1,0)(0.5,0.5)
\psline{-}(0.5,-0.5)(1,0)(1.5,-0.5)(1,-1)(0.5,-0.5)
\psline{-}(1.5,0.5)(2,1)(2.5,0.5)(2,0)(1.5,0.5)
\psline{-}(1.5,1.5)(2,2)(2.5,1.5)(2,1)(1.5,1.5)
\psline{-}(1.5,-0.5)(2,0)(2.5,-0.5)(2,-1)(1.5,-0.5)
\psline{-}(1.5,-1.5)(2,-1)(2.5,-1.5)(2,-2)(1.5,-1.5)
\psline{-}(2.5,2.5)(3,3)(3.5,2.5)(3,2)(2.5,2.5)
\psline{-}(2.5,1.5)(3,2)(3.5,1.5)(3,1)(2.5,1.5)
\psline{-}(2.5,0.5)(3,1)(3.5,0.5)(3,0)(2.5,0.5)
\psline{-}(2.5,-0.5)(3,0)(3.5,-0.5)(3,-1)(2.5,-0.5)
\psline{-}(2.5,-1.5)(3,-1)(3.5,-1.5)(3,-2)(2.5,-1.5)
\psline{-}(2.5,-2.5)(3,-2)(3.5,-2.5)(3,-3)(2.5,-2.5)
\psbezier[linewidth=0.5pt](1,0.5)(0.5,0.5)(0.25,0.75)(0.,1.5)
\rput(-0.65,1.75){$_{2r+d-3}$}
\psset{linecolor=myc2}
%\psline{-}(3.75,3.25)(3.75,3.75)
\psline{-}(3.25,2.75)(3.25,3.25)
\psline{-}(2.75,2.75)(2.75,4.25)
\psline{-}(2.25,2.25)(2.25,4.25)
\psline{-}(1.75,1.75)(1.75,4.25)
\psline{-}(1.25,1.25)(1.25,4.25)
\psline{-}(0.75,0.75)(0.75,3.25)
\psline{-}(0.25,0.25)(0.25,0.75)%
\psarc{-}(0,0.75){0.25}{0}{180}
%\psline{-}(3.75,-3.25)(3.75,-3.75)
\psline{-}(3.25,-2.75)(3.25,-3.25)
\psline{-}(2.75,-2.75)(2.75,-3.25)
\psline{-}(2.25,-2.25)(2.25,-4.25)
\psline{-}(1.75,-1.75)(1.75,-4.25)
\psline{-}(1.25,-1.25)(1.25,-4.25)
\psline{-}(0.75,-0.75)(0.75,-4.25)
\psline{-}(0.25,-0.25)(0.25,-3.25)
\psline{-}(-0.25,0.75)(-0.25,-3.25)
%\psarc{-}(0,3.75){0.25}{0}{90}
\psbezier{-}(0.75,3.25)(0.75,3.625)(0.5,3.75)(0,3.75)
%\psbezier{-}(1.25,3.25)(1.25,3.7125)(0.8333,4.0)(0,4.0)
%\psbezier{-}(1.75,3.25)(1.75,4.0)(1.166,4.25)(0,4.25)
\psbezier{-}(3.25,3.25)(3.25,3.625)(3.5,3.75)(4,3.75)
%\psbezier{-}(2.75,3.25)(2.75,3.8125)(3.1666,4.0)(4,4.0)
%\psbezier{-}(2.25,3.25)(2.25,4.0)(2.8333,4.25)(4,4.25)
%\psarc{-}(4,3.75){0.25}{90}{180}
%
\psarc{-}(-0.5,-3.25){0.25}{-90}{0}
\psbezier{-}(0.25,-3.25)(0.25,-3.625)(0.0,-3.75)(-0.5,-3.75)
\psbezier{-}(2.75,-3.25)(2.75,-3.625)(3.0,-3.75)(3.5,-3.75)
\psarc{-}(3.5,-3.25){0.25}{180}{-90}
%\psbezier{-}(0.25,3.75)()
\psset{linecolor=myc}
\psarc{-}(3.0,2.0){0.353553}{-45}{45}
\psarc{-}(3.0,1.0){0.353553}{-45}{45}
\psarc{-}(3.0,0.0){0.353553}{-45}{45}
\psarc{-}(3.0,-1.0){0.353553}{-45}{45}
\psarc{-}(3.0,-2.0){0.353553}{-45}{45}
\psarc{-}(0.5,0.5){0.353553}{-135}{-45}\psarc{-}(0.5,-0.5){0.353553}{45}{135}
\psarc{-}(1.0,0.0){0.353553}{-135}{-45}\psarc{-}(1.0,-1.0){0.353553}{45}{135}
\psarc{-}(1.5,-0.5){0.353553}{-135}{-45}\psarc{-}(1.5,-1.5){0.353553}{45}{135}%
\psarc{-}(2.0,-1.0){0.353553}{-135}{-45}\psarc{-}(2.0,-2.0){0.353553}{45}{135}
\psarc{-}(2.5,-1.5){0.353553}{-135}{-45}\psarc{-}(2.5,-2.5){0.353553}{45}{135}
\psarc{-}(2.5,-2.5){0.353553}{-45}{45}\psarc{-}(3.5,-2.5){0.353553}{135}{-135}
\end{pspicture} \right) 
 \\ & = v^{-\delta} \left(v^N \frac{S_{r+d}}{S_{2r+d}} + v^{-N} \frac{S_{r}}{S_{2r+d}}\right) v^{\delta} \left(v^{-N} \frac{S_{r+d}}{S_{2r+d-1}} + v^{N} \frac{S_{r}}{S_{2r+d-1}}\right) K_{d,r-1} 
 \\ & = \frac{4\cos^2(\mu N) S_rS_{r+d} + (2 C_{r+d/2}S_{d/2})^2}{S_{2r+d}S_{2r+d-1}} K_{d,r-1},
\end{align*}
where we have summed over configurations of the lower diagonal row and computed twist weights as explained earlier. The last term in (\ref{eq:dessinrecKdr}) can be seen to give $-(S_{2r+d}/S_{2r+d-1}) K_{d,r-1}$ by the same argument as the one given for $d=0$. A simple exercise using trigonometric functions shows that the two contributions sum up to equation (\ref{eq:recKdr}). Because the $v^\delta$s have cancelled out, $K_{d,r}$ is independent of the positions of the contractible loops of the original diagram. With the initial condition $K_{d,0} = 1$, we find
\begin{equation}
K_{d,r} = \prod_{k=1}^r \frac{S_k S_{k+d}}{S_{2k+d}S_{2k+d-1}} (4 \cos^2(\mu N) - 4 C^2_{k+d/2}) = \prod_{k=1}^r \frac{S_k}{S_{r+d+k}} (4 \cos^2(\mu N) - 4 C^2_{k+d/2}). \label{eq:finKdr}
\end{equation}

\subsection{The determinant of the Gram matrix}\label{sec:endofproof}

The purpose of the previous paragraph was to compute the constants $K_{d,r}$. Note that the result \eqref{eq:finKdr} for $K_{d,r}$ gives the expression \eqref{eq:finK0r} for $K_{0,r}$ if $d$ is set to zero and $2 \cos\mu N$ is replaced by $\alpha$.
\begin{Proposition}\label{thm:detGram} The determinant of the Gram matrix is
\begin{equation}\label{eq:enfinGNd}
\det  \mathcal{ \tilde G}_N^d =  \prod_{k=1}^{(N-d)/2} (\alpha^2 -4 C_{k+d/2}^2)^{\left(\begin{smallmatrix}N\\  \frac{N-d}2-k  \end{smallmatrix}
\right)}
=\prod_{k=1}^{(N-d)/2} \big(\langle k+d/2\rangle\langle -k-d/2)\rangle\big)^{\left(\begin{smallmatrix}N\\  \frac{N-d}2-k  \end{smallmatrix}\right)}
\end{equation}
for all $d\ge 0$. In the second form $\langle x\rangle = (-u^2)^xv^N-(-u^2)^{-x}v^{-N}$ and, in the first, $\alpha = 2 \cos(\mu N) = v^N+ v^{-N}$ for $d>0$.

\end{Proposition}
\noindent{\scshape Proof\ \ }
The result of the last section is
\begin{equation*}
K_{d,r} =  \frac{\prod_{i=1}^{r} S_{i}}{\prod_{j=r+d+1}^{2r+d}S_{j}}\prod_{k=1}^{r} (\alpha^2 - 4C_{k+d/2}^2).
\end{equation*}
Using (\ref{eq:detG0}) and (\ref{eq:Gint}), one finds
\begin{align*}
\det \tilde {\mathcal G}_N^d &=   \prod_{r=0}^{(N-d)/2-1} \prod_{k=1}^{(N-d)/2-r} \left({S_{d+2r+k+1}}/{S_k}\right)^{\dim V_N^{d+2r+2k}}   \\ & \hspace{3cm}\times  
\prod_{r=1}^{(N-d)/2} \left( \prod_{i=1}^r ({S_i}/{S_{r+d+i}} ) \prod_{k=1}^r\big(\alpha^2 - 4 C_{k+d/2}^2\big)\right)^{\dim V_N^{d+2r}}.
\end{align*}
The product of the factors $(\alpha^2-4C^2_{k+d/2})$ yields
\begin{align*}
\prod_{r=1}^{(N-d)/2} \prod_{k=1}^r(\alpha^2 - 4 C_{k+d/2}^2)^{\dim V_N^{d+2r}} &= \prod_{k=1}^{(N-d)/2} \prod_{r=k}^{(N-d)/2}(\alpha^2 - 4 C_{k+d/2}^2)^{\dim V_N^{d+2r}} \\ &=  \prod_{k=1}^{(N-d)/2} (\alpha^2 - 4 C_{k+d/2}^2)^{\sum_{r=k}^{(N-d)/2}\dim V_N^{d+2r}} \\ & =  \prod_{k=1}^{(N-d)/2} (\alpha^2 - 4 C_{k+d/2}^2)^{\dim \tilde V_N^{d+2k}}  =  \prod_{k=1}^{(N-d)/2} (\alpha^2 - 4 C_{k+d/2}^2)^{ \left(\begin{smallmatrix}N\\  (N-d)/2-k  \end{smallmatrix}\right)}. 
\end{align*}
One must therefore show that the rest is $1$. For this, the order of products is inverted in each of the subfactors:
\begin{align*}
\prod_{r=1}^{(N-d)/2} \prod_{i=1}^r S_i^{\dim V_N^{d+2r}} &=  \prod_{i=1}^{(N-d)/2}S_i^{\sum_{r=i}^{(N-d)/2}\dim V_N^{d+2r}}, \\
\prod_{r=1}^{(N-d)/2} \prod_{j=r+d+1}^{2r+d} S_j^{\dim V_N^{d+2r}} & = \left( \prod_{j=d+2}^{(N+d)/2} \prod_{r=\lceil (j-d)/2 \rceil}^{j-d-1} S_j^{\dim V_N^{d+2r}} \right) \left( \prod_{j=(N+d)/2+1}^{N} \prod_{r= \lceil (j-d)/2 \rceil}^{(N-d)/2} S_j^{\dim V_N^{d+2r}} \right) \\
& =  \left( \prod_{j=d+2}^{(N+d)/2} S_j^{\sum_{r=\lceil (j-d)/2 \rceil}^{j-d-1} \dim V_N^{d+2r}} \right) \left( \prod_{j=(N+d)/2+1}^{N}  S_j^{\sum_{r= \lceil (j-d)/2\rceil}^{(N-d)/2}\dim V_N^{d+2r}} \right), \\
\prod_{r=0}^{(N-d)/2-1} \prod_{k=1}^{(N-d-2r)/2} S_{d+2r+k+1}^{\dim V_N^{d+2r+2k}}  & = \prod_{r=0}^{(N-d)/2-1} \prod_{k'=2r+d+2}^{(N+d)/2+r+1} S_{k'}^{\dim V_N^{2k'-2r-d-2}} \\ 
& \hspace{-2cm}= \left( \prod_{k=d+2}^{(N+d)/2} \prod_{r=0}^{\lfloor(k-d-2)/2\rfloor} S_{k}^{\dim V_N^{2k-2r-d-2}}  \right)  \left( \prod_{k=(N+d)/2+1}^{N} \prod_{r=k-1-(N+d)/2}^{\lfloor(k-d-2)/2\rfloor} S_{k}^{\dim V_N^{2k-2r-d-2}} \right)\\
& \hspace{-2cm}= \left( \prod_{k=d+2}^{(N+d)/2} S_{k}^{ \sum_{r=0}^{\lfloor(k-d-2)/2\rfloor}\dim V_N^{2k-2r-d-2}}  \right)  \left( \prod_{k=(N+d)/2+1}^{N}  S_{k}^{\sum_{r=k-1-(N+d)/2}^{\lfloor(k-d-2)/2\rfloor}\dim V_N^{2k-2r-d-2}} \right)\\
& \hspace{-2cm}= \left( \prod_{k=d+2}^{(N+d)/2} S_{k}^{ \sum_{s= \lceil (k-d)/2 \rceil }^{k-d-1}\dim V_N^{d+2s}}  \right)  \left( \prod_{k=(N+d)/2+1}^{N}  S_{k}^{\sum_{s=\lceil (k-d)/2 \rceil}^{(N-d)/2}\dim V_N^{d+2s}} \right),\\
\prod_{r=0}^{(N-d)/2-1} \prod_{k=1}^{(N-d-2r)/2} S_k ^{\dim V_N^{d+2r+2k}}   &= \prod_{k=1}^{(N-d)/2} \prod_{r=0}^{(N-d)/2-k} S_k ^{\dim V_N^{d+2r+2k}} =  \prod_{k=1}^{(N-d)/2}  S_k ^{\sum_{r=0}^{(N-d)/2-k} \dim V_N^{d+2r+2k}} \\ 
& =  \prod_{k=1}^{(N-d)/2}  S_k ^{\sum_{s=k}^{(N-d)/2} \dim V_N^{d+2s}}. 
\end{align*}
It is then clear that everything cancels out. The second form in equation \eqref{eq:enfinGNd} follows from a straightforward trigonometric manipulation of the first.
\hfill$\square$

\medskip

The second form of the determinant \eqref{eq:enfinGNd} shows that its zeroes all lie on curves $(-u^2)^{2x}v^{2N}=1$ and that the structure of the action $\omega_d$ depends only on the twist parameter through its $2N$-th power. This is related to the observation made in section \ref{sec:extendedXXZ} that the two actions on the XXZ models defined in \cite{PasquierSaleur} and here are tied by their parameter as $e^{i\varphi}=v^{2N}$. The theorem \ref{thm:enfin} to be proven in the next section will go further in showing that these two actions are generically isomorphic to that of $\omega_d$ on $\tilde V_N^d$. 

\subsection{The determinant of $I_N^d(u,v)$}\label{sub:ind}

In theorem \ref{sec:homoGram}, we found that $\det I_N^d(u,v) \det I_N^d(u,v^{-1}) = \det \mathcal {\tilde G}_N^d$ with $\beta = u^2 + u^{-2}$, $\alpha = v^N + v^{-N}$ and the twist parameter is $v$. In this section, we show how to calculate $\det I_N^d(u,v)$.

We first introduce paths and the height function.  The set $P_y^N$ of paths with endpoint $y$ is the set of $\vec{x}=\{ x_1, x_2, ..., x_N\}$, where each step $x_i$ is either $+1$ or $-1$ and $\sum_{i=1}^N x_i =y $. The height $H(\vec x)$ of a path $\vec{x}$ is $H(\vec x) = \sum_{j=1}^N h_j$ with $h_j = \sum_{i=1}^j x_i$. Clearly $H(\vec{x}) = (N+1)y-\sum_{j=1}^N j x_j$. There are two natural bijections between $\tilde B_N^d$ on the one hand and $P^N_d$ or $P^N_{-d}$ on the other hand. (See figure \ref{fig:pathexample} for an example.) If $w \in \tilde B_N^d$, then $\mathcal B^\pm(w)$ is the path $\vec x=\{x_i,1\le i\le N\}$ where $x_i$ is $+1$ if a bubble starts at $i$ in $w$, and $-1$ if a bubble ends at $i$. Finally, if position $i$ is a defect, then $x_i=+1$ in $\mathcal B^+(w)$ and $-1$ in $\mathcal B^-(w)$. For $d=0$, $\mathcal B^+(w)= \mathcal B^-(w) \equiv \mathcal B(w)$. The fact that both $\mathcal B^\pm$ are bijections is straightforward.

\begin{figure}[h!]
\begin{center}
\begin{pspicture}(0,-0)(3.5,1.8)
\psset{linewidth=1pt}
\psset{unit=1.4}
\psdots[dotsize=0.08](0.1,0)(0.3,0.0)(0.5,0.0)(0.7,0.0)(0.9,0.0)(1.1,0.0)(1.3,0.0)(1.5,0.0)(1.7,0.0)(1.9,0.0)(2.1,0.0)(2.3,0.0)
\psset{linecolor=myc2}
\psline{-}(0.9,0)(0.9,0.4)
\psline{-}(1.9,0)(1.9,0.4)
\psarc{-}(1.4,0){0.1}{0}{180}
\psarc{-}(0.4,0){0.1}{0}{180}
\psarc{-}(0.0,0){0.1}{0}{90}
\psarc{-}(2.4,0){0.1}{90}{180}
\psbezier{-}(1.7,0)(1.7,0.3)(1.1,0.3)(1.1,0)
\psbezier{-}(2.1,0)(2.1,0.2)(2.35,0.25)(2.4,0.25)
\psbezier{-}(0.7,0)(0.7,0.3)(0.05,0.3)(0,0.25)
\end{pspicture}
\qquad \quad
\begin{pspicture}(0,-0)(2.5,-1.0)
\psset{linewidth=1pt}
\psline[arrows=->,arrows=->, linewidth= 0.5pt](-1,-0.75)(-0.375,-0.4375)
\rput(-0.75,-0.375){$_{\mathcal B^+}$}
\rput(-0.75,-1.375){$_{\mathcal B^-}$}
\psline[arrows=->,arrows=->, linewidth= 0.5pt](-1,-1.0)(-0.375,-1.3175)
\psline[arrows=->,arrows=->, linewidth= 0.5pt](0,0)(2.6,0)
\psline[arrows=->,arrows=->, linewidth= 0.5pt](0,-0.5)(0,0.5)
\psline{-}(0,0)(0.2,-0.2)(0.4,0.0)(0.6,-0.2)(0.8,-0.4)(1,-0.2)(1.2,0.0)(1.4,0.2)(1.6,0.0)(1.8,-0.2)(2,0.0)(2.2,0.2)(2.4,0.4)
\psdots[dotsize=0.10](0,0)(0.2,-0.2)(0.4,0.0)(0.6,-0.2)(0.8,-0.4)(1,-0.2)(1.2,0.0)(1.4,0.2)(1.6,0.0)(1.8,-0.2)(2,0.0)(2.2,0.2)(2.4,0.4)
\psline[arrows=->,arrows=->, linewidth= 0.5pt](0,-1.5)(2.6,-1.5)
\psline[arrows=->,arrows=->, linewidth= 0.5pt](0,-2.)(0,-1)
\psline{-}(0,-1.5)(0.2,-1.7)(0.4,-1.5)(0.6,-1.7)(0.8,-1.9)(1,-2.1)(1.2,-1.9)(1.4,-1.7)(1.6,-1.9)(1.8,-2.1)(2,-2.3)(2.2,-2.1)(2.4,-1.9)
\psdots[dotsize=0.10](0,-1.5)(0.2,-1.7)(0.4,-1.5)(0.6,-1.7)(0.8,-1.9)(1,-2.1)(1.2,-1.9)(1.4,-1.7)(1.6,-1.9)(1.8,-2.1)(2,-2.3)(2.2,-2.1)(2.4,-1.9)
\end{pspicture}
\vspace{1cm}
\end{center}
\caption{A depiction of the two bijections for a link state with $N=12$, $d=2$ and $r=2$. The link state $w$ has $\psi(w) = \{(2,3), (6,9), (7,8), (11,16), (12,13)\}$, $\sum_{(i,j) \in \psi(w)} j-i = 11$, $H(\mathcal B^+(w)) = -2$ and $H(\mathcal B^-(w)) = -24$.}\label{fig:pathexample}
\end{figure}
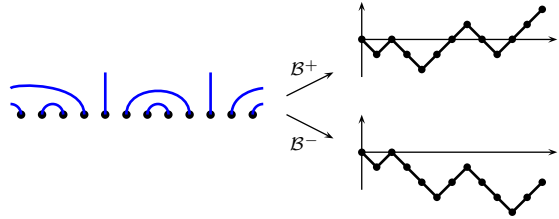

\begin{Lemme}
Let $w \in \tilde B_N^{d,r}$. Then $\sum_{(i,j) \in \psi(w)} (j-i) = \frac12\big(H(\mathcal B^+(w))+H(\mathcal B^-(w))\big) + Nr.$
\label{sec:usingH}
\end{Lemme}

\noindent{\scshape Proof\ \ } We start by considering the case $d=0$. For $r=0$, every $i$ and $j$ forming a pair $(i,j) \in \psi(w)$ are in the range $1,...,N$. A bubble that closes at position $+j$ (i.e.~$x_j = -1$) contributes $j$ to the sum $-\sum ix_i$, and one that opens at $j$ ($x_j = +1$) contributes $-j$. Therefore $H(\vec x)=\sum_{(i,j) \in \psi(w)} (j-i)$. For $r>0$, some bubbles close at positions $j\ge N$ and contribute $j$ to $\sum_{(i,j) \in \psi(w)} (j-i)$ but only $j-N$ to $H(w)$. For every one of these $r$ bubbles, we must add to  $-\sum_{j=1}^N jx_j$ a factor of $N$, which yields the correct result.

For $d>0$ and $r=0$, the $\sum_{(i,j) \in \psi(w)} (j-i)$ has contribution $j$ for a bubble that closes at $j$, $-j$ for a bubble that opens at $j$, and $0$ for a defect at position $j$. This sum is therefore $-\sum_j jy_j$ where $\vec y = (y_1, y_2, ..., y_N)$ is a not a path of $P^N_d$ or $P^N_{-d}$, but rather the $y_j$s are in $\{1,0,-1 \}$ and obtained from $w$ by setting $y_j = + 1$ or $-1$ if a bubble starts or ends at $j$, and $0$ if a defect is at position $j$. In fact, $\vec y =\frac12 (\mathcal B^+(w) + \mathcal B^-(w))$ does this exactly. Finally, generalizing to $r>0$ is no harder than in the case $d=0$.
\hfill$\square$
\begin{Theoreme}\label{prop:detInd}The determinant of the linear map $\tilde i_N^d$, expressed between the vectors of $\tilde B_N^d$ and the spin basis, is, up to a sign,
\begin{equation}\label{eq:detINd}
\det I_N^d(u,v) = \prod_{k=1}^{(N-d)/2} \Big(2i \sin\big(\Lambda (k+d/2) - \mu N\big)\Big)^{\left( \begin{smallmatrix}N\\  \frac{N-d}2-k  \end{smallmatrix} \right)}=
\prod_{k=1}^{(N-d)/2} \langle k+d/2\rangle^{\left( \begin{smallmatrix}N\\  \frac{N-d}2-k  \end{smallmatrix} \right)}
\end{equation}
where $v=e^{i\mu}$, $u = e^{i \lambda/2}$ and $\Lambda=\pi-\lambda$ and, in the second form, $\langle x\rangle = (-u^2)^xv^N-(-u^2)^{-x}v^{-N}$.
\end{Theoreme}
\noindent{\scshape Proof\ \ }
The Gram determinant allows for the following factorization:
\begin{equation*}\det \mathcal{ \tilde G}_N^d =  \prod_{k=1}^{(N-d)/2} \Big(2i \sin\big(\Lambda (k+d/2) - \mu N\big)\Big)^{\left( \begin{smallmatrix}N\\  \frac{N-d}2-k  \end{smallmatrix}\right)} \prod_{k=1}^{(N-d)/2} \Big(2i \sin\big(\Lambda (k+d/2) + \mu N\big)\Big)^{\left( \begin{smallmatrix}N\\  \frac{N-d}2-k  \end{smallmatrix}
\right)}
\end{equation*}
and the proposition is that $\det I_N^d(u,v)$ is the first product and $\det I_N^d(u,v^{-1})$ the second. This is compatible with the symmetry $v \leftrightarrow v^{-1}$ that corresponds to $\mu \leftrightarrow -\mu$. In fact, $\det I_N^d(u,v)$ is a polynomial in $u$ and $v$ (and their negative powers), and for each $\sin\big(\Lambda (k+d/2) + \mu N\big) \sin\big(\Lambda (k+d/2) - \mu N \big)$, one factor must contribute to $\det I_N^d(u,v)$ and the other to  $\det I_N^d(u,v^{-1})$. To understand how they are distributed, we look at the $u, v \rightarrow \infty$ limit. From (\ref{eq:homomorphisme}), we find, in this limit,
\begin{equation*}
\tilde i_N^d(w) \rightarrow \Big( \prod_{(i,j)\in \psi(w)} \hspace{-0.25cm}v^{j-i}u \  \sigma_j^-\Big) |0\rangle = u^{|\psi(w)|}v^{\sum_{(i,j) \in \psi(w)} (j-i)} \Big(\prod_{(i,j)\in \psi(w)} \hspace{-0.3cm} \sigma^-_j \Big) |0 \rangle.
\end{equation*}
It is easy to show that there is a one-to-one correspondence between states of the form $(\prod_{(i,j)\in \psi(w)}\sigma_j^-) |0 \rangle$ and the link states $w \in\tilde B_N^d$. Therefore, up to sign,
\begin{equation*}
\det I_N^d \underset{u,v \rightarrow \infty}{\longrightarrow} \prod_{w\in \tilde B_N^d} u^{|\psi(w)|}v^{\sum_{(i,j) \in \psi(w)} (j-i)} = u^{X_1}v^{X_2},
\end{equation*}
\begin{align}
\textrm{with} \qquad X_1 &= \sum_{w \in \tilde B_N^d}  |\psi(w)|  =  |\tilde B_N^d| \frac{N-d}2 =  \begin{pmatrix}N\\  \frac{N-d}2  \end{pmatrix}  \frac{N-d}2 
\label{eq:I1}\\
\textrm{and\ } \qquad  X_2 & = \sum_{w \in \tilde B_N^d}  \sum_{(i,j) \in \psi(w)} (j-i) = \sum_{r=0}^{(N-d)/2}\sum_{w \in \tilde B_N^{d,r}} \Big( {\textstyle\frac12}\big(H(\mathcal B^+(w))+H( \mathcal B^-(w))\big) + Nr\Big)  \nonumber \\
&= \sum_{r=0}^{(N-d)/2} N r |\tilde B_N^{d,r}|  = \sum_{r=0}^{(N-d)/2} N r | B_N^{d+2r}| = \sum_{r=0}^{(N-d)/2} N r \left( \begin{pmatrix} N \\ \frac{N-d}2-r \end{pmatrix} -  \begin{pmatrix} N \\ \frac{N-d}2-r -1 \end{pmatrix} \right) \nonumber \\
& =  \sum_{s=0}^{(N-d)/2-1} N \begin{pmatrix} N \\ s \end{pmatrix} \label{eq:I2}
\end{align}
where, for $X_2$, the second equality follows from lemma \ref{sec:usingH}. For the third, we used the fact that
\begin{equation}
\sum_{w \in \tilde B_N^d} \left(H(\mathcal B^+(w)) + H(\mathcal B^-(w))\right)= 0.
\label{eq:sommechemins}\end{equation}
Indeed, in terms of paths, this sum can be rewritten as $\sum_{\vec{x}\in P^N_d \cup P^N_{-d}} H(\vec{x})$. The sum is thus over all paths using edges drawn in Figure \ref{fig:interval}, a step in the north-east (south-east) direction corresponding to a positive $x_i$ (negative $x_i$). Paths in $P^N_d$ reach the upper dot, those in $P^N_{-d}$ the lower one. Because the shaded domain is symmetric under a horizontal mirror, each path $\vec x\in P^N_d$ has a partner $-\vec x\in P^N_{-d}$ such that $H(\vec x)+H(-\vec x)=0$ and the sum is $0$.
\begin{figure}[h!]
\begin{center}
\psset{unit=2}
\begin{pspicture}(0,-2.5)(3.5,2.5)
\psset{linewidth=1pt}
\pspolygon*[linestyle=dashed,linecolor=lightgray](0,0)(2,2)(3,1)(2,0)(3,-1)(2,-2)(0,0)
\psline[arrows=->,arrows=->](0,0)(3.5,0)
\psline[arrows=->,arrows=->](0,-2.4)(0,2.4)
\psline{-}(3,-0.05)(3,0.05)
\psline{-}(-0.05,1)(0.05,1)
\psline{-}(-0.05,2)(0.05,2)
\psline{-}(-0.05,-1)(0.05,-1)
\psline{-}(-0.05,-2)(0.05,-2)
\rput(3,-0.2){$N$}
\rput(-0.3,1){$d$}
\rput(-0.3,2){$\frac{N-d}2$}
\psdots[linewidth=0.02](3,1)(3,-1)
\psset{linewidth=0.5pt}
\psline(0,0)(2,2)(3,1)(2,0)(3,-1)(2,-2)(0,0)
\psline(0.125,-0.125)(2.125,1.875)
\psline(0.25,-0.25)(2.25,1.75)
\psline(0.375,-0.375)(2.375,1.625)
\psline(0.5,-0.5)(2.5,1.5)
\psline(0.625,-0.625)(2.625,1.375)
\psline(0.75,-0.75)(2.75,1.25)
\psline(0.875,-0.875)(2.875,1.125)
\psline(1,-1)(2,0)
\psline(1.125,-1.125)(2.125,-0.125)
\psline(1.25,-1.25)(2.25,-0.25)
\psline(1.375,-1.375)(2.375,-0.375)
\psline(1.5,-1.5)(2.5,-0.5)
\psline(1.625,-1.625)(2.625,-0.625)
\psline(1.75,-1.75)(2.75,-0.75)
\psline(1.875,-1.875)(2.875,-0.875)
\psline(0.125,0.125)(2.125,-1.875)
\psline(0.25,0.25)(2.25,-1.75)
\psline(0.375,0.375)(2.375,-1.625)
\psline(0.5,0.5)(2.5,-1.5)
\psline(0.625,0.625)(2.625,-1.375)
\psline(0.75,0.75)(2.75,-1.25)
\psline(0.875,0.875)(2.875,-1.125)
\psline(1,1)(2,0)
\psline(1.125,1.125)(2.125,0.125)
\psline(1.25,1.25)(2.25,0.25)
\psline(1.375,1.375)(2.375,0.375)
\psline(1.5,1.5)(2.5,0.5)
\psline(1.625,1.625)(2.625,0.625)
\psline(1.75,1.75)(2.75,0.75)
\psline(1.875,1.875)(2.875,0.875)
\psset{linecolor=myc2}
\psset{linecolor=myc}
\end{pspicture}
\caption{The domain containing all paths in the sum (\ref{eq:sommechemins}).}\label{fig:interval}
\end{center}
\end{figure}
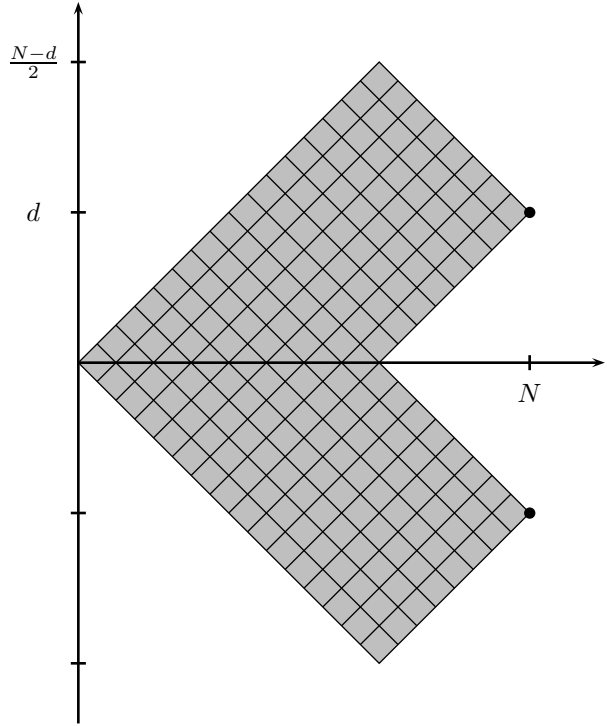
We now compare this result with the limiting behavior of the proposed $\det I^d_N(u,v)$. For $u, v \rightarrow \infty$, 
\begin{equation*}
\det I_N^d(u,v) = \prod_{k=1}^{(N-d)/2} (u^{2k+d}v^{N} -(-1)^d u^{-2k-d}v^{-N})^{\left( \begin{smallmatrix}N\\  \frac{N-d}2-k  \end{smallmatrix} \right)} \underset{u,v \rightarrow \infty}{\longrightarrow} u^{X_1'} v^{X_2'},
\end{equation*}
up to a sign. The constants $X_1'$ and $X_2'$ are
\begin{equation*}
 X_1' = \sum_{k=1}^{(N-d)/2} (2 k+d)  \begin{pmatrix}N\\  \frac{N-d}2-k  \end{pmatrix} , \qquad
X_2' =  \sum_{k=1}^{(N-d)/2} N \begin{pmatrix}N\\  \frac{N-d}2-k  \end{pmatrix} = \sum_{s=0}^{(N-d)/2-1} N  \begin{pmatrix}N\\  s  \end{pmatrix} .
\end{equation*}
$X_2'$ already coincides with $X_2$ and a simple exercise with combinatorial coefficients shows that $X_1$ and $X_1'$ also do. Any other choice of distribution of the factors $\sin\big(\Lambda(k+d/2) \pm \mu N\big)$ between $I_N^d(u,v)$ and $I_N^d(u,v^{-1})$ would have changed either $X_1'$, $X_2'$ or both, and the choice in \eqref{eq:detINd} is the only possible one.
\hfill$\square$ \\

Let $u=e^{i(\pi - \Lambda)/2}$ and $v=e^{i\mu}$ be fixed. A pair $(N,d)$ is {\em critical} if it belongs to
$$\left\{ (N,d)\,|\,  \sin \big(\Lambda(k+d/2)-\mu N\big)=0  \textrm{\ for some }k, 1\le k\le (N-d)/2\right\}.$$
Clearly the criticality of $(N,d)$ depends on $\mu$ and $\Lambda$ and the matrix $I_N^d$ (or the map $\tilde i_N^d$) is singular if and only if $(N,d)$ is critical. 
Equivalently, for $N$ and $d$ fixed, the map $\tilde i_N^d(u,v)$ is singular if and only if the point $(u,v)$ lies on one of the {\em critical curves} defined by $\langle k+d/2\rangle= (-u^2)^{k+d/2}v^N-(-u^2)^{-(k+d/2)}v^{-N}=0$, $1\le k\le (N-d)/2$.

\begin{Theoreme}\label{thm:enfin} Let $u=e^{i\lambda/2}$ and $v=e^{i\mu}$ be fixed. The map $\tau:\mathcal EPTL_N(\beta,\alpha)\rightarrow \textrm{End}\big((\mathbb C^2)^{\otimes N}\big)$ defined by $e_i\mapsto \bar e_i$ and $\Omega^{\pm 1}\mapsto\bar\Omega^{\pm 1}$ is a representation of $\mathcal EPTL_N(\beta,\alpha)$ with $\beta=u^2+u^{-2}$ and $\alpha=v^N+v^{-N}$. Moreover, if $(N,d)$ is not critical, then $\tilde i_N^d:\tilde V_N^d\rightarrow \left.(\mathbb C^2)^{\otimes N}\right|_{S^z=d/2}$ is an isomorphism between modules over $\mathcal EPTL_N(\beta,\alpha)$. \end{Theoreme}

\noindent{\scshape Proof\ \ } We left out in section \ref{sec:XXZprime} the question of whether the matrices $\bar e_i$s and $\bar\Omega^{\pm 1}$ verified equations \eqref{eq:nontrivial} and \eqref{eq:conjuOmega}. Clearly the matrix elements of $\left((\bar\Omega^{\pm 1}\bar e_N)^{N-1}-\bar\Omega^{\pm N}(\bar\Omega^{\pm 1}\bar e_N)\right)$ and $\left(\bar E\bar\Omega^{\pm 1}\bar E-(v^N+v^{-N})\bar E\right)$ are polynomials in $u, u^{-1}, v$ and $v^{-1}$. For all non-critical values of $(N,d)$, these matrices are zero since then $\det I_N^d\neq 0$ and $\bar\Omega^{\pm 1}=\tilde i_N^d \circ \Omega^{\pm 1}\circ(\tilde i_N^d)^{-1}$  on $\left.(\mathbb C^2)^{\otimes N}\right|_{S^z=d/2}$.  (See the end of section \ref{sec:linkAndSpin}.) Since the critical conditions $\Lambda(k+d/2)-\mu N \in \pi\mathbb Z$ represent a finite number of surfaces in the parameter space $(\mathbb C^\times)^2$ of $(u,v)$, then these matrices with polynomial elements vanish everywhere. So equations \eqref{eq:nontrivial} and \eqref{eq:conjuOmega} and all other defining relations are verified by the $\bar e_i$s and $\bar\Omega^{\pm 1}$. The fact that $\tilde i_N^d$ is an isomorphism of modules follows from the previous discussion and theorem \ref{prop:detInd}.\hfill$\square$

%%%%%%%%%%%%%%%%
%
\section{Conclusion}
%
%%%%%%%%%%%%%%%%

Two representations of the enlarged periodic Temperley-Lieb algebra $\mathcal EPTL_N(\beta, \alpha)$ were studied in this paper: the link representation $\omega_d$ for the loop models and the representation $\tau$ for the XXZ spin chain. The representation $\omega_d$ plays a role in Fortuin-Kasteleyn models and logarithmic minimal models, depends upon a twist parameter $v$ and is labeled by the number of defects $d$. The representation $\tau$ of the XXZ models is constructed from spin modules, depends upon two parameters $u$ and $v$ and is a direct sum of smaller representations acting on given eigenspaces of the total spin $S^z$. The main result of this paper is the construction of an intertwiner $\tilde i_N^d$ between the representation $\omega_d$ and the restriction of the 
representation $\tau$ to $S^z = \frac d 2$, with $\beta = u^2 + u^{-2}$ and $\alpha = v^{N}+ v^{-N}$.

Properties of $\tilde i_N^d$ were studied through its relation with the Gram
matrix $\tilde {\mathcal G}_N^d$, given by theorem \ref{sec:homoGram}. The critical condition for the pair $(N,d)$, namely that $\sin (\Lambda(k+d/2)-\mu N)$ be zero for some $k$, can be used to define criticality for pairs $(u,v)$ if the pair $(N,d)$ is fixed. The determinant of the transformation is
then found to be non-zero except on a finite number of critical curves in the $(u,v)$ plane. If $(u,v)$ is not critical, $\tilde i_N^d$ is an isomorphism between link and spin modules and any element of $\mathcal EPTL_N(\beta, \alpha)$ will have identical eigenvalues in both representations. 
The solutions of algebraic equations depending on a complex parameter are known to be holomorphic on $\mathbb C$ except at a finite number of points \cite{Kato}. This analyticity in one parameter (say $u$, the parameter $v$ being kept fixed) allows one to extend this coincidence of eigenvalues to the critical cases. The intertwiner $\tilde i_N^d$ may also be able to probe the Jordan structure of the transfer matrices in the representation $\omega_d$. 

In lattice regularizations of logarithmic minimal models, the transfer matrix $T_N(\lambda, \mu)$ and Hamiltonian $\mathcal H$ are expected to have non-trivial Jordan structures. One way to identify logarithmic theories is thus to search for Jordan cells arising in finite lattice models. For the representations $\omega_d$, our calculation of the determinant of the intertwiner gives some
insight on the possible existence of Jordan cells for $\omega_d (\mathcal H)$. If $\alpha$, $\beta$ and $v$ are such that the pair $(u,v)$ is not critical, the matrix $\omega_d(c)$ and the restriction of the matrix $\tau(c)$ to the sector $S^z = \frac d 2$ are related by a similarity transformation for any $c \in \mathcal EPTL_N(\beta, \alpha)$. Because $H = \tau(\mathcal H)$ is hermitian and diagonalizable if $u$ and $v$ are on the unit circle, $\omega_d(\mathcal H)$ may only have Jordan cells if the the pair $(u,v)$ is critical. One may then ask in which representations and for which values of $\alpha, \beta$ and $v$ the Jordan cells appear. We will return to this question in \cite{AMDYSAinprep}.

\section*{Acknowledgements} 

We would like to thank David Ridout and J\o{}rgen Rasmusen for helpful discussions and encouragement, and Hubert Saleur for bringing to our attention his work \cite{MartinSaleur} with Paul Martin. AMD holds a scholarship and YSA a grant of the Canadian Natural Sciences and Engineering Research Council. This support is gratefully acknowledged.

%%%%%%%%%%%%%%%%%%%%%
%
 

\begin{thebibliography}{99} 
%
%%%%%%%%%%%%%%%%%%%%%


\bibitem{Onsager} L.~Onsager, {\em A 2d model with an order-disorder transition}, Phys. Rev. {\bf 65} (1944) 117--149.

 \bibitem{TL} H.N.V.~Temperley, E.H.~Lieb, {\em Relations between the `percolation' and `colouring' problem and other graph-theoretical problems associated with regular planar lattices: some exact results for the `percolation' problem}, Proc. Roy. Soc. {\bf A322} (1971) 251--280. 
 
\bibitem{Baxter}  R.J.~Baxter, {\em Exactly solved models in statistical mechanics}, Academic Press (1982) 486 p. 
 
\bibitem{Lieb} E.H.~Lieb, {\em Solution of the dimer problem by the transfer matrix method}, J. Math. Phys. {\bf 8} (1967) 2339--2341.

\bibitem{PRZ}  P.A.~Pearce, J.~Rasmussen, J.-B.~Zuber, {\em Logarithmic minimal models}, J. Stat. Mech. {\bf P11017} (2006), arXiv:hep-th/0607232.

\bibitem{JacobsenSaleur} J.L.~Jacobsen, H.~Saleur, {\em Combinatorial aspects of boundary loop models}, J. Stat. Mech. {\bf P01021} (2008) 37 p., arXiv:math-th/07090912.

\bibitem{AMDSA} A.~Morin-Duchesne, Y.~Saint-Aubin, {\em The Jordan Structure of Two Dimensional Loop Models}, J. Stat. Mech. {\bf P04007}  (2011) 65 p., arXiv:1101.2885v4 [math-ph].

\bibitem{PRV} P.A.~Pearce, J.~Rasmussen, S.P.~Villani, {\em Solvable Critical Dense Polymers on the Cylinder}, J. Stat. Mech. {\bf P02010}  (2010) 43 p., arXiv:hep-th/0910.4444v3.

\bibitem{MartinSaleur} P.~Martin, H.~Saleur, {\em On an algebraic approach to higher dimensional statistical mechanics}, Commun.~Math.~Phys. {\bf 158} (1993) 155--190, arXiv:hep-th/9208061v1.

\bibitem{Lehrer} J.J.~Graham, G.I.~Lehrer, {\em The representation theory of affine Temperley-Lieb algebras}, Enseign. Math. {\bf 44} (1998) 173--218.

\bibitem{Greenseul} R.M.~Green, {\em On representations of affine Temperley-Lieb algebras}, Algebras and Modules II, CMS Conference Proceedings {\bf 24}, Amer. Math. Soc. (1998) 365--386.

\bibitem{ErdmannGreen} K.~Erdmann, R.M.~Green, {\em On representations of affine Temperley-Lieb algebras, II}, Pac.~Jour.~Math {\bf 191} (1999) 243--274, arXiv:math/9811017v1 [math.RT].

\bibitem{Pinson} T. H. Pinson, {\em Critical Percolation on the Torus}, J. Stat. Phys. {\bf 75} (1994) 1167--1177.

\bibitem{Arguin} L.-P.~Arguin, {\em Homology of Fortuin-Kasteleyn clusters of Potts models on the torus}, J. Stat. Phys. {\bf 109} (2002) 301--310, arXiv:hep-th/0111193.

\bibitem{RichardJacobsen} J.-F.~Richard, J.L.~Jacobsen, {\em Character decomposition of Potts model partition functions. I. Cyclic geometry}, Nucl. Phys.~{\bf 750} (2006) 250--264, arXiv:math-ph/0605016v1; {\em Eigenvalue amplitudes of the Potts model on a torus}, Nucl. Phys.~{\bf 769} (2007) 256--274, arXiv:math-ph/0608055v1.

\bibitem{AMDtoutseul} A.~Morin-Duchesne, {\em A proof of selection rules for critical dense polymers}, J. Phys. A: Math. Theor. {\bf 44 495003} (2011) 32 p., arXiv:math-ph/1109.6397.

\bibitem{GLcell} J.J.~Graham, G.I.~Lehrer, {\em Cellular algebras}, Inv.~Math.~{\bf 123} (1996) 1--34.

\bibitem{Westbury} B.W. Westbury, {\em The representation theory of the Temperley-Lieb algebras}, Math. Z. {\bf 219} (1995) 539--565.

\bibitem{DRYSA} D.~Ridout, Y.~Saint-Aubin, {\em The structure of the Temperley-Lieb algebra through its standard modules} (2012) arXiv:1204.4505v2 [math-ph], submitted for publication.

\bibitem{PasquierSaleur} V.~Pasquier, H.~Saleur, {\em Common structures between finite systems and conformal field theories through quantum groups}, Nucl.~Phys. B {\bf 330} (1990) 523--556.

\bibitem{GreenFan} R.M.~Green, C.K.~Fan, {\em On the affine Temperley-Lieb algebras}, J. London Math. Soc. {\bf 2 60} (1999) 366--380, arXiv:q-alg/9706003v1.

\bibitem{BPOB} R.E.~Behrend, P.A.~Pearce, D.L.~O'Brien, {\em Interaction-Round-A-Face Models with Fixed Boundary Conditions: The ABF Fusion Hierarchy}, J.~Stat.~Phys. {\bf 84} (1995) 1--48, arXiv:cond-mat/9511081.

\bibitem{BKW} R.J.~Baxter, S.B.~Kelland, F.Y.~Wu, {\em Equivalence of the Potts model or Whitney polynomial with an ice-type model}, J.~Phys.~A: Math.~Gen. {\bf 9} (1976) 397--406.

\bibitem{Martin} P.~Martin, {\em Potts Models and Related Problems in Statistical Mechanics}, World Scientific  (1991) 360 p.

\bibitem{AMDYSA0} A.~Morin-Duchesne, Y.~Saint-Aubin, {\em Critical exponents for the homology of Fortuin-Kasteleyn clusters on a torus}, Phys.~Rev. {\bf E80}, 021130 (2009) 13 p., arXiv:0812.2925v2 [cond-mat.stat-mech].

\bibitem{PDF} P.~Di Francesco, {\em Meander Determinants}, Commun. Math. Phys. {\bf 191} (1998) 543--583, arXiv:hep-th/9612026v1.

\bibitem{ChenPRZ} Qi Chen, J.H.~Przytycki, {\em The Gram determinant of the type B Termperley-Lieb algebra}, Adv.~in Appl.~Math. {\bf 43} (2008) 156--161, arXiv:0802.1083v2 [math.GT].

\bibitem{GoodmanWenzl} F.M.~Goodman, H.~Wenzl, {\em The Temperley-Lieb algebra at roots of unity}, Pacific J. Math. {\bf 161} (1993) 307--334.

\bibitem{Wenzl} H.~Wenzl, {\em Hecke algebras of type $A_n$ and subfactors}, Inv.~Math. {\bf 92} (1988) 349--383.

\bibitem{KauffmanLins} L.H.~Kauffman, S.~Lins, {\em Temperley-Lieb recoupling theory and invariant of 3-manifolds}, Princeton Univ.~Press (1994) 312 p.

\bibitem{Kato} T.~Kato, {\em Perturbation Theory for Linear Operators}, Springer (1966) 619 p.

\bibitem{AMDYSAinprep} A.~Morin-Duchesne, Y.~Saint-Aubin, {\em Jordan cells of periodic loop models} (2013)
in preparation.

\end{thebibliography}
\end{document}